\documentclass[longauth]{aaEC}

\usepackage{graphicx}
\usepackage{xcolor}
\usepackage{txfonts}
\usepackage{euclid}
\usepackage{lastpage}
\usepackage{placeins}

\usepackage[pdfencoding=auto,psdextra]{hyperref}
\hypersetup{
    colorlinks=true,
    linkcolor=blue,
    filecolor=magenta,      
    urlcolor=blue,
    citecolor=blue
}
\usepackage[switch, modulo]{lineno}

\usepackage[nameinlink,capitalise]{cleveref}
\crefname{section}{Sect.}{Sects.}
\Crefname{section}{Section}{Sections}
\crefname{figure}{Fig.}{Figs.}
\Crefname{figure}{Figure}{Figures}
\crefname{equation}{Eq.}{Eqs.}
\Crefname{equation}{Equation}{Equations}
\crefname{table}{Table}{Tables}
\crefname{appendix}{Appendix}{Appendices}

\begin{document} 

\title{Euclid Quick Data Release (Q1)}
\subtitle{Extending the quest for little red dots to $z<4$}   

\titlerunning{Little red dots with $z<4$ from \Euclid}

\newcommand{\orcid}[1]{} 

\author{Euclid Collaboration: L.~Bisigello\orcid{0000-0003-0492-4924}\thanks{\email{laura.bisigello@inaf.it}}\inst{\ref{aff1}}
\and G.~Rodighiero\orcid{0000-0002-9415-2296}\inst{\ref{aff2},\ref{aff1}}
\and S.~Fotopoulou\orcid{0000-0002-9686-254X}\inst{\ref{aff3}}
\and F.~Ricci\orcid{0000-0001-5742-5980}\inst{\ref{aff4},\ref{aff5}}
\and K.~Jahnke\orcid{0000-0003-3804-2137}\inst{\ref{aff6}}
\and A.~Feltre\orcid{0000-0001-6865-2871}\inst{\ref{aff7}}
\and V.~Allevato\orcid{0000-0001-7232-5152}\inst{\ref{aff8}}
\and F.~Shankar\orcid{0000-0001-8973-5051}\inst{\ref{aff9}}
\and P.~Cassata\orcid{0000-0002-6716-4400}\inst{\ref{aff2},\ref{aff1}}
\and E.~Dalla~Bont\`a\orcid{0000-0001-9931-8681}\inst{\ref{aff2},\ref{aff1},\ref{aff10}}
\and G.~Gandolfi\orcid{0000-0003-3248-5666}\inst{\ref{aff11},\ref{aff1}}
\and G.~Girardi\orcid{0009-0005-6156-4066}\inst{\ref{aff2},\ref{aff1}}
\and M.~Giulietti\orcid{0000-0002-1847-4496}\inst{\ref{aff12}}
\and A.~Grazian\orcid{0000-0002-5688-0663}\inst{\ref{aff1}}
\and C.~C.~Lovell\orcid{0000-0001-7964-5933}\inst{\ref{aff13}}
\and R.~Maiolino\orcid{0000-0002-4985-3819}\inst{\ref{aff14}}
\and T.~Matamoro~Zatarain\orcid{0009-0007-2976-293X}\inst{\ref{aff3}}
\and M.~Mezcua\orcid{0000-0003-4440-259X}\inst{\ref{aff15},\ref{aff16}}
\and I.~Prandoni\orcid{0000-0001-9680-7092}\inst{\ref{aff12}}
\and D.~Roberts\orcid{0009-0009-7662-0445}\inst{\ref{aff9}}
\and W.~Roster\orcid{0000-0002-9149-6528}\inst{\ref{aff17}}
\and M.~Salvato\orcid{0000-0001-7116-9303}\inst{\ref{aff17}}
\and M.~Siudek\orcid{0000-0002-2949-2155}\inst{\ref{aff18},\ref{aff15}}
\and F.~Tarsitano\orcid{0000-0002-5919-0238}\inst{\ref{aff19}}
\and Y.~Toba\orcid{0000-0002-3531-7863}\inst{\ref{aff20},\ref{aff21},\ref{aff22}}
\and A.~Vietri\orcid{0000-0003-4032-0853}\inst{\ref{aff2}}
\and L.~Wang\orcid{0000-0002-6736-9158}\inst{\ref{aff23},\ref{aff24}}
\and G.~Zamorani\orcid{0000-0002-2318-301X}\inst{\ref{aff25}}
\and M.~Baes\orcid{0000-0002-3930-2757}\inst{\ref{aff26}}
\and S.~Belladitta\orcid{0000-0003-4747-4484}\inst{\ref{aff6},\ref{aff25}}
\and A.~Nersesian\orcid{0000-0001-6843-409X}\inst{\ref{aff27},\ref{aff26}}
\and L.~Spinoglio\orcid{0000-0001-8840-1551}\inst{\ref{aff28}}
\and X.~Lopez~Lopez\orcid{0009-0008-5194-5908}\inst{\ref{aff29},\ref{aff25}}
\and N.~Aghanim\orcid{0000-0002-6688-8992}\inst{\ref{aff30}}
\and B.~Altieri\orcid{0000-0003-3936-0284}\inst{\ref{aff31}}
\and A.~Amara\inst{\ref{aff32}}
\and S.~Andreon\orcid{0000-0002-2041-8784}\inst{\ref{aff33}}
\and N.~Auricchio\orcid{0000-0003-4444-8651}\inst{\ref{aff25}}
\and H.~Aussel\orcid{0000-0002-1371-5705}\inst{\ref{aff34}}
\and C.~Baccigalupi\orcid{0000-0002-8211-1630}\inst{\ref{aff35},\ref{aff36},\ref{aff37},\ref{aff38}}
\and M.~Baldi\orcid{0000-0003-4145-1943}\inst{\ref{aff39},\ref{aff25},\ref{aff40}}
\and A.~Balestra\orcid{0000-0002-6967-261X}\inst{\ref{aff1}}
\and S.~Bardelli\orcid{0000-0002-8900-0298}\inst{\ref{aff25}}
\and A.~Basset\inst{\ref{aff41}}
\and P.~Battaglia\orcid{0000-0002-7337-5909}\inst{\ref{aff25}}
\and R.~Bender\orcid{0000-0001-7179-0626}\inst{\ref{aff17},\ref{aff42}}
\and A.~Biviano\orcid{0000-0002-0857-0732}\inst{\ref{aff36},\ref{aff35}}
\and A.~Bonchi\orcid{0000-0002-2667-5482}\inst{\ref{aff43}}
\and E.~Branchini\orcid{0000-0002-0808-6908}\inst{\ref{aff44},\ref{aff45},\ref{aff33}}
\and M.~Brescia\orcid{0000-0001-9506-5680}\inst{\ref{aff46},\ref{aff8}}
\and J.~Brinchmann\orcid{0000-0003-4359-8797}\inst{\ref{aff47},\ref{aff48}}
\and S.~Camera\orcid{0000-0003-3399-3574}\inst{\ref{aff49},\ref{aff50},\ref{aff51}}
\and G.~Ca\~nas-Herrera\orcid{0000-0003-2796-2149}\inst{\ref{aff52},\ref{aff53},\ref{aff54}}
\and V.~Capobianco\orcid{0000-0002-3309-7692}\inst{\ref{aff51}}
\and C.~Carbone\orcid{0000-0003-0125-3563}\inst{\ref{aff55}}
\and J.~Carretero\orcid{0000-0002-3130-0204}\inst{\ref{aff56},\ref{aff57}}
\and S.~Casas\orcid{0000-0002-4751-5138}\inst{\ref{aff58}}
\and M.~Castellano\orcid{0000-0001-9875-8263}\inst{\ref{aff5}}
\and G.~Castignani\orcid{0000-0001-6831-0687}\inst{\ref{aff25}}
\and S.~Cavuoti\orcid{0000-0002-3787-4196}\inst{\ref{aff8},\ref{aff59}}
\and K.~C.~Chambers\orcid{0000-0001-6965-7789}\inst{\ref{aff60}}
\and A.~Cimatti\inst{\ref{aff61}}
\and C.~Colodro-Conde\inst{\ref{aff62}}
\and G.~Congedo\orcid{0000-0003-2508-0046}\inst{\ref{aff63}}
\and C.~J.~Conselice\orcid{0000-0003-1949-7638}\inst{\ref{aff64}}
\and L.~Conversi\orcid{0000-0002-6710-8476}\inst{\ref{aff65},\ref{aff31}}
\and Y.~Copin\orcid{0000-0002-5317-7518}\inst{\ref{aff66}}
\and F.~Courbin\orcid{0000-0003-0758-6510}\inst{\ref{aff67},\ref{aff68}}
\and H.~M.~Courtois\orcid{0000-0003-0509-1776}\inst{\ref{aff69}}
\and M.~Cropper\orcid{0000-0003-4571-9468}\inst{\ref{aff70}}
\and A.~Da~Silva\orcid{0000-0002-6385-1609}\inst{\ref{aff71},\ref{aff72}}
\and H.~Degaudenzi\orcid{0000-0002-5887-6799}\inst{\ref{aff19}}
\and G.~De~Lucia\orcid{0000-0002-6220-9104}\inst{\ref{aff36}}
\and A.~M.~Di~Giorgio\orcid{0000-0002-4767-2360}\inst{\ref{aff28}}
\and C.~Dolding\orcid{0009-0003-7199-6108}\inst{\ref{aff70}}
\and H.~Dole\orcid{0000-0002-9767-3839}\inst{\ref{aff30}}
\and F.~Dubath\orcid{0000-0002-6533-2810}\inst{\ref{aff19}}
\and C.~A.~J.~Duncan\orcid{0009-0003-3573-0791}\inst{\ref{aff64}}
\and X.~Dupac\inst{\ref{aff31}}
\and S.~Dusini\orcid{0000-0002-1128-0664}\inst{\ref{aff73}}
\and A.~Ealet\orcid{0000-0003-3070-014X}\inst{\ref{aff66}}
\and S.~Escoffier\orcid{0000-0002-2847-7498}\inst{\ref{aff74}}
\and M.~Farina\orcid{0000-0002-3089-7846}\inst{\ref{aff28}}
\and R.~Farinelli\inst{\ref{aff25}}
\and F.~Faustini\orcid{0000-0001-6274-5145}\inst{\ref{aff43},\ref{aff5}}
\and S.~Ferriol\inst{\ref{aff66}}
\and F.~Finelli\orcid{0000-0002-6694-3269}\inst{\ref{aff25},\ref{aff75}}
\and M.~Frailis\orcid{0000-0002-7400-2135}\inst{\ref{aff36}}
\and E.~Franceschi\orcid{0000-0002-0585-6591}\inst{\ref{aff25}}
\and S.~Galeotta\orcid{0000-0002-3748-5115}\inst{\ref{aff36}}
\and K.~George\orcid{0000-0002-1734-8455}\inst{\ref{aff42}}
\and W.~Gillard\orcid{0000-0003-4744-9748}\inst{\ref{aff74}}
\and B.~Gillis\orcid{0000-0002-4478-1270}\inst{\ref{aff63}}
\and C.~Giocoli\orcid{0000-0002-9590-7961}\inst{\ref{aff25},\ref{aff40}}
\and P.~G\'omez-Alvarez\orcid{0000-0002-8594-5358}\inst{\ref{aff76},\ref{aff31}}
\and J.~Gracia-Carpio\inst{\ref{aff17}}
\and B.~R.~Granett\orcid{0000-0003-2694-9284}\inst{\ref{aff33}}
\and F.~Grupp\inst{\ref{aff17},\ref{aff42}}
\and S.~Gwyn\orcid{0000-0001-8221-8406}\inst{\ref{aff77}}
\and S.~V.~H.~Haugan\orcid{0000-0001-9648-7260}\inst{\ref{aff78}}
\and H.~Hoekstra\orcid{0000-0002-0641-3231}\inst{\ref{aff54}}
\and W.~Holmes\inst{\ref{aff79}}
\and I.~M.~Hook\orcid{0000-0002-2960-978X}\inst{\ref{aff80}}
\and F.~Hormuth\inst{\ref{aff81}}
\and A.~Hornstrup\orcid{0000-0002-3363-0936}\inst{\ref{aff82},\ref{aff83}}
\and P.~Hudelot\inst{\ref{aff84}}
\and M.~Jhabvala\inst{\ref{aff85}}
\and E.~Keih\"anen\orcid{0000-0003-1804-7715}\inst{\ref{aff86}}
\and S.~Kermiche\orcid{0000-0002-0302-5735}\inst{\ref{aff74}}
\and A.~Kiessling\orcid{0000-0002-2590-1273}\inst{\ref{aff79}}
\and B.~Kubik\orcid{0009-0006-5823-4880}\inst{\ref{aff66}}
\and M.~K\"ummel\orcid{0000-0003-2791-2117}\inst{\ref{aff42}}
\and M.~Kunz\orcid{0000-0002-3052-7394}\inst{\ref{aff87}}
\and H.~Kurki-Suonio\orcid{0000-0002-4618-3063}\inst{\ref{aff88},\ref{aff89}}
\and Q.~Le~Boulc'h\inst{\ref{aff90}}
\and A.~M.~C.~Le~Brun\orcid{0000-0002-0936-4594}\inst{\ref{aff91}}
\and D.~Le~Mignant\orcid{0000-0002-5339-5515}\inst{\ref{aff92}}
\and P.~Liebing\inst{\ref{aff70}}
\and S.~Ligori\orcid{0000-0003-4172-4606}\inst{\ref{aff51}}
\and P.~B.~Lilje\orcid{0000-0003-4324-7794}\inst{\ref{aff78}}
\and V.~Lindholm\orcid{0000-0003-2317-5471}\inst{\ref{aff88},\ref{aff89}}
\and I.~Lloro\orcid{0000-0001-5966-1434}\inst{\ref{aff93}}
\and G.~Mainetti\orcid{0000-0003-2384-2377}\inst{\ref{aff90}}
\and D.~Maino\inst{\ref{aff94},\ref{aff55},\ref{aff95}}
\and E.~Maiorano\orcid{0000-0003-2593-4355}\inst{\ref{aff25}}
\and O.~Mansutti\orcid{0000-0001-5758-4658}\inst{\ref{aff36}}
\and S.~Marcin\inst{\ref{aff96}}
\and O.~Marggraf\orcid{0000-0001-7242-3852}\inst{\ref{aff97}}
\and M.~Martinelli\orcid{0000-0002-6943-7732}\inst{\ref{aff5},\ref{aff98}}
\and N.~Martinet\orcid{0000-0003-2786-7790}\inst{\ref{aff92}}
\and F.~Marulli\orcid{0000-0002-8850-0303}\inst{\ref{aff29},\ref{aff25},\ref{aff40}}
\and R.~Massey\orcid{0000-0002-6085-3780}\inst{\ref{aff99}}
\and S.~Maurogordato\inst{\ref{aff100}}
\and E.~Medinaceli\orcid{0000-0002-4040-7783}\inst{\ref{aff25}}
\and S.~Mei\orcid{0000-0002-2849-559X}\inst{\ref{aff101},\ref{aff102}}
\and M.~Melchior\inst{\ref{aff103}}
\and Y.~Mellier\inst{\ref{aff104},\ref{aff84}}
\and M.~Meneghetti\orcid{0000-0003-1225-7084}\inst{\ref{aff25},\ref{aff40}}
\and E.~Merlin\orcid{0000-0001-6870-8900}\inst{\ref{aff5}}
\and G.~Meylan\inst{\ref{aff105}}
\and A.~Mora\orcid{0000-0002-1922-8529}\inst{\ref{aff106}}
\and M.~Moresco\orcid{0000-0002-7616-7136}\inst{\ref{aff29},\ref{aff25}}
\and L.~Moscardini\orcid{0000-0002-3473-6716}\inst{\ref{aff29},\ref{aff25},\ref{aff40}}
\and R.~Nakajima\orcid{0009-0009-1213-7040}\inst{\ref{aff97}}
\and C.~Neissner\orcid{0000-0001-8524-4968}\inst{\ref{aff107},\ref{aff57}}
\and S.-M.~Niemi\inst{\ref{aff52}}
\and J.~W.~Nightingale\orcid{0000-0002-8987-7401}\inst{\ref{aff108}}
\and C.~Padilla\orcid{0000-0001-7951-0166}\inst{\ref{aff107}}
\and S.~Paltani\orcid{0000-0002-8108-9179}\inst{\ref{aff19}}
\and F.~Pasian\orcid{0000-0002-4869-3227}\inst{\ref{aff36}}
\and K.~Pedersen\inst{\ref{aff109}}
\and W.~J.~Percival\orcid{0000-0002-0644-5727}\inst{\ref{aff110},\ref{aff111},\ref{aff112}}
\and V.~Pettorino\inst{\ref{aff52}}
\and S.~Pires\orcid{0000-0002-0249-2104}\inst{\ref{aff34}}
\and G.~Polenta\orcid{0000-0003-4067-9196}\inst{\ref{aff43}}
\and M.~Poncet\inst{\ref{aff41}}
\and L.~A.~Popa\inst{\ref{aff113}}
\and L.~Pozzetti\orcid{0000-0001-7085-0412}\inst{\ref{aff25}}
\and F.~Raison\orcid{0000-0002-7819-6918}\inst{\ref{aff17}}
\and R.~Rebolo\inst{\ref{aff114},\ref{aff115},\ref{aff62}}
\and A.~Renzi\orcid{0000-0001-9856-1970}\inst{\ref{aff2},\ref{aff73}}
\and J.~Rhodes\orcid{0000-0002-4485-8549}\inst{\ref{aff79}}
\and G.~Riccio\inst{\ref{aff8}}
\and E.~Romelli\orcid{0000-0003-3069-9222}\inst{\ref{aff36}}
\and M.~Roncarelli\orcid{0000-0001-9587-7822}\inst{\ref{aff25}}
\and E.~Rossetti\orcid{0000-0003-0238-4047}\inst{\ref{aff39}}
\and H.~J.~A.~Rottgering\orcid{0000-0001-8887-2257}\inst{\ref{aff54}}
\and B.~Rusholme\orcid{0000-0001-7648-4142}\inst{\ref{aff116}}
\and R.~Saglia\orcid{0000-0003-0378-7032}\inst{\ref{aff42},\ref{aff17}}
\and Z.~Sakr\orcid{0000-0002-4823-3757}\inst{\ref{aff117},\ref{aff118},\ref{aff119}}
\and D.~Sapone\orcid{0000-0001-7089-4503}\inst{\ref{aff120}}
\and B.~Sartoris\orcid{0000-0003-1337-5269}\inst{\ref{aff42},\ref{aff36}}
\and J.~A.~Schewtschenko\orcid{0000-0002-4913-6393}\inst{\ref{aff63}}
\and M.~Schirmer\orcid{0000-0003-2568-9994}\inst{\ref{aff6}}
\and P.~Schneider\orcid{0000-0001-8561-2679}\inst{\ref{aff97}}
\and T.~Schrabback\orcid{0000-0002-6987-7834}\inst{\ref{aff121}}
\and M.~Scodeggio\inst{\ref{aff55}}
\and A.~Secroun\orcid{0000-0003-0505-3710}\inst{\ref{aff74}}
\and G.~Seidel\orcid{0000-0003-2907-353X}\inst{\ref{aff6}}
\and S.~Serrano\orcid{0000-0002-0211-2861}\inst{\ref{aff16},\ref{aff122},\ref{aff15}}
\and P.~Simon\inst{\ref{aff97}}
\and C.~Sirignano\orcid{0000-0002-0995-7146}\inst{\ref{aff2},\ref{aff73}}
\and G.~Sirri\orcid{0000-0003-2626-2853}\inst{\ref{aff40}}
\and L.~Stanco\orcid{0000-0002-9706-5104}\inst{\ref{aff73}}
\and J.~Steinwagner\orcid{0000-0001-7443-1047}\inst{\ref{aff17}}
\and P.~Tallada-Cresp\'{i}\orcid{0000-0002-1336-8328}\inst{\ref{aff56},\ref{aff57}}
\and A.~N.~Taylor\inst{\ref{aff63}}
\and H.~I.~Teplitz\orcid{0000-0002-7064-5424}\inst{\ref{aff123}}
\and I.~Tereno\inst{\ref{aff71},\ref{aff124}}
\and S.~Toft\orcid{0000-0003-3631-7176}\inst{\ref{aff125},\ref{aff126}}
\and R.~Toledo-Moreo\orcid{0000-0002-2997-4859}\inst{\ref{aff127}}
\and F.~Torradeflot\orcid{0000-0003-1160-1517}\inst{\ref{aff57},\ref{aff56}}
\and I.~Tutusaus\orcid{0000-0002-3199-0399}\inst{\ref{aff118}}
\and L.~Valenziano\orcid{0000-0002-1170-0104}\inst{\ref{aff25},\ref{aff75}}
\and J.~Valiviita\orcid{0000-0001-6225-3693}\inst{\ref{aff88},\ref{aff89}}
\and T.~Vassallo\orcid{0000-0001-6512-6358}\inst{\ref{aff42},\ref{aff36}}
\and G.~Verdoes~Kleijn\orcid{0000-0001-5803-2580}\inst{\ref{aff24}}
\and A.~Veropalumbo\orcid{0000-0003-2387-1194}\inst{\ref{aff33},\ref{aff45},\ref{aff44}}
\and Y.~Wang\orcid{0000-0002-4749-2984}\inst{\ref{aff123}}
\and J.~Weller\orcid{0000-0002-8282-2010}\inst{\ref{aff42},\ref{aff17}}
\and A.~Zacchei\orcid{0000-0003-0396-1192}\inst{\ref{aff36},\ref{aff35}}
\and F.~M.~Zerbi\inst{\ref{aff33}}
\and I.~A.~Zinchenko\orcid{0000-0002-2944-2449}\inst{\ref{aff42}}
\and E.~Zucca\orcid{0000-0002-5845-8132}\inst{\ref{aff25}}
\and M.~Ballardini\orcid{0000-0003-4481-3559}\inst{\ref{aff128},\ref{aff129},\ref{aff25}}
\and M.~Bolzonella\orcid{0000-0003-3278-4607}\inst{\ref{aff25}}
\and E.~Bozzo\orcid{0000-0002-8201-1525}\inst{\ref{aff19}}
\and C.~Burigana\orcid{0000-0002-3005-5796}\inst{\ref{aff12},\ref{aff75}}
\and R.~Cabanac\orcid{0000-0001-6679-2600}\inst{\ref{aff118}}
\and A.~Cappi\inst{\ref{aff25},\ref{aff100}}
\and D.~Di~Ferdinando\inst{\ref{aff40}}
\and J.~A.~Escartin~Vigo\inst{\ref{aff17}}
\and L.~Gabarra\orcid{0000-0002-8486-8856}\inst{\ref{aff130}}
\and M.~Huertas-Company\orcid{0000-0002-1416-8483}\inst{\ref{aff62},\ref{aff18},\ref{aff131},\ref{aff132}}
\and J.~Mart\'{i}n-Fleitas\orcid{0000-0002-8594-569X}\inst{\ref{aff106}}
\and S.~Matthew\orcid{0000-0001-8448-1697}\inst{\ref{aff63}}
\and M.~Maturi\orcid{0000-0002-3517-2422}\inst{\ref{aff117},\ref{aff133}}
\and N.~Mauri\orcid{0000-0001-8196-1548}\inst{\ref{aff61},\ref{aff40}}
\and A.~Pezzotta\orcid{0000-0003-0726-2268}\inst{\ref{aff17}}
\and M.~P\"ontinen\orcid{0000-0001-5442-2530}\inst{\ref{aff88}}
\and C.~Porciani\orcid{0000-0002-7797-2508}\inst{\ref{aff97}}
\and I.~Risso\orcid{0000-0003-2525-7761}\inst{\ref{aff134}}
\and V.~Scottez\inst{\ref{aff104},\ref{aff135}}
\and M.~Sereno\orcid{0000-0003-0302-0325}\inst{\ref{aff25},\ref{aff40}}
\and M.~Tenti\orcid{0000-0002-4254-5901}\inst{\ref{aff40}}
\and M.~Viel\orcid{0000-0002-2642-5707}\inst{\ref{aff35},\ref{aff36},\ref{aff38},\ref{aff37},\ref{aff136}}
\and M.~Wiesmann\orcid{0009-0000-8199-5860}\inst{\ref{aff78}}
\and Y.~Akrami\orcid{0000-0002-2407-7956}\inst{\ref{aff137},\ref{aff138}}
\and I.~T.~Andika\orcid{0000-0001-6102-9526}\inst{\ref{aff139},\ref{aff140}}
\and S.~Anselmi\orcid{0000-0002-3579-9583}\inst{\ref{aff73},\ref{aff2},\ref{aff141}}
\and M.~Archidiacono\orcid{0000-0003-4952-9012}\inst{\ref{aff94},\ref{aff95}}
\and F.~Atrio-Barandela\orcid{0000-0002-2130-2513}\inst{\ref{aff142}}
\and C.~Benoist\inst{\ref{aff100}}
\and K.~Benson\inst{\ref{aff70}}
\and D.~Bertacca\orcid{0000-0002-2490-7139}\inst{\ref{aff2},\ref{aff1},\ref{aff73}}
\and M.~Bethermin\orcid{0000-0002-3915-2015}\inst{\ref{aff143}}
\and A.~Blanchard\orcid{0000-0001-8555-9003}\inst{\ref{aff118}}
\and L.~Blot\orcid{0000-0002-9622-7167}\inst{\ref{aff144},\ref{aff141}}
\and M.~L.~Brown\orcid{0000-0002-0370-8077}\inst{\ref{aff64}}
\and S.~Bruton\orcid{0000-0002-6503-5218}\inst{\ref{aff145}}
\and A.~Calabro\orcid{0000-0003-2536-1614}\inst{\ref{aff5}}
\and F.~Caro\inst{\ref{aff5}}
\and C.~S.~Carvalho\inst{\ref{aff124}}
\and T.~Castro\orcid{0000-0002-6292-3228}\inst{\ref{aff36},\ref{aff37},\ref{aff35},\ref{aff136}}
\and Y.~Charles\inst{\ref{aff92}}
\and F.~Cogato\orcid{0000-0003-4632-6113}\inst{\ref{aff29},\ref{aff25}}
\and T.~Contini\orcid{0000-0003-0275-938X}\inst{\ref{aff118}}
\and A.~R.~Cooray\orcid{0000-0002-3892-0190}\inst{\ref{aff146}}
\and O.~Cucciati\orcid{0000-0002-9336-7551}\inst{\ref{aff25}}
\and S.~Davini\orcid{0000-0003-3269-1718}\inst{\ref{aff45}}
\and F.~De~Paolis\orcid{0000-0001-6460-7563}\inst{\ref{aff147},\ref{aff148},\ref{aff149}}
\and G.~Desprez\orcid{0000-0001-8325-1742}\inst{\ref{aff24}}
\and A.~D\'iaz-S\'anchez\orcid{0000-0003-0748-4768}\inst{\ref{aff150}}
\and J.~J.~Diaz\inst{\ref{aff62}}
\and S.~Di~Domizio\orcid{0000-0003-2863-5895}\inst{\ref{aff44},\ref{aff45}}
\and J.~M.~Diego\orcid{0000-0001-9065-3926}\inst{\ref{aff151}}
\and A.~Enia\orcid{0000-0002-0200-2857}\inst{\ref{aff39},\ref{aff25}}
\and Y.~Fang\inst{\ref{aff42}}
\and A.~G.~Ferrari\orcid{0009-0005-5266-4110}\inst{\ref{aff40}}
\and P.~G.~Ferreira\orcid{0000-0002-3021-2851}\inst{\ref{aff130}}
\and A.~Finoguenov\orcid{0000-0002-4606-5403}\inst{\ref{aff88}}
\and A.~Fontana\orcid{0000-0003-3820-2823}\inst{\ref{aff5}}
\and F.~Fontanot\orcid{0000-0003-4744-0188}\inst{\ref{aff36},\ref{aff35}}
\and A.~Franco\orcid{0000-0002-4761-366X}\inst{\ref{aff148},\ref{aff147},\ref{aff149}}
\and K.~Ganga\orcid{0000-0001-8159-8208}\inst{\ref{aff101}}
\and J.~Garc\'ia-Bellido\orcid{0000-0002-9370-8360}\inst{\ref{aff137}}
\and T.~Gasparetto\orcid{0000-0002-7913-4866}\inst{\ref{aff36}}
\and V.~Gautard\inst{\ref{aff152}}
\and E.~Gaztanaga\orcid{0000-0001-9632-0815}\inst{\ref{aff15},\ref{aff16},\ref{aff13}}
\and F.~Giacomini\orcid{0000-0002-3129-2814}\inst{\ref{aff40}}
\and F.~Gianotti\orcid{0000-0003-4666-119X}\inst{\ref{aff25}}
\and G.~Gozaliasl\orcid{0000-0002-0236-919X}\inst{\ref{aff153},\ref{aff88}}
\and M.~Guidi\orcid{0000-0001-9408-1101}\inst{\ref{aff39},\ref{aff25}}
\and C.~M.~Gutierrez\orcid{0000-0001-7854-783X}\inst{\ref{aff154}}
\and A.~Hall\orcid{0000-0002-3139-8651}\inst{\ref{aff63}}
\and W.~G.~Hartley\inst{\ref{aff19}}
\and S.~Hemmati\orcid{0000-0003-2226-5395}\inst{\ref{aff116}}
\and C.~Hern\'andez-Monteagudo\orcid{0000-0001-5471-9166}\inst{\ref{aff115},\ref{aff62}}
\and H.~Hildebrandt\orcid{0000-0002-9814-3338}\inst{\ref{aff155}}
\and J.~Hjorth\orcid{0000-0002-4571-2306}\inst{\ref{aff109}}
\and J.~J.~E.~Kajava\orcid{0000-0002-3010-8333}\inst{\ref{aff156},\ref{aff157}}
\and Y.~Kang\orcid{0009-0000-8588-7250}\inst{\ref{aff19}}
\and V.~Kansal\orcid{0000-0002-4008-6078}\inst{\ref{aff158},\ref{aff159}}
\and D.~Karagiannis\orcid{0000-0002-4927-0816}\inst{\ref{aff128},\ref{aff160}}
\and K.~Kiiveri\inst{\ref{aff86}}
\and C.~C.~Kirkpatrick\inst{\ref{aff86}}
\and S.~Kruk\orcid{0000-0001-8010-8879}\inst{\ref{aff31}}
\and J.~Le~Graet\orcid{0000-0001-6523-7971}\inst{\ref{aff74}}
\and L.~Legrand\orcid{0000-0003-0610-5252}\inst{\ref{aff161},\ref{aff162}}
\and M.~Lembo\orcid{0000-0002-5271-5070}\inst{\ref{aff128},\ref{aff129}}
\and F.~Lepori\orcid{0009-0000-5061-7138}\inst{\ref{aff163}}
\and G.~Leroy\orcid{0009-0004-2523-4425}\inst{\ref{aff164},\ref{aff99}}
\and G.~F.~Lesci\orcid{0000-0002-4607-2830}\inst{\ref{aff29},\ref{aff25}}
\and J.~Lesgourgues\orcid{0000-0001-7627-353X}\inst{\ref{aff58}}
\and L.~Leuzzi\orcid{0009-0006-4479-7017}\inst{\ref{aff29},\ref{aff25}}
\and T.~I.~Liaudat\orcid{0000-0002-9104-314X}\inst{\ref{aff165}}
\and A.~Loureiro\orcid{0000-0002-4371-0876}\inst{\ref{aff166},\ref{aff167}}
\and J.~Macias-Perez\orcid{0000-0002-5385-2763}\inst{\ref{aff168}}
\and G.~Maggio\orcid{0000-0003-4020-4836}\inst{\ref{aff36}}
\and M.~Magliocchetti\orcid{0000-0001-9158-4838}\inst{\ref{aff28}}
\and E.~A.~Magnier\orcid{0000-0002-7965-2815}\inst{\ref{aff60}}
\and C.~Mancini\orcid{0000-0002-4297-0561}\inst{\ref{aff55}}
\and F.~Mannucci\orcid{0000-0002-4803-2381}\inst{\ref{aff7}}
\and R.~Maoli\orcid{0000-0002-6065-3025}\inst{\ref{aff169},\ref{aff5}}
\and C.~J.~A.~P.~Martins\orcid{0000-0002-4886-9261}\inst{\ref{aff170},\ref{aff47}}
\and L.~Maurin\orcid{0000-0002-8406-0857}\inst{\ref{aff30}}
\and M.~Miluzio\inst{\ref{aff31},\ref{aff171}}
\and P.~Monaco\orcid{0000-0003-2083-7564}\inst{\ref{aff172},\ref{aff36},\ref{aff37},\ref{aff35}}
\and C.~Moretti\orcid{0000-0003-3314-8936}\inst{\ref{aff38},\ref{aff136},\ref{aff36},\ref{aff35},\ref{aff37}}
\and G.~Morgante\inst{\ref{aff25}}
\and S.~Nadathur\orcid{0000-0001-9070-3102}\inst{\ref{aff13}}
\and K.~Naidoo\orcid{0000-0002-9182-1802}\inst{\ref{aff13}}
\and A.~Navarro-Alsina\orcid{0000-0002-3173-2592}\inst{\ref{aff97}}
\and S.~Nesseris\orcid{0000-0002-0567-0324}\inst{\ref{aff137}}
\and F.~Passalacqua\orcid{0000-0002-8606-4093}\inst{\ref{aff2},\ref{aff73}}
\and K.~Paterson\orcid{0000-0001-8340-3486}\inst{\ref{aff6}}
\and L.~Patrizii\inst{\ref{aff40}}
\and A.~Pisani\orcid{0000-0002-6146-4437}\inst{\ref{aff74},\ref{aff173}}
\and D.~Potter\orcid{0000-0002-0757-5195}\inst{\ref{aff163}}
\and S.~Quai\orcid{0000-0002-0449-8163}\inst{\ref{aff29},\ref{aff25}}
\and M.~Radovich\orcid{0000-0002-3585-866X}\inst{\ref{aff1}}
\and P.-F.~Rocci\inst{\ref{aff30}}
\and S.~Sacquegna\orcid{0000-0002-8433-6630}\inst{\ref{aff147},\ref{aff148},\ref{aff149}}
\and M.~Sahl\'en\orcid{0000-0003-0973-4804}\inst{\ref{aff174}}
\and D.~B.~Sanders\orcid{0000-0002-1233-9998}\inst{\ref{aff60}}
\and E.~Sarpa\orcid{0000-0002-1256-655X}\inst{\ref{aff38},\ref{aff136},\ref{aff37}}
\and C.~Scarlata\orcid{0000-0002-9136-8876}\inst{\ref{aff175}}
\and J.~Schaye\orcid{0000-0002-0668-5560}\inst{\ref{aff54}}
\and A.~Schneider\orcid{0000-0001-7055-8104}\inst{\ref{aff163}}
\and D.~Sciotti\orcid{0009-0008-4519-2620}\inst{\ref{aff5},\ref{aff98}}
\and E.~Sellentin\inst{\ref{aff176},\ref{aff54}}
\and A.~Shulevski\orcid{0000-0002-1827-0469}\inst{\ref{aff177},\ref{aff24},\ref{aff54},\ref{aff178}}
\and L.~C.~Smith\orcid{0000-0002-3259-2771}\inst{\ref{aff179}}
\and K.~Tanidis\orcid{0000-0001-9843-5130}\inst{\ref{aff130}}
\and C.~Tao\orcid{0000-0001-7961-8177}\inst{\ref{aff74}}
\and G.~Testera\inst{\ref{aff45}}
\and R.~Teyssier\orcid{0000-0001-7689-0933}\inst{\ref{aff173}}
\and S.~Tosi\orcid{0000-0002-7275-9193}\inst{\ref{aff44},\ref{aff134}}
\and A.~Troja\orcid{0000-0003-0239-4595}\inst{\ref{aff2},\ref{aff73}}
\and M.~Tucci\inst{\ref{aff19}}
\and C.~Valieri\inst{\ref{aff40}}
\and A.~Venhola\orcid{0000-0001-6071-4564}\inst{\ref{aff180}}
\and D.~Vergani\orcid{0000-0003-0898-2216}\inst{\ref{aff25}}
\and G.~Verza\orcid{0000-0002-1886-8348}\inst{\ref{aff181}}
\and P.~Vielzeuf\orcid{0000-0003-2035-9339}\inst{\ref{aff74}}
\and A.~Viitanen\orcid{0000-0001-9383-786X}\inst{\ref{aff86},\ref{aff5}}
\and N.~A.~Walton\orcid{0000-0003-3983-8778}\inst{\ref{aff179}}
\and J.~R.~Weaver\orcid{0000-0003-1614-196X}\inst{\ref{aff182}}
\and E.~Soubrie\orcid{0000-0001-9295-1863}\inst{\ref{aff30}}
\and D.~Scott\orcid{0000-0002-6878-9840}\inst{\ref{aff183}}}
										   
\institute{INAF-Osservatorio Astronomico di Padova, Via dell'Osservatorio 5, 35122 Padova, Italy\label{aff1}
\and
Dipartimento di Fisica e Astronomia "G. Galilei", Universit\`a di Padova, Via Marzolo 8, 35131 Padova, Italy\label{aff2}
\and
School of Physics, HH Wills Physics Laboratory, University of Bristol, Tyndall Avenue, Bristol, BS8 1TL, UK\label{aff3}
\and
Department of Mathematics and Physics, Roma Tre University, Via della Vasca Navale 84, 00146 Rome, Italy\label{aff4}
\and
INAF-Osservatorio Astronomico di Roma, Via Frascati 33, 00078 Monteporzio Catone, Italy\label{aff5}
\and
Max-Planck-Institut f\"ur Astronomie, K\"onigstuhl 17, 69117 Heidelberg, Germany\label{aff6}
\and
INAF-Osservatorio Astrofisico di Arcetri, Largo E. Fermi 5, 50125, Firenze, Italy\label{aff7}
\and
INAF-Osservatorio Astronomico di Capodimonte, Via Moiariello 16, 80131 Napoli, Italy\label{aff8}
\and
School of Physics \& Astronomy, University of Southampton, Highfield Campus, Southampton SO17 1BJ, UK\label{aff9}
\and
Jeremiah Horrocks Institute, University of Central Lancashire, Preston, PR1 2HE, UK\label{aff10}
\and
Dipartimento di Fisica e Astronomia ``G. Galilei", Universit\`a di Padova, Vicolo dell'Osservatorio 3, 35122 Padova, Italy\label{aff11}
\and
INAF, Istituto di Radioastronomia, Via Piero Gobetti 101, 40129 Bologna, Italy\label{aff12}
\and
Institute of Cosmology and Gravitation, University of Portsmouth, Portsmouth PO1 3FX, UK\label{aff13}
\and
Cavendish Laboratory, University of Cambridge, JJ Thomson Avenue, Cambridge, CB3 0HE, UK\label{aff14}
\and
Institute of Space Sciences (ICE, CSIC), Campus UAB, Carrer de Can Magrans, s/n, 08193 Barcelona, Spain\label{aff15}
\and
Institut d'Estudis Espacials de Catalunya (IEEC),  Edifici RDIT, Campus UPC, 08860 Castelldefels, Barcelona, Spain\label{aff16}
\and
Max Planck Institute for Extraterrestrial Physics, Giessenbachstr. 1, 85748 Garching, Germany\label{aff17}
\and
Instituto de Astrof\'isica de Canarias (IAC); Departamento de Astrof\'isica, Universidad de La Laguna (ULL), 38200, La Laguna, Tenerife, Spain\label{aff18}
\and
Department of Astronomy, University of Geneva, ch. d'Ecogia 16, 1290 Versoix, Switzerland\label{aff19}
\and
Department of Physical Sciences, Ritsumeikan University, Kusatsu, Shiga 525-8577, Japan\label{aff20}
\and
National Astronomical Observatory of Japan, 2-21-1 Osawa, Mitaka, Tokyo 181-8588, Japan\label{aff21}
\and
Academia Sinica Institute of Astronomy and Astrophysics (ASIAA), 11F of ASMAB, No.~1, Section 4, Roosevelt Road, Taipei 10617, Taiwan\label{aff22}
\and
SRON Netherlands Institute for Space Research, Landleven 12, 9747 AD, Groningen, The Netherlands\label{aff23}
\and
Kapteyn Astronomical Institute, University of Groningen, PO Box 800, 9700 AV Groningen, The Netherlands\label{aff24}
\and
INAF-Osservatorio di Astrofisica e Scienza dello Spazio di Bologna, Via Piero Gobetti 93/3, 40129 Bologna, Italy\label{aff25}
\and
Sterrenkundig Observatorium, Universiteit Gent, Krijgslaan 281 S9, 9000 Gent, Belgium\label{aff26}
\and
STAR Institute, University of Li{\`e}ge, Quartier Agora, All\'ee du six Ao\^ut 19c, 4000 Li\`ege, Belgium\label{aff27}
\and
INAF-Istituto di Astrofisica e Planetologia Spaziali, via del Fosso del Cavaliere, 100, 00100 Roma, Italy\label{aff28}
\and
Dipartimento di Fisica e Astronomia "Augusto Righi" - Alma Mater Studiorum Universit\`a di Bologna, via Piero Gobetti 93/2, 40129 Bologna, Italy\label{aff29}
\and
Universit\'e Paris-Saclay, CNRS, Institut d'astrophysique spatiale, 91405, Orsay, France\label{aff30}
\and
ESAC/ESA, Camino Bajo del Castillo, s/n., Urb. Villafranca del Castillo, 28692 Villanueva de la Ca\~nada, Madrid, Spain\label{aff31}
\and
School of Mathematics and Physics, University of Surrey, Guildford, Surrey, GU2 7XH, UK\label{aff32}
\and
INAF-Osservatorio Astronomico di Brera, Via Brera 28, 20122 Milano, Italy\label{aff33}
\and
Universit\'e Paris-Saclay, Universit\'e Paris Cit\'e, CEA, CNRS, AIM, 91191, Gif-sur-Yvette, France\label{aff34}
\and
IFPU, Institute for Fundamental Physics of the Universe, via Beirut 2, 34151 Trieste, Italy\label{aff35}
\and
INAF-Osservatorio Astronomico di Trieste, Via G. B. Tiepolo 11, 34143 Trieste, Italy\label{aff36}
\and
INFN, Sezione di Trieste, Via Valerio 2, 34127 Trieste TS, Italy\label{aff37}
\and
SISSA, International School for Advanced Studies, Via Bonomea 265, 34136 Trieste TS, Italy\label{aff38}
\and
Dipartimento di Fisica e Astronomia, Universit\`a di Bologna, Via Gobetti 93/2, 40129 Bologna, Italy\label{aff39}
\and
INFN-Sezione di Bologna, Viale Berti Pichat 6/2, 40127 Bologna, Italy\label{aff40}
\and
Centre National d'Etudes Spatiales -- Centre spatial de Toulouse, 18 avenue Edouard Belin, 31401 Toulouse Cedex 9, France\label{aff41}
\and
Universit\"ats-Sternwarte M\"unchen, Fakult\"at f\"ur Physik, Ludwig-Maximilians-Universit\"at M\"unchen, Scheinerstrasse 1, 81679 M\"unchen, Germany\label{aff42}
\and
Space Science Data Center, Italian Space Agency, via del Politecnico snc, 00133 Roma, Italy\label{aff43}
\and
Dipartimento di Fisica, Universit\`a di Genova, Via Dodecaneso 33, 16146, Genova, Italy\label{aff44}
\and
INFN-Sezione di Genova, Via Dodecaneso 33, 16146, Genova, Italy\label{aff45}
\and
Department of Physics "E. Pancini", University Federico II, Via Cinthia 6, 80126, Napoli, Italy\label{aff46}
\and
Instituto de Astrof\'isica e Ci\^encias do Espa\c{c}o, Universidade do Porto, CAUP, Rua das Estrelas, PT4150-762 Porto, Portugal\label{aff47}
\and
Faculdade de Ci\^encias da Universidade do Porto, Rua do Campo de Alegre, 4150-007 Porto, Portugal\label{aff48}
\and
Dipartimento di Fisica, Universit\`a degli Studi di Torino, Via P. Giuria 1, 10125 Torino, Italy\label{aff49}
\and
INFN-Sezione di Torino, Via P. Giuria 1, 10125 Torino, Italy\label{aff50}
\and
INAF-Osservatorio Astrofisico di Torino, Via Osservatorio 20, 10025 Pino Torinese (TO), Italy\label{aff51}
\and
European Space Agency/ESTEC, Keplerlaan 1, 2201 AZ Noordwijk, The Netherlands\label{aff52}
\and
Institute Lorentz, Leiden University, Niels Bohrweg 2, 2333 CA Leiden, The Netherlands\label{aff53}
\and
Leiden Observatory, Leiden University, Einsteinweg 55, 2333 CC Leiden, The Netherlands\label{aff54}
\and
INAF-IASF Milano, Via Alfonso Corti 12, 20133 Milano, Italy\label{aff55}
\and
Centro de Investigaciones Energ\'eticas, Medioambientales y Tecnol\'ogicas (CIEMAT), Avenida Complutense 40, 28040 Madrid, Spain\label{aff56}
\and
Port d'Informaci\'{o} Cient\'{i}fica, Campus UAB, C. Albareda s/n, 08193 Bellaterra (Barcelona), Spain\label{aff57}
\and
Institute for Theoretical Particle Physics and Cosmology (TTK), RWTH Aachen University, 52056 Aachen, Germany\label{aff58}
\and
INFN section of Naples, Via Cinthia 6, 80126, Napoli, Italy\label{aff59}
\and
Institute for Astronomy, University of Hawaii, 2680 Woodlawn Drive, Honolulu, HI 96822, USA\label{aff60}
\and
Dipartimento di Fisica e Astronomia "Augusto Righi" - Alma Mater Studiorum Universit\`a di Bologna, Viale Berti Pichat 6/2, 40127 Bologna, Italy\label{aff61}
\and
Instituto de Astrof\'{\i}sica de Canarias, V\'{\i}a L\'actea, 38205 La Laguna, Tenerife, Spain\label{aff62}
\and
Institute for Astronomy, University of Edinburgh, Royal Observatory, Blackford Hill, Edinburgh EH9 3HJ, UK\label{aff63}
\and
Jodrell Bank Centre for Astrophysics, Department of Physics and Astronomy, University of Manchester, Oxford Road, Manchester M13 9PL, UK\label{aff64}
\and
European Space Agency/ESRIN, Largo Galileo Galilei 1, 00044 Frascati, Roma, Italy\label{aff65}
\and
Universit\'e Claude Bernard Lyon 1, CNRS/IN2P3, IP2I Lyon, UMR 5822, Villeurbanne, F-69100, France\label{aff66}
\and
Institut de Ci\`{e}ncies del Cosmos (ICCUB), Universitat de Barcelona (IEEC-UB), Mart\'{i} i Franqu\`{e}s 1, 08028 Barcelona, Spain\label{aff67}
\and
Instituci\'o Catalana de Recerca i Estudis Avan\c{c}ats (ICREA), Passeig de Llu\'{\i}s Companys 23, 08010 Barcelona, Spain\label{aff68}
\and
UCB Lyon 1, CNRS/IN2P3, IUF, IP2I Lyon, 4 rue Enrico Fermi, 69622 Villeurbanne, France\label{aff69}
\and
Mullard Space Science Laboratory, University College London, Holmbury St Mary, Dorking, Surrey RH5 6NT, UK\label{aff70}
\and
Departamento de F\'isica, Faculdade de Ci\^encias, Universidade de Lisboa, Edif\'icio C8, Campo Grande, PT1749-016 Lisboa, Portugal\label{aff71}
\and
Instituto de Astrof\'isica e Ci\^encias do Espa\c{c}o, Faculdade de Ci\^encias, Universidade de Lisboa, Campo Grande, 1749-016 Lisboa, Portugal\label{aff72}
\and
INFN-Padova, Via Marzolo 8, 35131 Padova, Italy\label{aff73}
\and
Aix-Marseille Universit\'e, CNRS/IN2P3, CPPM, Marseille, France\label{aff74}
\and
INFN-Bologna, Via Irnerio 46, 40126 Bologna, Italy\label{aff75}
\and
FRACTAL S.L.N.E., calle Tulip\'an 2, Portal 13 1A, 28231, Las Rozas de Madrid, Spain\label{aff76}
\and
NRC Herzberg, 5071 West Saanich Rd, Victoria, BC V9E 2E7, Canada\label{aff77}
\and
Institute of Theoretical Astrophysics, University of Oslo, P.O. Box 1029 Blindern, 0315 Oslo, Norway\label{aff78}
\and
Jet Propulsion Laboratory, California Institute of Technology, 4800 Oak Grove Drive, Pasadena, CA, 91109, USA\label{aff79}
\and
Department of Physics, Lancaster University, Lancaster, LA1 4YB, UK\label{aff80}
\and
Felix Hormuth Engineering, Goethestr. 17, 69181 Leimen, Germany\label{aff81}
\and
Technical University of Denmark, Elektrovej 327, 2800 Kgs. Lyngby, Denmark\label{aff82}
\and
Cosmic Dawn Center (DAWN), Denmark\label{aff83}
\and
Institut d'Astrophysique de Paris, UMR 7095, CNRS, and Sorbonne Universit\'e, 98 bis boulevard Arago, 75014 Paris, France\label{aff84}
\and
NASA Goddard Space Flight Center, Greenbelt, MD 20771, USA\label{aff85}
\and
Department of Physics and Helsinki Institute of Physics, Gustaf H\"allstr\"omin katu 2, 00014 University of Helsinki, Finland\label{aff86}
\and
Universit\'e de Gen\`eve, D\'epartement de Physique Th\'eorique and Centre for Astroparticle Physics, 24 quai Ernest-Ansermet, CH-1211 Gen\`eve 4, Switzerland\label{aff87}
\and
Department of Physics, P.O. Box 64, 00014 University of Helsinki, Finland\label{aff88}
\and
Helsinki Institute of Physics, Gustaf H{\"a}llstr{\"o}min katu 2, University of Helsinki, Helsinki, Finland\label{aff89}
\and
Centre de Calcul de l'IN2P3/CNRS, 21 avenue Pierre de Coubertin 69627 Villeurbanne Cedex, France\label{aff90}
\and
Laboratoire d'etude de l'Univers et des phenomenes eXtremes, Observatoire de Paris, Universit\'e PSL, Sorbonne Universit\'e, CNRS, 92190 Meudon, France\label{aff91}
\and
Aix-Marseille Universit\'e, CNRS, CNES, LAM, Marseille, France\label{aff92}
\and
SKA Observatory, Jodrell Bank, Lower Withington, Macclesfield, Cheshire SK11 9FT, UK\label{aff93}
\and
Dipartimento di Fisica "Aldo Pontremoli", Universit\`a degli Studi di Milano, Via Celoria 16, 20133 Milano, Italy\label{aff94}
\and
INFN-Sezione di Milano, Via Celoria 16, 20133 Milano, Italy\label{aff95}
\and
University of Applied Sciences and Arts of Northwestern Switzerland, School of Computer Science, 5210 Windisch, Switzerland\label{aff96}
\and
Universit\"at Bonn, Argelander-Institut f\"ur Astronomie, Auf dem H\"ugel 71, 53121 Bonn, Germany\label{aff97}
\and
INFN-Sezione di Roma, Piazzale Aldo Moro, 2 - c/o Dipartimento di Fisica, Edificio G. Marconi, 00185 Roma, Italy\label{aff98}
\and
Department of Physics, Institute for Computational Cosmology, Durham University, South Road, Durham, DH1 3LE, UK\label{aff99}
\and
Universit\'e C\^{o}te d'Azur, Observatoire de la C\^{o}te d'Azur, CNRS, Laboratoire Lagrange, Bd de l'Observatoire, CS 34229, 06304 Nice cedex 4, France\label{aff100}
\and
Universit\'e Paris Cit\'e, CNRS, Astroparticule et Cosmologie, 75013 Paris, France\label{aff101}
\and
CNRS-UCB International Research Laboratory, Centre Pierre Binetruy, IRL2007, CPB-IN2P3, Berkeley, USA\label{aff102}
\and
University of Applied Sciences and Arts of Northwestern Switzerland, School of Engineering, 5210 Windisch, Switzerland\label{aff103}
\and
Institut d'Astrophysique de Paris, 98bis Boulevard Arago, 75014, Paris, France\label{aff104}
\and
Institute of Physics, Laboratory of Astrophysics, Ecole Polytechnique F\'ed\'erale de Lausanne (EPFL), Observatoire de Sauverny, 1290 Versoix, Switzerland\label{aff105}
\and
Aurora Technology for European Space Agency (ESA), Camino bajo del Castillo, s/n, Urbanizacion Villafranca del Castillo, Villanueva de la Ca\~nada, 28692 Madrid, Spain\label{aff106}
\and
Institut de F\'{i}sica d'Altes Energies (IFAE), The Barcelona Institute of Science and Technology, Campus UAB, 08193 Bellaterra (Barcelona), Spain\label{aff107}
\and
School of Mathematics, Statistics and Physics, Newcastle University, Herschel Building, Newcastle-upon-Tyne, NE1 7RU, UK\label{aff108}
\and
DARK, Niels Bohr Institute, University of Copenhagen, Jagtvej 155, 2200 Copenhagen, Denmark\label{aff109}
\and
Waterloo Centre for Astrophysics, University of Waterloo, Waterloo, Ontario N2L 3G1, Canada\label{aff110}
\and
Department of Physics and Astronomy, University of Waterloo, Waterloo, Ontario N2L 3G1, Canada\label{aff111}
\and
Perimeter Institute for Theoretical Physics, Waterloo, Ontario N2L 2Y5, Canada\label{aff112}
\and
Institute of Space Science, Str. Atomistilor, nr. 409 M\u{a}gurele, Ilfov, 077125, Romania\label{aff113}
\and
Consejo Superior de Investigaciones Cientificas, Calle Serrano 117, 28006 Madrid, Spain\label{aff114}
\and
Universidad de La Laguna, Departamento de Astrof\'{\i}sica, 38206 La Laguna, Tenerife, Spain\label{aff115}
\and
Caltech/IPAC, 1200 E. California Blvd., Pasadena, CA 91125, USA\label{aff116}
\and
Institut f\"ur Theoretische Physik, University of Heidelberg, Philosophenweg 16, 69120 Heidelberg, Germany\label{aff117}
\and
Institut de Recherche en Astrophysique et Plan\'etologie (IRAP), Universit\'e de Toulouse, CNRS, UPS, CNES, 14 Av. Edouard Belin, 31400 Toulouse, France\label{aff118}
\and
Universit\'e St Joseph; Faculty of Sciences, Beirut, Lebanon\label{aff119}
\and
Departamento de F\'isica, FCFM, Universidad de Chile, Blanco Encalada 2008, Santiago, Chile\label{aff120}
\and
Universit\"at Innsbruck, Institut f\"ur Astro- und Teilchenphysik, Technikerstr. 25/8, 6020 Innsbruck, Austria\label{aff121}
\and
Satlantis, University Science Park, Sede Bld 48940, Leioa-Bilbao, Spain\label{aff122}
\and
Infrared Processing and Analysis Center, California Institute of Technology, Pasadena, CA 91125, USA\label{aff123}
\and
Instituto de Astrof\'isica e Ci\^encias do Espa\c{c}o, Faculdade de Ci\^encias, Universidade de Lisboa, Tapada da Ajuda, 1349-018 Lisboa, Portugal\label{aff124}
\and
Cosmic Dawn Center (DAWN)\label{aff125}
\and
Niels Bohr Institute, University of Copenhagen, Jagtvej 128, 2200 Copenhagen, Denmark\label{aff126}
\and
Universidad Polit\'ecnica de Cartagena, Departamento de Electr\'onica y Tecnolog\'ia de Computadoras,  Plaza del Hospital 1, 30202 Cartagena, Spain\label{aff127}
\and
Dipartimento di Fisica e Scienze della Terra, Universit\`a degli Studi di Ferrara, Via Giuseppe Saragat 1, 44122 Ferrara, Italy\label{aff128}
\and
Istituto Nazionale di Fisica Nucleare, Sezione di Ferrara, Via Giuseppe Saragat 1, 44122 Ferrara, Italy\label{aff129}
\and
Department of Physics, Oxford University, Keble Road, Oxford OX1 3RH, UK\label{aff130}
\and
Universit\'e PSL, Observatoire de Paris, Sorbonne Universit\'e, CNRS, LERMA, 75014, Paris, France\label{aff131}
\and
Universit\'e Paris-Cit\'e, 5 Rue Thomas Mann, 75013, Paris, France\label{aff132}
\and
Zentrum f\"ur Astronomie, Universit\"at Heidelberg, Philosophenweg 12, 69120 Heidelberg, Germany\label{aff133}
\and
INAF-Osservatorio Astronomico di Brera, Via Brera 28, 20122 Milano, Italy, and INFN-Sezione di Genova, Via Dodecaneso 33, 16146, Genova, Italy\label{aff134}
\and
ICL, Junia, Universit\'e Catholique de Lille, LITL, 59000 Lille, France\label{aff135}
\and
ICSC - Centro Nazionale di Ricerca in High Performance Computing, Big Data e Quantum Computing, Via Magnanelli 2, Bologna, Italy\label{aff136}
\and
Instituto de F\'isica Te\'orica UAM-CSIC, Campus de Cantoblanco, 28049 Madrid, Spain\label{aff137}
\and
CERCA/ISO, Department of Physics, Case Western Reserve University, 10900 Euclid Avenue, Cleveland, OH 44106, USA\label{aff138}
\and
Technical University of Munich, TUM School of Natural Sciences, Physics Department, James-Franck-Str.~1, 85748 Garching, Germany\label{aff139}
\and
Max-Planck-Institut f\"ur Astrophysik, Karl-Schwarzschild-Str.~1, 85748 Garching, Germany\label{aff140}
\and
Laboratoire Univers et Th\'eorie, Observatoire de Paris, Universit\'e PSL, Universit\'e Paris Cit\'e, CNRS, 92190 Meudon, France\label{aff141}
\and
Departamento de F{\'\i}sica Fundamental. Universidad de Salamanca. Plaza de la Merced s/n. 37008 Salamanca, Spain\label{aff142}
\and
Universit\'e de Strasbourg, CNRS, Observatoire astronomique de Strasbourg, UMR 7550, 67000 Strasbourg, France\label{aff143}
\and
Center for Data-Driven Discovery, Kavli IPMU (WPI), UTIAS, The University of Tokyo, Kashiwa, Chiba 277-8583, Japan\label{aff144}
\and
California Institute of Technology, 1200 E California Blvd, Pasadena, CA 91125, USA\label{aff145}
\and
Department of Physics \& Astronomy, University of California Irvine, Irvine CA 92697, USA\label{aff146}
\and
Department of Mathematics and Physics E. De Giorgi, University of Salento, Via per Arnesano, CP-I93, 73100, Lecce, Italy\label{aff147}
\and
INFN, Sezione di Lecce, Via per Arnesano, CP-193, 73100, Lecce, Italy\label{aff148}
\and
INAF-Sezione di Lecce, c/o Dipartimento Matematica e Fisica, Via per Arnesano, 73100, Lecce, Italy\label{aff149}
\and
Departamento F\'isica Aplicada, Universidad Polit\'ecnica de Cartagena, Campus Muralla del Mar, 30202 Cartagena, Murcia, Spain\label{aff150}
\and
Instituto de F\'isica de Cantabria, Edificio Juan Jord\'a, Avenida de los Castros, 39005 Santander, Spain\label{aff151}
\and
CEA Saclay, DFR/IRFU, Service d'Astrophysique, Bat. 709, 91191 Gif-sur-Yvette, France\label{aff152}
\and
Department of Computer Science, Aalto University, PO Box 15400, Espoo, FI-00 076, Finland\label{aff153}
\and
Instituto de Astrof\'\i sica de Canarias, c/ Via Lactea s/n, La Laguna 38200, Spain. Departamento de Astrof\'\i sica de la Universidad de La Laguna, Avda. Francisco Sanchez, La Laguna, 38200, Spain\label{aff154}
\and
Ruhr University Bochum, Faculty of Physics and Astronomy, Astronomical Institute (AIRUB), German Centre for Cosmological Lensing (GCCL), 44780 Bochum, Germany\label{aff155}
\and
Department of Physics and Astronomy, Vesilinnantie 5, 20014 University of Turku, Finland\label{aff156}
\and
Serco for European Space Agency (ESA), Camino bajo del Castillo, s/n, Urbanizacion Villafranca del Castillo, Villanueva de la Ca\~nada, 28692 Madrid, Spain\label{aff157}
\and
ARC Centre of Excellence for Dark Matter Particle Physics, Melbourne, Australia\label{aff158}
\and
Centre for Astrophysics \& Supercomputing, Swinburne University of Technology,  Hawthorn, Victoria 3122, Australia\label{aff159}
\and
Department of Physics and Astronomy, University of the Western Cape, Bellville, Cape Town, 7535, South Africa\label{aff160}
\and
DAMTP, Centre for Mathematical Sciences, Wilberforce Road, Cambridge CB3 0WA, UK\label{aff161}
\and
Kavli Institute for Cosmology Cambridge, Madingley Road, Cambridge, CB3 0HA, UK\label{aff162}
\and
Department of Astrophysics, University of Zurich, Winterthurerstrasse 190, 8057 Zurich, Switzerland\label{aff163}
\and
Department of Physics, Centre for Extragalactic Astronomy, Durham University, South Road, Durham, DH1 3LE, UK\label{aff164}
\and
IRFU, CEA, Universit\'e Paris-Saclay 91191 Gif-sur-Yvette Cedex, France\label{aff165}
\and
Oskar Klein Centre for Cosmoparticle Physics, Department of Physics, Stockholm University, Stockholm, SE-106 91, Sweden\label{aff166}
\and
Astrophysics Group, Blackett Laboratory, Imperial College London, London SW7 2AZ, UK\label{aff167}
\and
Univ. Grenoble Alpes, CNRS, Grenoble INP, LPSC-IN2P3, 53, Avenue des Martyrs, 38000, Grenoble, France\label{aff168}
\and
Dipartimento di Fisica, Sapienza Universit\`a di Roma, Piazzale Aldo Moro 2, 00185 Roma, Italy\label{aff169}
\and
Centro de Astrof\'{\i}sica da Universidade do Porto, Rua das Estrelas, 4150-762 Porto, Portugal\label{aff170}
\and
HE Space for European Space Agency (ESA), Camino bajo del Castillo, s/n, Urbanizacion Villafranca del Castillo, Villanueva de la Ca\~nada, 28692 Madrid, Spain\label{aff171}
\and
Dipartimento di Fisica - Sezione di Astronomia, Universit\`a di Trieste, Via Tiepolo 11, 34131 Trieste, Italy\label{aff172}
\and
Department of Astrophysical Sciences, Peyton Hall, Princeton University, Princeton, NJ 08544, USA\label{aff173}
\and
Theoretical astrophysics, Department of Physics and Astronomy, Uppsala University, Box 515, 751 20 Uppsala, Sweden\label{aff174}
\and
Minnesota Institute for Astrophysics, University of Minnesota, 116 Church St SE, Minneapolis, MN 55455, USA\label{aff175}
\and
Mathematical Institute, University of Leiden, Einsteinweg 55, 2333 CA Leiden, The Netherlands\label{aff176}
\and
ASTRON, the Netherlands Institute for Radio Astronomy, Postbus 2, 7990 AA, Dwingeloo, The Netherlands\label{aff177}
\and
Center for Advanced Interdisciplinary Research, Ss. Cyril and Methodius University in Skopje, Macedonia\label{aff178}
\and
Institute of Astronomy, University of Cambridge, Madingley Road, Cambridge CB3 0HA, UK\label{aff179}
\and
Space physics and astronomy research unit, University of Oulu, Pentti Kaiteran katu 1, FI-90014 Oulu, Finland\label{aff180}
\and
Center for Computational Astrophysics, Flatiron Institute, 162 5th Avenue, 10010, New York, NY, USA\label{aff181}
\and
Department of Astronomy, University of Massachusetts, Amherst, MA 01003, USA\label{aff182}
\and
Department of Physics and Astronomy, University of British Columbia, Vancouver, BC V6T 1Z1, Canada\label{aff183}}    

\date{Received 14 March 2025; Accepted 19 July 2025}
 
  \abstract
  {
  Recent \textit{James Webb} Space Telescope (JWST) observations have revealed an interesting population of sources with a compact morphology and a characteristic `v-shaped' continuum, namely blue at rest-frame $\lambda<4000\,\AA$ and red at longer wavelengths. The nature of these sources, called `little red dots' (LRDs), is still highly debated, since it is unclear if they host active galactic nuclei (AGN) and their number seems to drastically drop at $z<4$.
  We take advantage of the $63\,\rm \deg^2$ covered by the quick \Euclid Quick Data Release (Q1) to extend the search for LRDs to brighter magnitudes and to lower redshifts than what has been possible with JWST. This is fundamental to have a broader view of the evolution of this peculiar galaxy population.
  The selection is performed by fitting the available photometric data (\Euclid, the \textit{Spitzer} Infrared Array Camera (IRAC), and ground-based $griz$ data) with two power laws, to retrieve both the rest-frame optical and UV slopes consistently over a large redshift range (i.e, $z<7.6$). We then exclude extended objects and possible line emitters, and perform a careful visual inspection to remove any imaging artefacts. The final selection includes 3341 LRD candidates from $z=0.33$ to $z=3.6$, with 29 detected also in IRAC. 
  The resulting rest-frame UV luminosity function, in contrast with previous JWST studies, shows that the number density of LRD candidates increases from high-redshift down to $z=1.5$--$2.5$ and decreases at even lower redshifts. However, less evolution is apparent focusing on the subsample of more robust LRD candidates having also IRAC detections, which however is affected by low statistics and limited by the IRAC resolution. The comparison with previous quasar (QSO) UV luminosity functions shows that LRDs are not the dominant AGN population at $z<4$ and $M_{\rm UV}<-21$. Follow-up studies of these LRD candidates are pivotal to confirm their nature, probe their physical properties and check for their compatibility with JWST sources, given that the different spatial resolution and wavelength coverage of \Euclid and JWST could select different samples of compact sources.
  }


   \maketitle
%

\section{Introduction}
Supermassive black holes (SMBHs) of tens of millions of Solar masses appear to be ubiquitous at the centres of local galaxies \citep[e.g.,][]{Magorrian1998,Gultekin2009}. Moreover, a close co-evolution linking SMBHs to their host galaxies is suggested by tight scaling relations observed between the SMBH masses and different galactic properties \citep[e.g.,][]{Magorrian1998,Silk1998,Gebhardt2000,Ferrarese2002,Mullaney2012,Delvecchio2022}. However, theoretical models calibrated against present-day scaling relations seem to produce a wide range of SMBH properties at higher redshifts, due to differences in the implementation of supernova and BH feedback and sub-grid physics \citep[e.g.,][]{Habouzit2020,Habouzit2021}. It is therefore important to extend the analysis of SMBHs and their host galaxies to a wide range of times.

Observational evidence for massive accreting BHs shining as AGN at increasingly higher redshifts puts strong constraints on their formation scenarios and the mass of their seeds. Indeed, forming a $10^{9}\, M_{\odot}$ SMBH by $z=7$ requires either a heavy seed ($M_{\rm BH}\sim10^{5}\, M_{\odot}$), a light seed ($M_{\rm BH}\sim10^{2}\, M_{\odot}$) accreting for some time at super-Eddington or hyper-Eddington rates \citep[e.g.,][]{Wyithe2012,Alexander2014,Inayoshi2016,Begelman2017,Pacucci2017,Pacucci2022,Maiolino2024c}, or primordial BHs formed as a result of fluctuations in the early Universe \citep{Hawking1971,Carr1974,Dayal2024}. It is, however, unclear if the currently observed high-$z$ AGN can be considered representative of the whole AGN population. The study of low-luminosity AGN and low-mass SMBHs is therefore key to placing constraints on the mass distribution for seeds of high-$z$ AGN.  

The epoch of reionisation marks the transition phase at which the first sources of ultraviolet (UV) radiation were able, after the so-called `Dark Ages', to ionise hydrogen atoms in the surrounding intergalactic medium for the first time \citep{Barkana2001,Dayal2018}. However, it is still largely debated which sources were responsible for this process. Many studies identify low-mass metal-poor star-forming galaxies at high redshift as the main driver of cosmic reionisation \citep[e.g.,][]{Atek2024,Simmonds2024,Dayal2024b}, while others show evidence that faint AGN can contribute significantly to reionisation \citep[e.g.,][]{Asthana2024}, and in some cases dominate reionisation \citep[e.g.,][]{Madau2024,Grazian2024}. Quantifying the number density of such faint AGN is therefore fundamental to constrain their contribution to reionising the Universe.

The first years of observations with JWST have revealed a new intriguing population of compact red sources characterised by a peculiar `v-shaped' spectral energy distribution (SED), namely a blue rest-frame UV continuum and a steep red slope in the rest-frame optical \citep[e.g.,][]{Kocevski2023,Kocevski2024,Harikane2023,Matthee2024,Greene2024,Labbe2023,Labbe2023b,Killi2024,Furtak2023}. These so-called `little red dots' (LRDs), mainly observed at $z\geq4$, can be easily selected by photometric observations, given their compact morphology and peculiar SED shape. However, particular care needs to be taken to remove contaminating populations, like brown dwarfs that correspond to 21\% of colour-selected JWST LRD candidates, following \citet[][]{Langeroodi2023}. 

The nature of these LRDs is heavily debated. Indeed, their steep rest-frame optical slopes are consistent with either a reddened AGN continuum or emission from dusty star formation \citep[e.g.,][]{Kocevski2023,Barro2023,Labbe2023,Akins2024}, with evidence supporting both scenarios. For example, spectroscopic follow-up studies of LRDs have shown that around 80\% of them show broad hydrogen (H\,$\alpha$ and H\,$\beta$) emission \citep[e.g.,][]{Kocevski2023,Kocevski2024,Kokorev2024,Killi2024,Matthee2024,Furtak2023,Greene2024,Wang_2024}, with line widths of generally $\rm FWHM \leq 1000\,km\,s^{-1}$, but with some going as up as $3000\rm\,km\,s^{-1}$. These line widths would far exceed that of typical star-forming galaxies at lower redshift \citep[e.g.,][]{Fumagalli2012}, supporting the AGN scenario. In this picture, we have a direct view of the broad-line region and the accretion disk of the AGN, but with foreground dust attenuation originating either from a dusty interstellar medium (ISM) or from nuclear dust \citep[see, e.g.,][]{Netzer2015,Hickox2018}. This scenario would make LRDs similar to the `red quasars' observed at lower redshifts \citep[e.g.,][]{Webster1995,Richards2003,Urrutia2008,Glikman2012,Glikman2015}, considering also that some of these red quasar show a similar excess of UV light \citep{Wethers2018,Stepney2024}. While the majority of LRDs may host an AGN, they represent a sub-sample of the overall AGN population. Indeed, out of the large number of broad line AGN discovered by JWST, only 10--30\% have SEDs typical of LRDs \citep{Hainline2024}.

The SMBH masses inferred for these LRDs lie in the range $M_{\rm BH}=10^{6}$--$10^{8}\,\rm M_{\odot}$, while their stellar masses range from $10^{7}$ to $10^{11}\,\rm M_{\odot}$, showing that a fraction of them are over-massive relative to their host galaxies’ stellar masses when compared to the local $M_{\rm BH}$--$M_{*}$ relation \citep[e.g.,][]{Maiolino2024b,Harikane2023}, while they are consistent with the local relation between $M_{\rm BH}$ and stellar velocity dispersion \citep{Maiolino2024b}. This finding indicates they may be going through super-Eddington accretion or originate from heavy BH seeds. 
\citet{Bellovary2025} has instead recently hypothesised that LRDs could be runaway-collapse globular clusters with tidal disruption events, which could explain their compact size, UV luminosities, and large number densities.

In the scenario where LRDs are totally powered by dusty starbursts, they would correspond to very massive galaxies ($M_{*}=10^{9}$--$10^{11}\,\rm M_{\odot}$). However, in this picture, we would end up with an excess of massive early galaxies, some of which in tension with cosmological models and some so compact to be unstable against supernovae feedback \citep{Wang_2024,Akins2024}, supporting the AGN scenario. However, mid-infrared (mid-IR) observations at 5--25$\,\micron\,$ have shown a remarkably flat continuum, favouring SED models consistent with a dusty, compact starburst and only a mild contribution from an obscured AGN \citep{Williams2024,PerezGonzalez2024} or an AGN torus deficient of hot dust \citep{Leung2024,Barro2024}. Up to now, only two LRDs have been detected in the far-IR, providing some indications of warm dust emission \citep{Barro2024,Juodzbalis2024}, while many other LRDs have not been detected in the far-IR limiting the possible presence of cool or warm dust linked to star-formation \citep{Labbe2025}. Moreover, several broad-line objects show a clear Balmer break in their spectra, implying that evolved stars may indeed contribute to the rest-frame optical \citep{Kokorev2024b}, though the precise contribution is highly degenerate \citep{Wang_2024}. 

In addition, the majority of LRDs are not strong X-ray emitters, being undetected (or marginally detected) even in very deep observations or using stacking analysis \citep{Ananna2024,Yue2024}. For this reason, some studies have hypothesised that the broad-line components could not be due to an AGN, but to outflows driven by star formation or inelastic Raman scattering of stellar UV continua by neutral hydrogen atoms \citep{Kokubo2024}. Alternatively, some studies have reported that the broadening could be consistent with the stellar velocity dispersion, if the galaxy is going through a short-lived phase when the central densities are much higher than at later times \citep{Baggen2024}. Other studies have instead suggested that the X-ray weakness could be due to a very steep X-ray spectrum, induced by the absorption by large, Compton thick columns, or a very high BH accretion rate \citep{Maiolino2024}. The last two scenarios would also explain the non-detection at radio frequencies \citep{Mazzolari2024}. The presence of dense neutral gas around the AGN accretion disc would also mimic a Balmer break, indicating that the rest-frame optical may not be due to evolved stars \citep{Inayoshi2024}. However, the absence of variability in the rest-frame UV, except for a few LRDs \citep{Zhang2024}, may point against a strong contribution by AGN \citep{Tee2024,Kokubo2024}.

The uncertainties on the nature of the LRDs are, at least partially, driven by their high-$z$ nature, which implies they are faint and require near-IR observations. While preliminary studies trying to find low-$z$ and local analogues have been performed by, for example, \citet{Noboriguchi2023}, \citet{Mezcua2024}, and \citet{Lin2024}, large near-IR surveys are necessary to follow the redshift evolution of LRDs, since their number density has been suggested to dramatically drop at $z<4$ \citep{Kocevski2024}. Moreover, such large near-IR surveys are key to probe their clustering on large scales with \citet{Tanaka2024} already suggesting that LRDs may show an excess of clustering at kpc scales. The large area observed and the near-IR coverage of \Euclid \citep{EuclidSkyOverview} is therefore ideal to search and study LRDs.

In this work we search for LRDs using the newly available \cite{Q1cite}, combined with publicly available \textit{Spitzer} Infrared Array Camera (IRAC) images and ground-based optical data. The area and depth of these \Euclid observations are ideal to have a first characterisation of the bright end of the LRD luminosity function and to extend the search to $z<4$. To select LRDs, we follow the approach by \citet{Kocevski2024} and select them using cuts on the rest-frame optical and UV slopes.
The paper is structured as follows. In \cref{sec:data} we present a summary of the \Euclid data products used in this work, as well as the ancillary IRAC images. In \cref{sec:selection} we outline our selection method and discuss our findings in \cref{sec:results}. The final conclusion and future prospects are reported in \cref{sec:summary}. Throughout the paper, we consider a $\Lambda$CDM cosmology with $H_0=70\,{\rm km}\,{\rm s}^{-1}\,{\rm Mpc}^{-1} $, $\Omega_{\rm m}=0.27$, $\Omega_\Lambda=0.73$. All magnitudes are reported in the AB system \citep{Oke1983}.

\section{Data description}\label{sec:data}
\subsection{Euclid data products}
An overview of the Q1 data release is described in \citet{Q1-TP001}, the Visible Camera (VIS) and Near-Infrared Spectrometer and Photometer (NISP) processing and data products are in \citet{Q1-TP002} and \citet{Q1-TP003}, respectively, while the photometric catalogue is discussed in \citet{Q1-TP004}. An overview of the \Euclid ESA mission's scientific objectives is reported in \cite{EuclidSkyOverview}. 

Briefly, Q1 includes $63\,\mathrm{deg}^2$ of the extragalactic sky, divided into three fields: $22\,\mathrm{deg}^2$ in the Euclid Deep Field North (EDF-N); $12\,\mathrm{deg}^2$ in the Euclid Deep Field Fornax (EDF-F); and $28\,\mathrm{deg}^2$ in the Euclid Deep Field South (EDF-S).
Each field has been observed in four photometric bands, one in the visible \citep[$\IE$,][]{EuclidSkyVIS}, and three in the near-IR \citep[NISP, $\YE$, $\JE$, and $\HE$ band, see][]{EuclidSkyNISP}. In addition, these observations are complemented with ground-based observations carried out with multiple instruments covering between $0.3\,\micron$ and $0.9\,\micron$, as part of the Ultraviolet Near-Infrared Optical Northern Survey (UNIONS, Gwyn et al. in prep.) or the Dark Energy Survey \citep{DES}. The complete list of filters available in each field and their corresponding observational depths are reported in \cref{tab:filters}.

In this work we consider aperture photometry measured for all bands on the images convolved to the worst spatial resolution (usually a ground based band). We consider an aperture with a diameter of 2 full width at half maximum (FWHM, median value of $1\arcsecf3$) and we correct it to total \citep[see][for more details]{Q1-TP004}. We also correct each flux for galactic extinction, using the position of each source and the relation by \citet{Gordon2023}, which heavily relies on the results by \citet{Gordon2009,Fitzpatrick2019,Gordon2021}, and \citet{Decleir2022}.

\begin{table}[]
    \centering
    \caption{The filters used in this work, with associated observed depths.}\label{tab:filters}
    \resizebox{0.49\textwidth}{!}{
    \begin{tabular}{lccccc}
    \hline \hline
        \noalign{\vskip 2pt}
        Band &  $\lambda_{\rm eff}$ [\micron] & EDF-F & EDF-N & EDF-S \\
        \hline
        \noalign{\vskip 2pt}
        CFHT/MegaCam $u$      & $0.372$ & $    $  & $23.4$ & $    $ \\
        HSC $g$               & $0.480$ & $    $  & $24.9$ & $    $ \\
        CFHT/MegaCam $r$      & $0.640$ & $    $  & $24.0$ & $    $ \\
        PAN-STARRS $i$        & $0.755$ & $    $  & $23.1$ & $    $ \\
        HSC $z$               & $0.891$ & $    $  & $23.3$ & $    $ \\
        Decam $g$             & $0.473$ & $24.6$  & $    $ & $24.7$ \\
        Decam $r$             & $0.642$ & $24.3$  & $    $ & $24.4$ \\
        Decam $i$             & $0.784$ & $23.8$  & $    $ & $23.8$ \\
        Decam $z$             & $0.926$ & $23.1$  & $    $ & $23.1$ \\
        VIS/\IE               & $0.715$ & $24.7$  & $24.7$ & $24.7$ \\
        NISP/\YE              & $1.085$ & $23.1$  & $23.2$ & $23.1$ \\ 
        NISP/\JE              & $1.375$ & $23.2$  & $23.3$ & $23.3$ \\
        NISP/\HE              & $1.773$ & $23.2$  & $23.2$ & $23.2$ \\
        IRAC/IRAC1 & $3.550$ & $24.0$  & $24.0$ & $23.1$ \\
        IRAC/IRAC2 & $4.493$ & $24.0$  & $24.0$ & $23.0$ \\
        IRAC/IRAC3 & $5.696$ & $21.2$ & $20.0$  & \\
        IRAC/IRAC4 & $7.799$ & $19.9$ & $21.1$  & \\
    \hline
    \end{tabular}}
    \tablefoot{Reported magnitudes are the $10\sigma$ observed depths. Optical and \Euclid magnitudes refer to an extended source in a $2\times\rm FWHM$ diameter aperture and correspond to the median depths of the observing tiles \citep{Q1-TP004} For IRAC1 and IRAC2 values see \cite{Moneti-EP17} and \cite{EP-McPartland}, depths correspond to average values in the fields derived considering $2\arcsec$ empty apertures. For the IRAC3 and IRAC channels we report the average depths derived from the catalogue, after correcting from aperture to total magnitudes.}
\end{table}

\subsection{IRAC photometry}\label{sec:IRAC}

\begin{table}[]
\caption{Area coverage in $\rm deg^{2}$ of the available IRAC images \citep{Moneti-EP17}.}
    \centering
    \begin{tabular}{cccc}
    \hline\hline
        \noalign{\vskip 2pt}
        Channel & EDF-F & EDF-N & EDF-S \\
         \hline
        \noalign{\vskip 2pt}
        IRAC1 & 10.52 & 11.74 & 23.60 \\
        IRAC2 & 11.05 & 11.54 & 23.14 \\
        IRAC3 & 7.78 & 0.61 & \dots \\
        IRAC4 & 7.77 & 0.62 & \dots \\
        \hline
    \end{tabular}
    \label{tab:iracarea}
\end{table}

We start from the collection of IRAC images described in \citet{Moneti-EP17} covering a fraction of the Euclid Deep Fields (EDFs) as part of the Cosmic Dawn survey \citep{EP-McPartland}. The images result from a collection of different programmes with non-uniform coverage, both in area and depth. We report the average $10\sigma$ depths in \cref{tab:filters} and the area coverage in \cref{tab:iracarea}. The IRAC3 and IRAC4 filters are available only for a small portion of the EDF-F and EDF-N, while they are not available for the EDF-S.

Starting from the public images, we remove the sky background using the \texttt{PHOTUTILS} python package \citep{Bradley2023}, deriving the median value using a $3 \rm\, pixel\times3\,pixel$ filter. Using the same package, we then use the position of all \Euclid sources to extract aperture photometry with $1\arcsec$ radius aperture, consistent with half the worst FWHM of IRAC, on all available IRAC images. In this preliminary work, no attempt is made to de-blend IRAC sources using \Euclid positions. For this reason a careful visual check is performed when selecting LRD candidates to remove objects affected by blending.

To derive the correction from aperture to total fluxes for all four IRAC filters, we compare this catalogue with a separate extraction, performed only for the EDF-N \citep{Bisigello2025a}. The extraction is performed using as detection image the co-added IRAC1 and IRAC2 images, weighted for each uncertainty map, and considering Kron apertures, derived with a scaling parameter of the unscaled Kron radius of 1.8 and a minimum value for the unscaled Kron radius of 2.5 pixels. We then match the two EDF-N catalogues, with aperture and Kron fluxes, to derive the aperture-to-total correction in each filter. We apply the same correction in all three fields.

We verify that the total fluxes are consistent with the catalogues described in \citet{EP-Zalesky}, which, however, cover only two out of three EDFs and include only the IRAC1 and IRAC2 filters. The comparison is shown in \cref{sec:IRAC_comparison}, showing that magnitudes derived in this work are on average brighter by 0.1 magnitudes in the EDF-F and by 0.3 magnitudes in the EDF-N, with respect to the magnitudes presented in the Cosmic Dawn Catalogue. The agreement is, however, within 0.1 magnitudes if we consider only bright objects (i.e., IRAC1 or $\rm IRAC2<21$).

\subsection{Photometric redshifts}
As a first estimate of the photometric redshift, we consider the median value and the two first modes derived in the main \Euclid pipeline, which are described in detail in \citet{Q1-TP005}. In addition, given that these estimates are limited to $z=6$, we consider also the redshift estimation derived for NISP-detected objects extending the redshift range up to $z=12$. In this case we consider the redshift of the highest peak of the probability distribution as the best estimate. As tested in \cref{sec:z_pipeline}, the redshift estimation for LRDs derived from the pipeline includes about 40\% of outliers, because LRD templates are not included in the pipeline at the moment. We will discuss later the method we use to improve over these estimates.

\section{Sample selection}\label{sec:selection}
\begin{table*}[]
    \caption{Number of sources retrieved by different steps of the LRD selection in the three EDFs. }
    \centering
    \begin{tabular}{lrrr}
    \hline
    \hline
    \noalign{\vskip 2pt}
     & EDF-F & EDF-N & EDF-S \\
     \hline
     \noalign{\vskip 2pt}
      Total sources & 5\,328\,489 & 11\,378\,352 & 13\,060\,965\\
      Reliable objects  & 3\,640\,908 & 7\,342\,804 & 8\,588\,063\\
      \hline
      \noalign{\vskip 2pt}
      IRAC-detected, $z\leq6$ & 2\,762\,173 & 2\,556\,970 & 4\,388\,869\\
      $N_{\rm filter,\rm S/N>3}\geq4$ & 1\,094\,377 (43\%) & 631\,576 (23\%) & 1\,850\,820 (43\%) \\
      v-shaped continuum & 891 (0.08\%) & 616 (0.09\%) & 3848 (0.2\%)\\
      Compact  & 42 (5\%) & 22 (4\%) & 173 (5\%)\\
      No emission lines & 20 (48\%) & 8 (36\%) & 86 (50\%) \\
      $\chi^{2}<100$ & 16 (80\%) & 7 (87\%) & 82 (94\%) \\
      Visual inspection & 8 (50\%) & 1 (14\%) & 20 (24\%) \\
      \hline
      \noalign{\vskip 2pt}
      No-IRAC, $z\leq2.1$  & 1\,021\,175 & 4\,520\,496 & 3\,935\,178\\
      $N_{\rm filter,\rm S/N>3}\geq4$ & 189\,275 (18\%) & 692\,437 (15\%) & 735\,675 (19\%)\\
      v-shaped continuum & 15\,838 (8\%) & 45\,797 (7\%) & 59\,035 (8\%)\\
      Compact & 624 (4\%) & 1611 (3\%) & 2804 (5\%)\\
      No emission lines & 558 (90\%) & 1344 (83\%) & 2560 (91\%) \\  
      $\chi^{2}<100$ & 546 (87\%) & 1233 (76\%) & 2509 (98\%) \\ 
      Visual inspection & 422 (77\%) & 970 (79\%) & 1920 (76\%) \\
      \hline
      \noalign{\vskip 2pt}
     Total $z>6$ candidates & 24\,050 & 71\,776 & 53\,131\\
     Reliable galaxies & 12\,325 (51\%) & 27\,535 (38\%) & 42\,564 (80\%) \\
     IRAC-detected & 3508 (28\%) & 5024 (18\%) & 4534 (11\%) \\
     $N_{\rm filter,\rm S/N>3}\geq4$ & 1883 (54\%) & 3098 (62\%) & 1617 (36\%) \\
     v-shaped continuum & 103 (5\%) & 128 (4\%) & 68 (4\%) \\
     Compact & 0 & 0 & 0 \\
     \hline
    \end{tabular}
    \label{tab:numbers}
    \tablefoot{For the definition of the different selections we refer to  \cref{sec:selection}. Percentages correspond to the percentage of objects selected in one line with respect to the line above.}
\end{table*}

Previous JWST photometric studies have identified LRDs mainly by applying some colour cuts to compact sources \citep[e.g.,][]{Barro2023,Labbe2023b,Kokorev2024}. However, this type of selection is difficult to apply directly to \Euclid, given the different filter set. In addition, a simple colour cut does not allow to have a selection of sources uniform across different redshifts, given that it is based on observed and not rest-frame properties. Therefore, we consider the alternative approach adopted by \citet{Kocevski2024} and select sources with compact morphology, red continuum in the rest-frame optical wavelengths, and blue continuum in the the rest-frame UV. The latter two quantities are directly derived by fitting the available photometric data. Below we report in details the entire classification procedure we apply, while in \cref{tab:numbers} we report the original numbers of sources in each field and how they change at different selection steps. 

\subsection{LRD selection procedure}
As a first conservative selection, we remove objects classified as stars ($\texttt{PHZ\_CLASSIFICATION}=1$) using the classification from the PHZ processing function \citep{Q1-TP005}. It is important to notice that this classification is based on photometry and not on compactness, which is fundamental to not remove any LRD candidates. We also retrieve from the \Euclid archive only galaxies with reliable photometry, defined as having $\texttt{ DET\_QUALITY\_FLAG}=0$, $\texttt{SPURIOUS\_FLAG}=0$, and $\texttt{FLAG}=0$ for any \Euclid filter. This selection allows us to remove objects in the proximity of bright stars, blended sources, saturated or bad pixels, and sources contaminated by close neighbours. We refer to the subsample of objects obtained after this selection step as reliable objects.

We proceed by removing objects outside the area covered by IRAC observations, as these bands are fundamental to extending the search of LRDs at $z>2$. For the same reason we keep only objects with signal-to-noise ${\rm S/N}>3$ in the IRAC1 or IRAC2 filter. We refer to these objects as IRAC-detected sources. Moreover, to fit at the same time both the UV and the optical rest-frame continuum we considered only objects with ${\rm S/N}>3$ in more than four filters. We also assure that there are at least two filters tracing the rest-frame UV continuum and two filters tracing the rest-frame optical one. 

\begin{figure}
    \centering
    \includegraphics[width=0.85\linewidth,trim={18 20 18 17},clip,keepaspectratio]{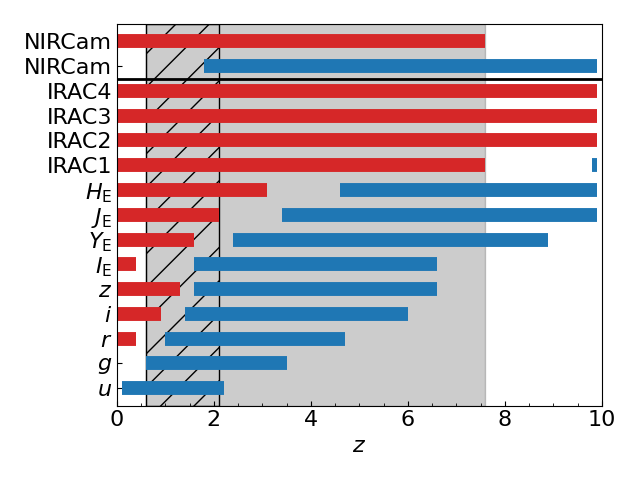}
    \caption{Redshifts in which the different \Euclid, IRAC, and ground-based filters trace the optical (red bars) or UV (blue bars) continuum. The grey shaded area indicates the redshift range in which we have at least four filters to derive the slopes necessary to select LRDs. The hatched area indicates the redshift range covered without IRAC. The first two bars, separated by an horizontal black solid line, indicate the redshift range in which at least two JWST NIRCam broad-band filters trace the rest-frame optical or UV continuum.}
    \label{fig:LRDs_bands}
\end{figure}

We continue by selecting only sources with the characteristic v-shaped continuum, following, as mentioned before, the approach by \citet{Kocevski2024}. In particular, the continuum slope $\beta$, defined such that $f_{\lambda}\propto\lambda^{\beta}$, is determined by performing a $\chi^{2}$ minimisation fit to the observed magnitudes
\begin{equation}
   m_{i} = -2.5~(\beta+2)~\logten {\paren{\frac{\lambda_{i}}{\lambda_{\rm break}}}}+c\;, 
\end{equation}

where $m_{i}$ is the AB magnitude measured in the $i$th filter with an effective wavelength of $\lambda_{i}$ and $\lambda_{\rm break}=3645\,\AA$ is the wavelength of the break of the v-shape continuum. This fit is performed to derive both the rest-frame UV and optical spectral slopes, $\beta_{\rm UV}$ and $\beta_{\rm opt}$. In  \cref{fig:LRDs_bands}, we report the filters used to fit $\beta_{\rm UV}$ and $\beta_{\rm opt}$ at different redshifts, considering both \Euclid, ground-based bands, and the IRAC1 and IRAC2 IRAC filters. We considered a filter to trace the rest-frame UV or optical part of the SED if the filter is totally redward or blueward $\lambda_{\rm break}$, therefore avoiding the redshift interval where the filter is at the break. Given the filters availability and considering the small area coverage by the two longest IRAC filters, the fit is mainly possible between $z=1$ and $z=7.6$, for the EDF-F and EDF-S, while it can be extended down to $z=0.6$ in the EDF-N, thanks to the presence of $u$-band observations. In the same plot we report for comparison the redshift range where at least two JWST NIRCam broad-band filters cover the rest-frame UV and optical parts of the SED. The absence of the shortest filters, which are not often included in JWST surveys, would push the blue limit to higher redshifts. Considering the NIRCam medium or narrow band filters or the MIRI bands could instead increase the redshift range in which the LRD analysis can be performed.

Considering that the redshifts of LRDs are not correctly recovered by the standard \Euclid pipeline (see \cref{sec:z_pipeline}), since at the moment LRD templates are not included, we include the redshift as a free parameter in the fit. We consider the median redshift from the pipeline as an initial guess and the first and second mode (i.e., first and second peak in the redshift probability distribution) redshifts as limits. For candidates at $z>6$, the photometric redshift is derived considering a secondary branch of the pipeline, therefore in this case we consider the first peak of the redshift probability distribution as the initial guess. The fit is performed with the \texttt{Scipy} package \citep{SciPy}. To take into account possible unknown uncertainties, we add in quadrature to the flux uncertainties 5\% of the flux. In the fit we consider all available filters, but we include fluxes with ${\rm S/N<3}$ as 0 with an error equal to two times flux uncertainties, to take into account unknown uncertainties. Uncertainties on the output properties are derived repeating the fit 100 times after randomising the fluxes, considering a Gaussian function centred on the measured value and with $\sigma$ equal to the flux uncertainties.
The performance of the new redshift estimation and the redshift of the pipeline are analysed in \cref{sec:z_pipeline}.

Following the selection done by \citet{Kocevski2024}, we select objects with
\begin{equation}
\begin{cases}
    \;\beta_{\rm opt}>0\;, \\
    \;\beta_{\rm UV}<-0.37\;,\\
    \;\beta_{\rm UV}>-2.8\;.\\
\end{cases}
\end{equation}
The third cut to the UV slope is applied to remove contamination by brown dwarfs. These objects have near-IR colours similar to reddened AGN, but they appear significantly bluer at shorter wavelengths \citep{Langeroodi2023}. In addition, to assure a v-shape SED, when the two longest wavelength filters have ${\rm S/N}>3$, we request that the flux of the filter at the longest wavelength is the highest.

In order to select only compact sources, we consider sources with $\mu_{\rm max}-m_{\rm point-like}<-2.6\,\rm mag\,arcsec^{-2}$, which corresponds to the difference between the peak surface brightness ($\mu_{\rm max}$) above the background in the detection band (\IE for VIS-detected objects and \JE+\YE+\HE for NISP-detected objects) and the expected magnitude for point-like sources ($m_{\rm point-like}$). The chosen threshold has been optimised in the \Euclid pipeline to select compact objects, like stars \citep{Q1-TP004}. 

Given that a red rest-frame optical continuum can be mimicked by the presence of strong nebular emission lines, we performed an additional selection to remove such objects. In particular, the H\,$\alpha+[\ion{N}{II}]$ complex is present in the \YE filter at $z=0.44$--0.85, in the \JE filter at $z=0.77$--1.39, in the \HE filter at $z=1.31$--2.09, and in the IRAC1 filter at $z=3.82$--4.99. In these redshift intervals, we impose that the flux of the contaminated band is lower than the flux of the next redward filter. A negative slope would indeed indicate the presence of strong nebular emission lines, even if the overall optical slope is consistent with our LRD slope selection. We notice that many previous JWST samples do not perform this selection and indeed LRD optical colours may be boosted by nebular emission lines \citep{Hainline2024}. However, a more complex analysis, which we leave for future studies, is necessary to understand if the underlying continuum of these strong line emitters is anyway consistent with LRD selection.   

We then look at the distribution of the $\chi^{2}$ and remove any object with $\chi^{2}\geq100$, with this threshold chosen looking at the overall distribution of $\chi^{2}$ and on random selection of SED. 

Finally, given that the number of sources is limited, we visually check the \Euclid and IRAC cutouts of any remaining object, to remove cases affected by blending issues in the IRAC bands, any remaining artefacts, or sources extended in the NISP filters. Future work, performing a detailed deblending analysis, could improve over this step. Unfortunately, this selection removes potential close pairs, which some LRDs may be part of (e.g., \citealt{Tanaka2024} found three dual LRD candidates over $0.54\deg^{2}$).

We also consider the photometric redshift estimation derived for NISP-detected sources that extend the redshift boundary at $z>6$. As for the previous sample, we select only reliable sources, using the flagging available in the catalogue, we impose a detection in the IRAC filters, a ${\rm S/N}>3$ in at least four bands, and we select only v-shaped and compact sources. The number of sources selected in the different steps are listed in \cref{tab:numbers}, but, in the end, we do not obtain any additional LRD candidates.

At $z=1$--2.1 the search for LRDs can be performed without the IRAC filters, since the \JE and \HE filters cover the rest-frame optical continuum (see  \cref{fig:LRDs_bands}). Therefore, we repeat the selections described above, limiting the analysis to objects with the pipeline redshift $z_{\rm pipeline}<3$, as a conservative cut, but removing any object detected in IRAC. As for the previous selection, we look for objects detected in at least four filters (two blueward and two redward $\lambda_{\rm break}$), with a v-shaped continuum, compact, with no evident contamination from strong nebular emission lines, and with $\chi^{2}<100$. We visually inspect the \Euclid cutouts of any remaining object to remove any left-over artefact or extended objects.

\begin{figure*}
    \centering
    \includegraphics[width=0.915\linewidth,trim={3 0 3 0},clip,keepaspectratio]{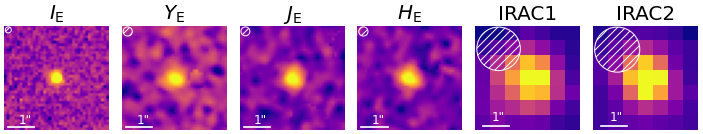}
    \includegraphics[width=0.92\linewidth,trim={3 0 3 25},clip,keepaspectratio]{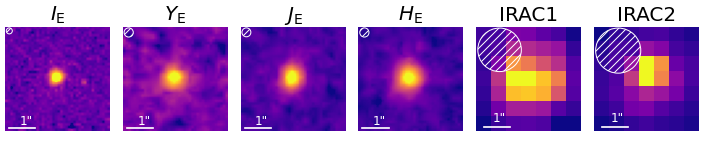}
    \caption{Two example LRD candidates, showing $4\arcsec \times4\arcsec$ cutouts in the four \Euclid filters and the two shortest IRAC channels. From left to right: \IE, \YE, \JE, \HE, IRAC1, and IRAC2. We reported the size of the PSF in the top left and the physical scale on the bottom left of each panel.}
    \label{fig:cutouts}
\end{figure*}

\begin{figure}
    \centering
    \includegraphics[width=0.85\linewidth,trim={15 15 15 15},clip,keepaspectratio]{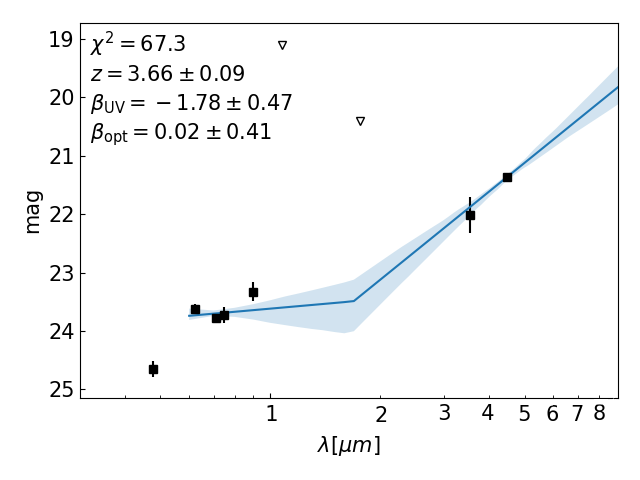}
    \includegraphics[width=0.85\linewidth,trim={15 15 15 15},clip,keepaspectratio]{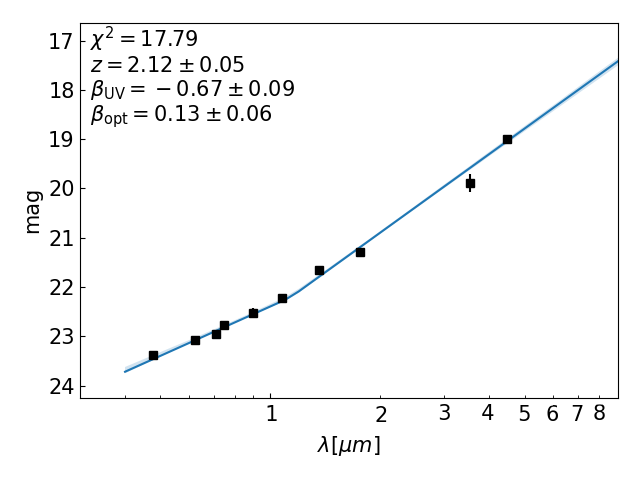}
    \caption{Fit with two power-laws of the photometric data of the two example LRD candidates shown in \cref{fig:cutouts}. We show fluxes with ${\rm S/N}>3$ as black squares, while $3\sigma$ upper limits are shown with empty triangles. The best fit is shown with a blue solid line, while the shaded region shown the 16\% and 84\% uncertainties. We report on the top left the $\chi^{2}$ and the output parameters.}
    \label{fig:SED}
\end{figure}

The final sample of LRD candidates includes 29 objects with IRAC detections, corresponding to a density of $\rm 0.8\,deg^{-2}$, and 3312 objects without IRAC and limited to $z\leq2.1$, corresponding to a density of $\rm 57.3\,deg^{-2}$. The total sample includes 3341 LRD candidates. The number of LRD candidates needs to be considered conservative, given the uncertainties in the rest-frame UV and optical slopes outline in \cref{sec:rand_slopes} and the conservative selection steps performed. The complete list of LRD candidates and their properties is reported in \cref{tab:prop}, while the cutouts and photometric fit of two LRDs are shown as examples in Figs \ref{fig:cutouts} and \ref{fig:SED}. When looking at the cutouts, it is important to take into account that the \Euclid point-spread-function (PSF) is slightly undersampled in \IE, having a pixel scale of 0\arcsecf1 and a FWHM=0\arcsecf158-0\arcsecf164 \cite{Q1-TP002}. The NISP images were instead interpolated to the same pixel scale of \IE, but originally they have a pixel scale of 0\arcsecf3 and a FWHM=0.35”.

The redshift and the magnitude distributions in the three fields is reported in \cref{sec:prop}, showing that the selection is reasonably uniform across the fields.

\subsection{Differences between the IRAC detected and IRAC undetected LRD candidates.}
The difference in density of LRD candidates with and without IRAC is probably driven by several factors. On one side, the number of LRD candidates with IRAC may be underestimated, because blending issues can affect IRAC fluxes, producing a boost in the contaminated band which results in a large $\chi^2$, and at the same time we perform a more strict visual check removing any possible blended source. Indeed, as can be seen in \cref{tab:numbers}, the fraction of sources removed by the $\chi^2$ cut and the visual inspection is larger for IRAC-detected sources. To understand the importance of blending, we verified that around 15\% of all reliable \Euclid sources have a neighbour within $2\arcsec$ (equal to the worst IRAC FWHM) and 5\% within $1\arcsec$. Given that there are hints of LRDs could have an excess of clustering at kpc scales \citep{Tanaka2024}, the effect of blending may be even more strong in LRDs than in the general galaxy population. In addition, the FWHM of IRAC is larger than the radius used for the photometry, so flux loss may effect the detection of sources.

On the other side, the number of LRD candidates selected without IRAC may be overestimated, as the H\,$\alpha+[\ion{N}{II}]$ complex is contained within the \HE filter at $z=1.31$--2.09. The presence of the IRAC bands can help to identify them. In addition, the wider wavelength coverage can simply improve the removal of any type of contaminants.

However, some differences may be intrinsic. Indeed, observations with the JWST mid-infrared instrument (MIRI) have shown that the rest-frame continuum at $\lambda \gtrsim0.7\,\rm\mu m$ becomes remarkably flat, indicating a mild contribution from an obscured AGN \citep{Williams2024,PerezGonzalez2024} or an AGN torus deficient of hot dust \citep{Leung2024,Barro2024}. At $z<2.8$ IRAC bands cover the same rest-frame wavelengths as MIRI at $z>5$ (i.e., $\lambda\geq0.7\,\rm \mu m$), covering the flatter part of the SED.

\section{Results}\label{sec:results}

\subsection{Comparison with JWST LRD catalogues}\label{sec:jwstcomp}
\begin{table}[]
\caption{FWHM of the \Euclid VIS/NISP, JWST NIRCam, and IRAC point spread functions.}
    \centering
    \resizebox{0.5\textwidth}{!}{
    \begin{tabular}{cccc}
    \hline \hline
    \noalign{\vskip 2pt}
        Filter & \multicolumn{3}{c}{FWHM} \\
         & [arcsec] & [kpc] at $z=1$ & [kpc] at $z=6$ \\
         \hline
         \noalign{\vskip 2pt}
        \Euclid/\IE & 0.16 & 1.3 & 0.9 \\
        \Euclid/\HE & $0.3$ & 2.5 & 1.8\\
        JWST/F070W & $0.029$ &  0.2 & 0.17\\
        JWST/F150W & $0.050$ & 0.4 & 0.29 \\
        JWST/F356W & $0.116$ & 0.96 & 0.67\\
        JWST/F444W & $0.145$ & 1.19 & 0.85\\
        IRAC/IRAC1 & $1.66$--$1.95$$^{a}$ & 13.7--16.1 & 9.7--11.4\\
        IRAC/IRAC2 & $1.72$--$2.02$$^{a}$ & 14.2--16.6 & 10.0--11.8\\
        \hline
        
    \end{tabular}}
    \tablefoot{Citations: \Euclid/\IE \citep{Q1-TP002}, \Euclid/\HE \citep{Q1-TP003}. JWST \citep{Rigby_2023}. $^{a}$ The two numbers refer to when \textit{Spitzer} was cryogenically cooled and when it was warm, respectively.}
    \label{tab:FWHM_comp}
\end{table}

\begin{figure}
    \centering
    \includegraphics[width=0.9\linewidth,trim={15 20 15 30},clip,keepaspectratio]{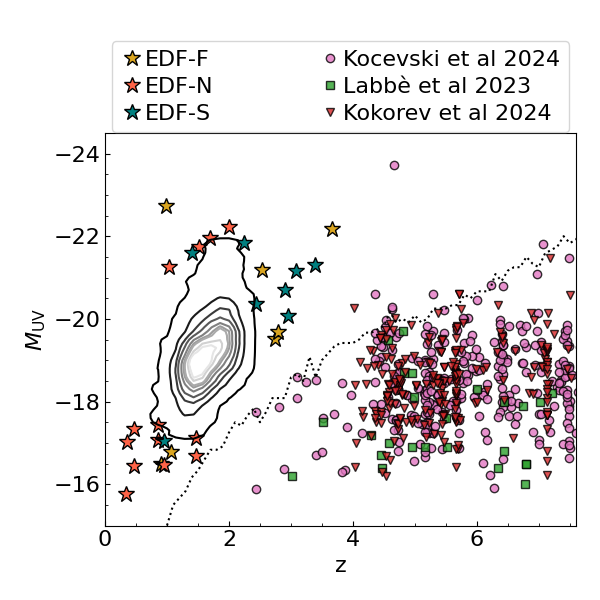}
    \caption{$M_{\rm UV}$ and redshift for all LRD candidates, show as contour lines, equally spaced from 10\% to 90\% with the last one representing 99\% of the distribution. The remaining 1\% of the sample is shown with stars, colour-coded based on their field. We also show three samples of LRDs selected with JWST observations \citep{Kocevski2024,Labbe2023,Kokorev2024,Akins2024}. The black dotted line show the 80\% completeness expected once the EDFs are at their final depth.}
    \label{fig:MUVz}    
\end{figure}

In \cref{fig:MUVz} we show a comparison of the redshift and $M_{\rm UV}$ at 1450\,\AA \, of our sample of LRD candidates, derived by fitting the available data with two power laws, as explained in detail in \cref{sec:z_pipeline}. We also compare our results with some of the previous samples derived with JWST data \citep{Kocevski2024, Labbe2023,Kokorev2024}. Indeed, our sample is complementary to the JWST ones derived up to now, since it covers brighter magnitudes, with only one JWST source being as bright as \Euclid selected sources. It is also clear that our sample extends to lower redshifts compared to the JWST one. In the same figure we also show the $M_{\rm UV}$ that we expect to reach at the end of the \Euclid mission. Indeed, the EDFs will be two magnitudes deeper when completed, which, even with IRAC data remaining unchanged, will allow us to reach $M_{\rm UV}=-20$ at $z=4$, extending the overlap with JWST, but also probing the bright end up to $z=8$.

It is necessary, however, to consider that, even if we select only compact sources, the different angular resolution of JWST and \Euclid could have an impact on the selection. In \cref{tab:FWHM_comp} we report the FWHM of the two missions at similar wavelengths, with the corresponding physical scales given at $z=1$ and $z=6$. The difference with respect to IRAC is even larger, reaching a factor of at least 10 in physical size. Therefore, follow-up campaigns will be necessary to confirm that the LRD candidates identified with \Euclid and IRAC are as compact as the JWST ones.

\begin{figure}
    \centering
    \includegraphics[width=0.9\linewidth,trim={20 20 20 20},clip,keepaspectratio]{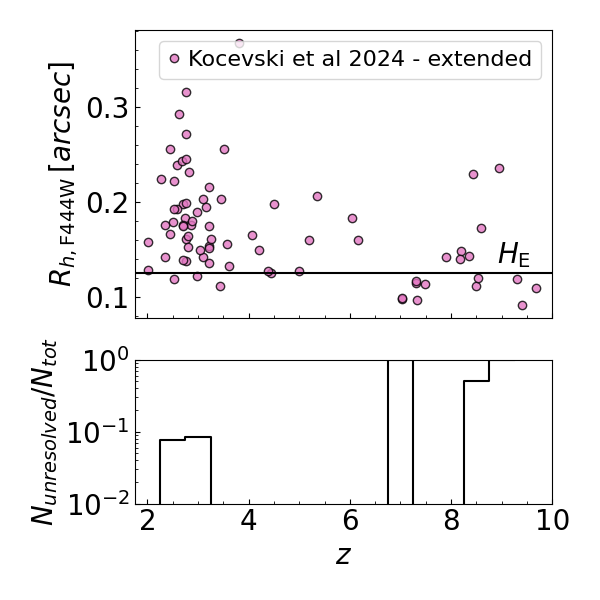}
    \caption{Top: Half-light ratio in the JWST/F444W filter vs. redshift for the sample of extended v-shape objects by \citet{Kocevski2024}. The horizontal solid line indicate the \HE band FWHM, scaled assuming a Gaussian function. Bottom: fraction of extended v-shape objects that is expected to be unresolved in the \HE band, as a function of redshift. }
    \label{fig:extendedLRD}
\end{figure}

To understand the level of contamination due to the different angular resolutions of JWST and \Euclid, we analysed the subsample of extended v-shape sources from \citet[][priv. com.]{Kocevski2024}. These sources are mainly dusty star-forming galaxies and correspond to 14\% of all v-shape sources, but their redshift distribution is skewed toward $z<4$ \citep[see Fig. 7 in][]{Kocevski2024}. In \cref{fig:extendedLRD} we compared the size of these extended v-shape sources, derived from the JWST/F444W filter, with the \Euclid angular resolution. In particular, both the JWST/F444W filter and the \HE filter trace the rest-frame optical of the LRDs up to $z=3.1$ and can therefore be directly compared. From this comparison it is evident that the majority of these sources are extended also for \Euclid, with only 8\% (15) of them that are below the \HE filter FWHM at $z=2$--4. Combining this finding with the redshift distribution of the LRD candidates by \citet{Kocevski2024}, we obtained that we would expect 15\% of contaminants at  $z=2$--4.

At lower redshift these contaminants are expected to contribute even less, but we do not have direct measurements as the sample by \citet{Kocevski2024} is limited to $z>2$. However, considering the redshift evolution of the size mass relation for late-type galaxies \citep{Wel2014}, we expect a median increase of the size of galaxies by 1.2 between $z=2.25$ and $z=1.75$ and 1.9 between $z=2.25$ and $z=0.75$. The smallest extended v-shape source at $z<4$ by \citet{Kocevski2024} is only a factor of 0.88 the \Euclid angular resolution in the \HE band, so we expect all sources at $z<2$ to be resolved. 

\subsection{Comparison with other \Euclid AGN catalogues}\label{sec:QSO}

Other works have focused on the selection of AGN using, mainly, Q1 photometric data and we therefore investigate the amount of overlap with our sample of LRD candidates. 
In particular, the \Euclid AGN catalogue presented by \citet{Q1-SP027} focuses on the selection of AGN, mainly blue QSOs, based on \Euclid photometry, but also ancillary data from UV to IR, as well as previous public spectroscopic data. \citet{Q1-SP003} instead did a careful matching of \Euclid sources with public X-ray surveys, identifying X-ray detected AGN. The details of these selections and the overlap with our samples of LRD candidates is reported in \cref{sec:QSO-appendix}. 
Overall, in the EDF-F and EDF-S, no sources are identified by all QSO selections, while 1898 (80\%) are selected by at least one criterion. In the EDF-N, 816 sources, which corresponds to 84\% of the EDF-N sample of LRD candidates, is selected by at least one QSO criterion, while no sources are selected by all criteria.

\subsection{LRDs luminosity function } \label{sec:LF}
To measure the UV luminosity function of our sample of LRD candidates, we used the 1/ $V_{\rm max}$ method \citep{Rowan-Robinson1968,Schmidt1968}. In particular, for each magnitude and redshift bin the luminosity function is defined as
\begin{equation}
\Phi(M)dM=\frac{1}{\Delta M}\sum_{i}^{N}\frac{1}{w_{i}\,V_{{\rm max,}i}}\;,
\end{equation}
where $\Delta M$ is the width of the magnitude bin, $w_{i}$ is the completeness correction for the $i$th object, and $V_{{\rm max,}i}$ is the maximum comoving volume at which object $i$ could have been detected. To calculate the latter, we need to consider the area covered by the observations, the minimum redshift of the bin and the maximum redshift at which each source could be observed. The latter is derived considering the v-shaped model of each source, described by two power laws with slopes $\beta_{{\rm UV},i}$ and $\beta_{{\rm opt},i}$, normalised to the absolute UV magnitude ($M_{{\rm UV},i}$). This model is then shifted from the minimum to the maximum redshift of the redshift bin, and at each redshift step it is convolved with the \Euclid and ancillary bands to estimate the expected fluxes. These fluxes are then used to derive the maximum redshift at which we would have at least four filters with ${\rm S/N}>3$.

\subsubsection{Area}

The area associated with each field has been derived considering the coverage map for the four \Euclid filters, combined to remove the area in which at least one filter has been masked. In addition, we also removed additional masked areas covering bright stars, both halo and ground bleeding trails, as well as extended bright foreground sources. In addition, we combined these masks with the coverage by IRAC to derive the area associated with IRAC-detected sources. The areas considered are listed in \cref{tab:area_LF}.

\begin{table}[]
    \centering
    \caption{Area in $\rm deg^{2}$ used to estimate the LRD luminosity function.}
    \begin{tabular}{cccc}
        \hline \hline
    \noalign{\vskip 2pt}
         & EDF-F & EDF-N & EDF-S \\
         \hline
         \noalign{\vskip 2pt}
       After masking & 11.85 & 19.76 & 26.16\\
       w/IRAC  & 2.57 & 13.46 & 21.10\\
       \hline
    \end{tabular}
    \label{tab:area_LF}
\end{table}

\subsubsection{Completeness limits and correction}

To derive the 80\% completeness limit of our sample, we derived the fraction of LRDs having at least four filters with ${\rm S/N}>3$ as a function of both redshift and $M_{\rm UV}$. In particular, we considered redshift bins of $\Delta z=0.25$ up to $z=6$, since no LRDs are found at higher redshifts. We also considered magnitude bins of $\Delta M_{\rm UV}=0.25$ from $M_{\rm UV}=-25$ to $M_{\rm UV}=-10$. In each redshift-magnitude bin, we randomly extracted 1000 $\beta_{\rm UV}$ and $\beta_{\rm opt}$ values from the observed distribution. We then derived the fraction of these mock LRDs that are observed in at least four filters with ${\rm S/N}>3$, considering the depths of the different fields reported in \cref{tab:filters}. We used this estimate to correct the derived luminosity function.

In addition, we also applied the corrections for wrong redshifts, derived from the analysis of the redshift recovery on the simulated sample in \cref{sec:z_pipeline}. This is a correction that varies from 0.6 for $z=1$--$1.5$, since the number of sources in this redshift range is expected to be overestimated, to 1.3 for $z=2$--$2.5$, where instead the number of objects is underestimated. A more detailed completeness correction, taking into account for example the compactness of the sources, will be estimated in future works, once the systematic effects of the telescope are better understood. 

Finally, we considered a correction of 15\% due to the contamination by extended sources that are unresolved by \Euclid (see \cref{sec:jwstcomp}). This contamination is expected to increase at $z>6$, but we found no LRD candidates at these high-$z$ values. At the same time, this correction is expected to decrease at $z<2$ due to the average increase of galaxy size with decreasing redshift, but we lack the data to properly quantify this decrease. Therefore, as a conservative approach, we applied the same correction at all redshifts.

\subsubsection{Uncertainties}

We estimated the uncertainties of the luminosity function by performing a bootstrap analysis of the sample, generating 100 random samples starting from the entire LRD sample. In addition, on every realisation, we randomized both the redshift and the absolute UV magnitude considering a Gaussian function centred on the best-fit value and with standard deviation equal to their respective uncertainties. We added in quadrature to these uncertainties the Poisson errors following the prescription by \citet{Gehrels1986}. We do not include any uncertainty due to cosmic variance given the large areas analysed in this work.

\subsubsection{Estimated luminosity functions}

\begin{figure*}[h!]
    \centering
    \includegraphics[width=0.45\linewidth,trim={0 40 0 0},clip,keepaspectratio]{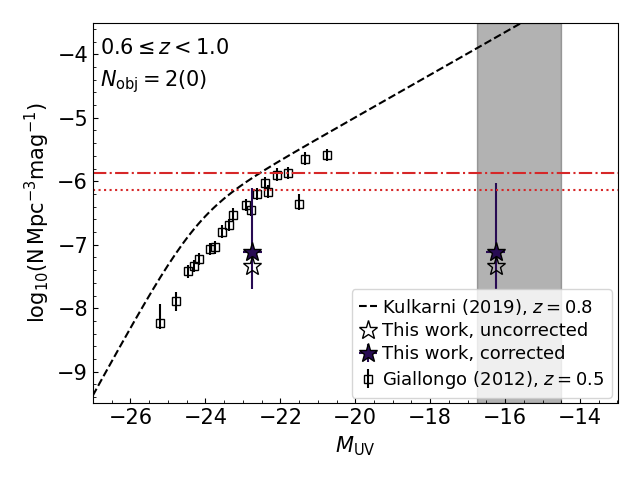}
    \includegraphics[width=0.413\linewidth,trim={40 40 0 0},clip,keepaspectratio]{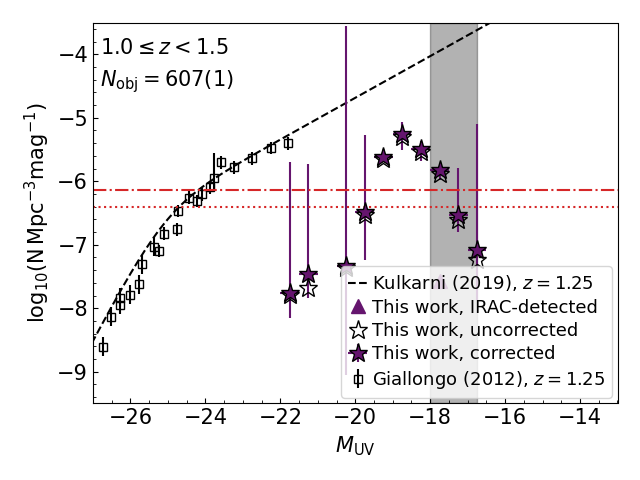}
    \includegraphics[width=0.45\linewidth,trim={0 40 0 0},clip,keepaspectratio]{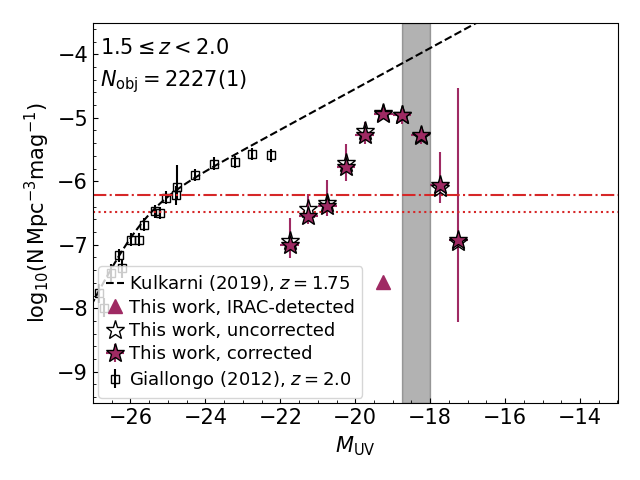}
    \includegraphics[width=0.413\linewidth,trim={40 40 0 0},clip,keepaspectratio]{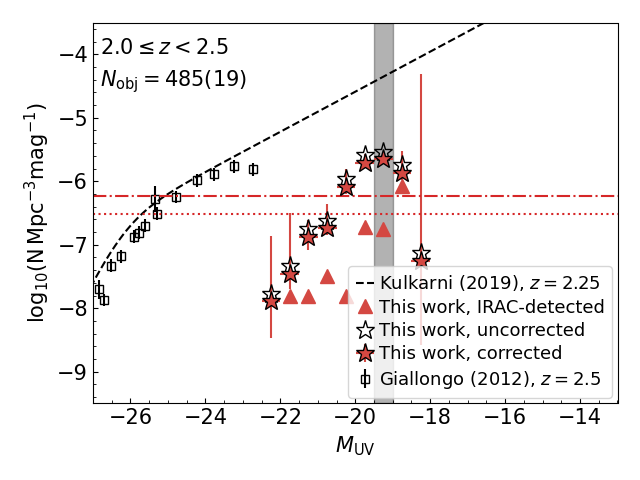}
    \includegraphics[width=0.45\linewidth]{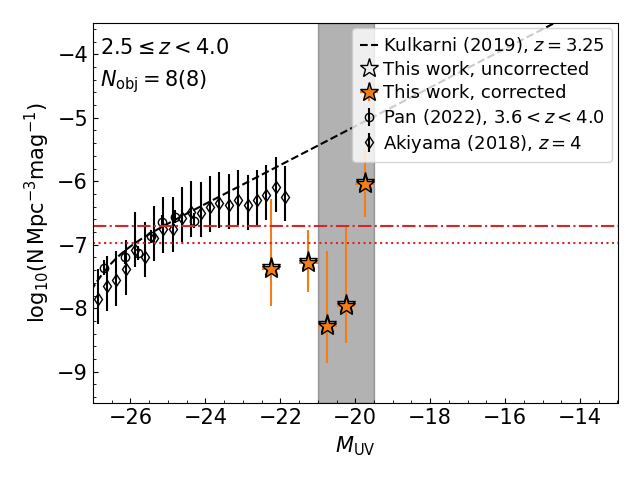}
    \caption{UV luminosity function of the LRD candidates from $z=0$ (top left) to $z=4$ (bottom right). Empty stars indicate the values before any corrections, while the coloured stars show the density after correcting for completeness and redshift estimation. Coloured triangles show the conservative estimation derived considering only IRAC-detected sources. The black vertical shaded regions show the range of the 80\% completeness in the three fields. In the top left we report the number of objects inside the redshift bin and, in brackets, the number of IRAC-detected sources in the bin. Empty symbols show UV luminosity functions of QSOs from \citet[][and reference therein]{Giallongo2012}, \citet{Akiyama2018}, and \citet{Pan2022}. We also report the UV luminosity function of QSOs modelled by \citet[][black dashed line]{Kulkarni2019}. For a comparison, the red horizontal dash-dotted line shows the minimum density probed by HST CANDELS survey \citep[$0.29\deg^{2}$][]{Grogin2011,Koekemoer2011}, while the red horizontal dotted line shows the minimum density probed by the JWST COSMOS-WEB survey \citep[0.54$\deg^{2}$][]{Casey2023}.}
    \label{fig:LF}    
\end{figure*}

\begin{figure}
    \centering
    \includegraphics[width=\linewidth,trim={15 15 15 16},clip,keepaspectratio]{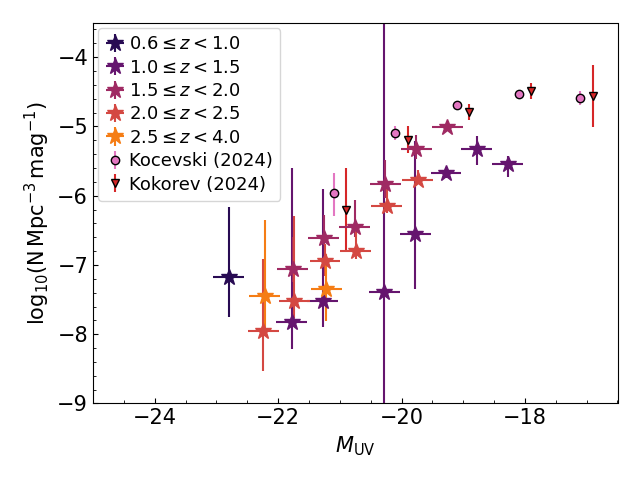}
    \caption{Comparison of the corrected UV luminosity function of LRD candidates derived in this work from $z=0$ to $z=4$, as well as values at $z=6$ from the literature \citep{Kocevski2024,Kokorev2024}.}
    \label{fig:LF_all}
\end{figure}

We derived the UV luminosity functions of LRDs as explained in the previous sections and considering the following redshift bins: $0.6\leq z <1.0$; $1.0\leq z <1.5$; $1.5\leq z <2.0$; $2.0\leq z <2.5$; and $2.5\leq z <4.0$. In the first redshift bin we considered only objects in EDF-N, as the other fields lack the $u$ band observations necessary to properly trace the UV slope. 11 objects (0.3\% of the sample) are not considered in the UV luminosity functions because they correspond to a redshift where the UV slope is not properly covered. The derived UV luminosity functions are shown in  \cref{fig:LF}, where we also compare our results with UV luminosity functions from the literature of both LRDs, based on JWST data \citep{Kokorev2024,Kocevski2024}, and QSO, including JWST data \citep{Maiolino2024b,Harikane2023,Grazian2024} or before JWST \citep{Giallongo2012,McGreer2013,Giallongo2015,Akiyama2018,Niida2020,Pan2022}. The values of the LRD luminosity function of this work are also reported in \cref{tab:LF}. When analysing these results, it is necessary to consider that the luminosity function based only on \Euclid data may be overestimated, as there may be contaminants that have not been properly included in the corrections. On the other hand, the luminosity function based on IRAC observations could be severely underestimated, given that we strictly remove any source affected by blending. \\

 In the highest redshift bin (i.e., $4.0\leq z <6.0$) no candidates are found, making comparisons with previous JWST results impossible. We must therefore wait for future \Euclid releases, to improve over these results. Indeed, the depth of the EDFs is expected to increase by 2 magnitudes until the end of the mission \citep{EuclidSkyOverview}. We predict that with the final depths of the EDFs \Euclid will be able to observe the brightest LRDs observed by JWST up to now (\cref{fig:MUVz}).

 At redshifts $z<4$ we can compare only to previous QSO UV luminosity functions, since observations of LRDs with JWST were mainly limited to $z>4$. The luminosity function of the LRDs is always well below the QSO luminosity functions at bright magnitudes by 1.2--2.6 dex, which however corresponds to 1--4$\sigma$ because of low number statistic. This difference generally indicates that LRDs are not the dominant AGN population at these magnitudes at $z<4$, assuming they are AGN. The LRD luminosity functions are instead closer to the QSO ones at at $z > 1$ and at the faintest magnitudes probed by this work. The closest point is at $1.5\leq z<2.0$, where the LRD luminosity function is only 0.7\,dex below the QSO luminosity function around $M_{\rm UV}=-19.25\,\rm mag$, but given the large statistic the uncertainties are small and this difference corresponds to more than 10$\sigma$. We cannot verify if the LRD luminosity function is closer to the QSO one at even fainter magnitudes because of incompleteness. 

 In the same figure we also report the maximum volume probed by one of the largest JWST survey so far, that is COSMOS-Web \citep{Casey2023} covering $0.6\,\rm deg^{2}$ (red dotted line in \cref{fig:LF}), and the Cosmic Assembly Near-infrared Deep Extragalactic Legacy Survey \citep[CANDELS][]{Grogin2011,Koekemoer2011} covering $0.29\,\rm deg^{2}$ with \textit{Hubble} Space Telescope (HST, red dash-dotted line in \cref{fig:LF}). At $z<1$ and $z>2.5$ the area of these two surveys are too small to observe a significant number of objects with luminosities similar to the candidates selected in this work. However, we would expect to observe some objects at the other redshifts: $17_{-7}^{+8}$ at $1.0\leq z<1.5$; $62_{-16}^{+16}$ at $1.5\leq z<2.0$; and $15_{-5}^{+4}$ at $2.0\leq z<2.5$ in CANDELS; $31_{-13}^{+15}$ at $1.0\leq z<1.5$; $116_{-30}^{+30}$ at $1.5\leq z<2.0$; and $29_{-8}^{+7}$ at $2.0\leq z<2.5$ in COSMOS-Web. However, at these redshifts, no LRD was identified with HST and few with JWST. For example, \citet{Kocevski2024} found 17 LRD candidates at $z<4$, while in the same area, given the LRD luminosity function found in this work, we would expect $56_{-10}^{+11}$ candidates. This difference ($2\sigma$ deviation) in number density may be due to contaminants or may be due to the different angular resolution of \Euclid and JWST, making future follow-up studies fundamental. However, it is important to notice that the subsample of more robust LRD candidates, that are the one observed also by IRAC, have a number density too low to be detected by any JWST or HST survey. 

 We also report the conservative luminosity function derived only with the IRAC-detected sources, which have more constraints but have also been more heavily cleaned to avoid blending issues. This estimations are compatible with the few LRD candidates at $z<4$ observed by JWST and HST.

A comparison between the LRDs UV luminosity functions derived in this work is shown in \cref{fig:LF_all}. From this plot it is possible to identify a tentative evolution of the luminosity function with redshift. Indeed, the density of LRD candidates increases from the high redshift down to $z=1.5$--$2.5$ and decreases at even lower redshifts. For example, at $1.5\leq z<2.0$ and $M_{\rm UV}=-20.25\,\rm mag$ we obtained a density of $\logten(N\,\rm Mpc^{-3}\,mag^{-1})=-5.83^{+0.35}_{-0.20}$. Our estimate at the same magnitude but at $2.0\leq z<2.5$ is slightly lower, that is $\logten(N\,\rm Mpc^{-3}\,mag^{-1})=-6.15^{+0.32}_{-0.10}$, but it may be incomplete given the necessity of using IRAC data at $z>2.1$. On the contrary, the drop visible at $z<1.5$ is not linked to observational biases, as we expect \Euclid data to have the same capacity of selecting LRDs down to $z=0.7$. To give a more quantitative estimate, at $M_{\rm UV}=-19.25\,\rm mag$ the LRD density is $\logten(N\,\rm Mpc^{-3}\,mag^{-1})=-5.67^{+0.07}_{-0.07}$ at $1.0\leq z<1.5$ while it is $\logten(N\,\rm Mpc^{-3}\,mag^{-1})=-5.01^{+0.02}_{-0.06}$ at $1.5\leq z<2.0$, with a drop of 0.67\,dex. The statistics in the lowest redshift bin are quite poor, since it is based on the EDF-N only, preventing us to further explore the possible redshift evolution. However, the density of LRD candidates with $M_{\rm UV}\geq -21\,\rm mag$ at $1\leq z <2.5$ is also similar to the one derived with JWST at $4\leq z <6$ \citep{Kocevski2024,Kokorev2024}, which is $\logten(N\,\rm Mpc^{-3}\,mag^{-1})=-5.19\pm0.20$ and $\logten(N\,\rm Mpc^{-3}\,mag^{-1})=-5.10\pm0.1$ respectively, supporting the idea that the high-$z$ evolution could be due to incompleteness in our LRD sample. This evolution is not present if we consider only the few LRD candidates detected in IRAC, as statistic is too low.

Overall, we may speculate that the LRDs number density seems to remain constant from $z=4$--6 down to $z=1.5$--2.5 and this resemble the observed evolution of the star formation rate density \citep[e.g.,][]{Madau2014,Gruppioni2020,Traina2024}, the BH accretion rate density \citep[e.g.,][]{Delvecchio2014} and in the molecular gas mass \citep[see][for a review]{Tacconi2020}. If we consider that LRDs may be very dense and dust-obscured starbursts or obscured SMBHs maybe accreting at super-Eddingthon rate, one could think that their frequency follow the trend visible for the more general population of star-forming galaxies or AGN, which is at the end regulated by the availability of cold gas.

Further analysis is needed to understand the level of contamination in our sample or why these LRD candidates have been missed by previous JWST studies.  A possibility could also be that these sources are not as compact as the ones observed with JWST, being compact when considering \Euclid spatial resolution, but not with JWST, which is 4--6 times better that the \Euclid's. Following the estimates from \cref{sec:jwstcomp} we considered 15\% of such contaminants in the estimation of the luminosity function, but detailed spectroscopic follow-ups are necessary to properly quantify the purity of our sample of LRD candidates. If the sample is dominate by contaminants, considering that the number densities seems to remain similar from $z=6$ to $z=2$, it could be that we are following different evolutionary stages of the same galaxy population. 

In the future, as the depth of the EDFs increases and the spectroscopic coverage improves, the identification of LRDs and the estimation of their properties will improve, making it possible to derive more conclusive results on the possible redshift evolution of the LRD luminosity function.

\begin{table*}[]
\renewcommand{\arraystretch}{1.2}
    \caption{UV luminosity function of LRD candidates.}
    \centering
    \begin{tabular}{cccccc}
\hline\hline
\noalign{\vskip 2pt}
$M_{\rm UV}$  & $0.6\leq z < 1.0$ & $1.0\leq z <1.5$  & $1.5\leq z <2.0$  & $2.0\leq z <2.5$ & $2.5\leq z <4.0$\\
\hline
\noalign{\vskip 2pt}
$-22.75$ &$ -7.18 _{- 0.57 }^{+ 1.01 }$ &                             &                               &                              & \\
$-22.25$ &                              &                             &                               &$ -7.95 _{- 0.57 }^{+ 1.03 }$  &$ -7.45 _{- 0.57 }^{+ 1.1 }$\\
$-21.75$ &                              &$ -7.82 _{- 0.40 }^{+ 2.22 }$ &$ -7.06 _{- 0.19 }^{+ 0.47 }$ &$ -7.52 _{- 0.25 }^{+ 1.23 }$  & \\
$-21.25$ &                              &$ -7.52 _{- 0.38 }^{+ 1.61 }$ &$ -6.61 _{- 0.11 }^{+ 0.34 }$ &$ -6.94 _{- 0.22 }^{+ 0.24 }$  &$ -7.35 _{- 0.45 }^{+ 0.51 }$\\
$-20.75$ &                              &                              &$ -6.45 _{- 0.15 }^{+ 0.39 }$ &$ -6.80 _{- 0.12 }^{+ 0.4 }$   &$ -8.34 _{- 0.57 }^{+ 1.19 }$\\
$-20.25$ &                              &$ -7.39 _{- 1.66 }^{+ 4.01 }$ &$ -5.83 _{- 0.20 }^{+ 0.35 }$ &$ -6.15 _{- 0.10 }^{+ 0.32 }$  &$ -8.03 _{- 0.57 }^{+ 1.40 }$\\
$-19.75$ &                              &$ -6.55 _{- 0.80 }^{+ 1.33 }$ &$ -5.33 _{- 0.14 }^{+ 0.21 }$ &$ -5.77 _{- 0.04 }^{+ 0.13 }$  &$ -6.11 _{- 0.52 }^{+ 0.58 }$ \\
$-19.25$ &                              &$ -5.67 _{- 0.07 }^{+ 0.07 }$ &$ -5.01 _{- 0.06 }^{+ 0.02 }$ &$ -5.72 _{- 0.06 }^{+ 0.10 }$ & \\
$-18.75$ &                              &$ -5.32 _{- 0.23 }^{+ 0.19 }$ &$ -5.02 _{- 0.14 }^{+ 0.11 }$ &$ -5.92 _{- 0.11 }^{+ 0.29 }$ \\
$-18.25$ &                              &$ -5.55 _{- 0.18 }^{+ 0.12 }$ &$ -5.33 _{- 0.14 }^{+ 0.09 }$ &$ -7.32 _{- 1.28 }^{+ 2.94 }$  \\
$-17.75$ &                              &$ -5.88 _{- 0.15 }^{+ 0.07 }$ &$ -6.12 _{- 0.25 }^{+ 0.45 }$ &\\
$-17.25$ &                              &$ -6.60 _{- 0.27 }^{+ 0.77 }$ &$ -6.98 _{- 1.35 }^{+ 2.39 }$ &\\
$-16.75$ &                              &$ -7.15 _{- 0.59 }^{+ 1.77 }$ &                             &\\
$-16.25$ &$ -7.18 _{- 0.57 }^{+ 1.07 }$ &                              &                             &\\

\noalign{\vskip 2pt}
\hline 
    \end{tabular}
    \label{tab:LF}
    \tablefoot{The first column show the central value of the UV absolute magnitude bins, which are 0.5 mag wide. The other columns show the logarithm of the number density in $\rm Mpc^{-3}\,mag^{-1}$ in different redshift bins, reported in the first row.}

\end{table*}

\section{Conclusions}\label{sec:summary}
In this work we have taken advantage of the \Euclid Q1 data covering around $63\,\rm deg^2$ to search for LRD candidates. The selection was performed exploiting \Euclid photometric data, ground-based ancillary data, and \text{Spitzer}/IRAC data. After a conservative selection, including conservative flagging and careful visual inspection of all candidates, we obtained a final sample of 3341 LRD candidates, 29 of which are detected also with IRAC. 

Even if we impose a {\rm S/N>3} in at least four filters, LRD candidates are relatively faint, with mean magnitudes of $\IE=25.5$, $\YE=24.3$, $\JE=24.0$, and $\HE=23.7$, which are fainter than the $5\sigma$ limits of the Q1 data release. Their photometric redshift has been estimated to range between $z=0.33$ and $z=3.6$, where previous JWST selections identified a rapid decline of LRD candidates. However, at $z>4$, the depth of the ancillary IRAC data limit our sample to only the brightest sources.

We also derived the rest-frame UV luminosity function, but no overlap is present in the parameter space of \Euclid and JWST, limiting direct comparison. Indeed, \Euclid is complementary to JWST observations because it can probe the bright-end of the UV luminosity function, thanks to the large area covered. The most puzzling result is that the UV luminosity function of LRD candidates increases from high $z$ down to $z=1.5$--$2.5$, below which it decreases again. This is in contrast with previous JWST results which derived a drastic drop of LRDs at $z<4$, with almost no candidates at $z<2$. However, the $z>4$ evolution of our luminosity function may be affected by incompleteness in our LRD candidates sample. It is important to note that no significant evolution is apparent focusing on the subsample of more robust LRD candidates having also IRAC detections, which however has poor statistics (only 29 objects) and is limited by the IRAC resolution. Clearly, more observations will be required to clarify this situation. 
Another interesting result is that the LRD UV luminosity function remains below the QSO one, except at $M_{\rm UV}>-21$, where they become compatible. This could indicate that, if LRDs are indeed AGN, they are anyway a sub-dominant AGN population.

Further analyses are necessary to validate the LRD candidates of this work and optimise the removal of possible contaminants. In addition, further critical studies need to probe their structure and check for their compatibility with JWST sources, given the different spatial resolution of the two telescopes. If a sizeable fraction of the LRD candidates identified in this work are confirmed by future studies, it will enable us to study the redshift evolution over a broad redshift and luminosity range thus helping shedding light on these mysterious sources. Future \Euclid data releases are expected to increase the number of LRD candidates, but also to increase the overlap in parameter space with previous JWST samples, as the \Euclid Wide survey will be wider than the Q1 at similar depth, while the \Euclid Deep survey will be considerably deeper. In addition, a direct comparison between JWST and \Euclid selections will be possible in future releases, as \Euclid will cover some of the area already observed with JWST.

\noindent
\textit{Data availability:} \cref{tab:prop} is only available in electronic form at the CDS via anonymous ftp to cdsarc.u-strasbg.fr (130.79.128.5) or via \href{http://cdsweb.u-strasbg.fr/cgi-bin/qcat?J/A+A/}{http://cdsweb.u-strasbg.fr/cgi-bin/qcat?J/A+A/}.

\begin{acknowledgements}
The research activities described in this paper were carried out with contribution of the Next Generation EU funds within the National Recovery and Resilience Plan (PNRR), Mission 4 -- Education and Research, Component 2 -- From Research to Business (M4C2), Investment Line 3.1 -- Strengthening and creation of Research Infrastructures, Project IR0000034--“STILES -- Strengthening the Italian Leadership in ELT and SKA”.
This work has benefited from the support of Royal Society Research Grant RGS{\textbackslash}R1\textbackslash231450.
This research was supported by the International Space Science Institute (ISSI) in Bern, through ISSI International Team project \#23-573 ``Active Galactic Nuclei in Next Generation Surveys''.
L.B., F.R., and V.A acknowledge the support from the INAF Large Grant ``AGN \& Euclid: a close entanglement'' Ob. Fu. 01.05.23.01.14. A.F. acknowledges the support from project ``VLT- MOONS'' CRAM 1.05.03.07, INAF Large Grant 2022 ``The metal circle: a new sharp view of the baryon cycle up to Cosmic Dawn with the latest generation IFU facilities'' and INAF Large Grant 2022 ``Dual and binary SMBH in the multi-messenger era''.
L.B. thanks A. Marasco for computational support and D. Kocevski for providing the data necessary for comparing the \Euclid LRD candidates with the JWST ones..
This research made use of Photutils, an Astropy package for detection and photometry of astronomical sources \citep{Bradley2023}.

\AckEC

\AckQone

\AckDatalabs

Based on data from UNIONS, a scientific collaboration using three Hawaii-based telescopes: CFHT, Pan-STARRS, Subaru \url{www.skysurvey.cc}\,. Based on data from the Dark Energy Camera (DECam) on the Blanco 4-m Telescope at CTIO in Chile \url{https://www.darkenergysurvey.org}\,. This work uses results from the ESA mission {\it Gaia}, whose data are being processed by the Gaia Data Processing and Analysis Consortium \url{https://www.cosmos.esa.int/gaia}\,.

This publication is based on observations made with the Spitzer Space Telescope, which is operated by the Jet Propulsion Laboratory, California Institute of Technology under a contract with NASA, and has made use of the NASA/IPAC Infrared Science Archive, which is funded by the National Aeronautics and Space Administration and operated by the California Institute of Technology. 

\end{acknowledgements}

\bibliographystyle{aa} 
\bibliography{my} 

\begin{thebibliography}{126}
\expandafter\ifx\csname natexlab\endcsname\relax\def\natexlab#1{#1}\fi

\bibitem[{{Abbott} {et~al.}(2018){Abbott}, {Abdalla}, {Allam}, {Amara}, {Annis}, {Asorey}, {Avila}, {Ballester}, {Banerji}, {Barkhouse}, {Baruah}, {Baumer}, {Bechtol}, {Becker}, {Benoit-L{\'e}vy}, {Bernstein}, {Bertin}, {Blazek}, {Bocquet}, {Brooks}, {Brout}, {Buckley-Geer}, {Burke}, {Busti}, {Campisano}, {Cardiel-Sas}, {Carnero Rosell}, {Carrasco Kind}, {Carretero}, {Castander}, {Cawthon}, {Chang}, {Chen}, {Conselice}, {Costa}, {Crocce}, {Cunha}, {D'Andrea}, {da Costa}, {Das}, {Daues}, {Davis}, {Davis}, {De Vicente}, {DePoy}, {DeRose}, {Desai}, {Diehl}, {Dietrich}, {Dodelson}, {Doel}, {Drlica-Wagner}, {Eifler}, {Elliott}, {Evrard}, {Farahi}, {Fausti Neto}, {Fernandez}, {Finley}, {Flaugher}, {Foley}, {Fosalba}, {Friedel}, {Frieman}, {Garc{\'\i}a-Bellido}, {Gaztanaga}, {Gerdes}, {Giannantonio}, {Gill}, {Glazebrook}, {Goldstein}, {Gower}, {Gruen}, {Gruendl}, {Gschwend}, {Gupta}, {Gutierrez}, {Hamilton}, {Hartley}, {Hinton}, {Hislop}, {Hollowood}, {Honscheid}, {Hoyle}, {Huterer}, {Jain}, {James}, {Jeltema},
  {Johnson}, {Johnson}, {Kacprzak}, {Kent}, {Khullar}, {Klein}, {Kovacs}, {Koziol}, {Krause}, {Kremin}, {Kron}, {Kuehn}, {Kuhlmann}, {Kuropatkin}, {Lahav}, {Lasker}, {Li}, {Li}, {Liddle}, {Lima}, {Lin}, {L{\'o}pez-Reyes}, {MacCrann}, {Maia}, {Maloney}, {Manera}, {March}, {Marriner}, {Marshall}, {Martini}, {McClintock}, {McKay}, {McMahon}, {Melchior}, {Menanteau}, {Miller}, {Miquel}, {Mohr}, {Morganson}, {Mould}, {Neilsen}, {Nichol}, {Nogueira}, {Nord}, {Nugent}, {Nunes}, {Ogando}, {Old}, {Pace}, {Palmese}, {Paz-Chinch{\'o}n}, {Peiris}, {Percival}, {Petravick}, {Plazas}, {Poh}, {Pond}, {Porredon}, {Pujol}, {Refregier}, {Reil}, {Ricker}, {Rollins}, {Romer}, {Roodman}, {Rooney}, {Ross}, {Rykoff}, {Sako}, {Sanchez}, {Sanchez}, {Santiago}, {Saro}, {Scarpine}, {Scolnic}, {Serrano}, {Sevilla-Noarbe}, {Sheldon}, {Shipp}, {Silveira}, {Smith}, {Smith}, {Smith}, {Soares-Santos}, {Sobreira}, {Song}, {Stebbins}, {Suchyta}, {Sullivan}, {Swanson}, {Tarle}, {Thaler}, {Thomas}, {Thomas}, {Troxel}, {Tucker}, {Vikram}, {Vivas},
  {Walker}, {Wechsler}, {Weller}, {Wester}, {Wolf}, {Wu}, {Yanny}, {Zenteno}, {Zhang}, {Zuntz}, {DES Collaboration}, {Juneau}, {Fitzpatrick}, \& {Nikutta}}]{DES}
{Abbott}, T.~M.~C., {Abdalla}, F.~B., {Allam}, S., {et~al.} 2018, \apjs, 239, 18

\bibitem[{{Akins} {et~al.}(2024){Akins}, {Casey}, {Lambrides}, {Allen}, {Andika}, {Brinch}, {Champagne}, {Cooper}, {Ding}, {Drakos}, {Faisst}, {Finkelstein}, {Franco}, {Fujimoto}, {Gentile}, {Gillman}, {Gozaliasl}, {Harish}, {Hayward}, {Hirschmann}, {Ilbert}, {Kartaltepe}, {Kocevski}, {Koekemoer}, {Kokorev}, {Liu}, {Long}, {McCracken}, {McKinney}, {Onoue}, {Paquereau}, {Renzini}, {Rhodes}, {Robertson}, {Shuntov}, {Silverman}, {Tanaka}, {Toft}, {Trakhtenbrot}, {Valentino}, \& {Zavala}}]{Akins2024}
{Akins}, H.~B., {Casey}, C.~M., {Lambrides}, E., {et~al.} 2024, arXiv e-prints, arXiv:2406.10341

\bibitem[{{Akiyama} {et~al.}(2018){Akiyama}, {He}, {Ikeda}, {Niida}, {Nagao}, {Bosch}, {Coupon}, {Enoki}, {Imanishi}, {Kashikawa}, {Kawaguchi}, {Komiyama}, {Lee}, {Matsuoka}, {Miyazaki}, {Nishizawa}, {Oguri}, {Ono}, {Onoue}, {Ouchi}, {Schulze}, {Silverman}, {Tanaka}, {Tanaka}, {Terashima}, {Toba}, \& {Ueda}}]{Akiyama2018}
{Akiyama}, M., {He}, W., {Ikeda}, H., {et~al.} 2018, \pasj, 70, S34

\bibitem[{{Alexander} \& {Natarajan}(2014)}]{Alexander2014}
{Alexander}, T. \& {Natarajan}, P. 2014, Science, 345, 1330

\bibitem[{{Ananna} {et~al.}(2024){Ananna}, {Bogd{\'a}n}, {Kov{\'a}cs}, {Natarajan}, \& {Hickox}}]{Ananna2024}
{Ananna}, T.~T., {Bogd{\'a}n}, {\'A}., {Kov{\'a}cs}, O.~E., {Natarajan}, P., \& {Hickox}, R.~C. 2024, \apjl, 969, L18

\bibitem[{{Assef} {et~al.}(2018){Assef}, {Stern}, {Noirot}, {Jun}, {Cutri}, \& {Eisenhardt}}]{Assef2018}
{Assef}, R.~J., {Stern}, D., {Noirot}, G., {et~al.} 2018, \apjs, 234, 23

\bibitem[{{Asthana} {et~al.}(2024){Asthana}, {Haehnelt}, {Kulkarni}, {Bolton}, {Gaikwad}, {Keating}, \& {Puchwein}}]{Asthana2024}
{Asthana}, S., {Haehnelt}, M.~G., {Kulkarni}, G., {et~al.} 2024, arXiv e-prints, arXiv:2409.15453

\bibitem[{{Atek} {et~al.}(2024){Atek}, {Labb{\'e}}, {Furtak}, {Chemerynska}, {Fujimoto}, {Setton}, {Miller}, {Oesch}, {Bezanson}, {Price}, {Dayal}, {Zitrin}, {Kokorev}, {Weaver}, {Brammer}, {Dokkum}, {Williams}, {Cutler}, {Feldmann}, {Fudamoto}, {Greene}, {Leja}, {Maseda}, {Muzzin}, {Pan}, {Papovich}, {Nelson}, {Nanayakkara}, {Stark}, {Stefanon}, {Suess}, {Wang}, \& {Whitaker}}]{Atek2024}
{Atek}, H., {Labb{\'e}}, I., {Furtak}, L.~J., {et~al.} 2024, \nat, 626, 975

\bibitem[{{Baggen} {et~al.}(2024){Baggen}, {van Dokkum}, {Brammer}, {de Graaff}, {Franx}, {Greene}, {Labb{\'e}}, {Leja}, {Maseda}, {Nelson}, {Rix}, {Wang}, \& {Weibel}}]{Baggen2024}
{Baggen}, J. F.~W., {van Dokkum}, P., {Brammer}, G., {et~al.} 2024, \apjl, 977, L13

\bibitem[{{Barkana} \& {Loeb}(2001)}]{Barkana2001}
{Barkana}, R. \& {Loeb}, A. 2001, \physrep, 349, 125

\bibitem[{{Barro} {et~al.}(2024){Barro}, {Perez-Gonzalez}, {Kocevski}, {McGrath}, {Leung}, {Cullen}, {Dunlop}, {Ellis}, {Finkelstein}, {Grogin}, {Illingworth}, {Kartaltepe}, {Koekemoer}, {Lucas}, {McLure}, \& {Yang}}]{Barro2024}
{Barro}, G., {Perez-Gonzalez}, P.~G., {Kocevski}, D.~D., {et~al.} 2024, arXiv e-prints, arXiv:2412.01887

\bibitem[{{Barro} {et~al.}(2023){Barro}, {Perez-Gonzalez}, {Kocevski}, {McGrath}, {Trump}, {Simons}, {Somerville}, {Yung}, {Arrabal Haro}, {Bagley}, {Cleri}, {Costantin}, {Davis}, {Dickinson}, {Finkelstein}, {Giavalisco}, {Gomez-Guijarro}, {Hathi}, {Hirschmann}, {Akins}, {Holwerda}, {Huertas-Company}, {Lucas}, {Papovich}, {Seille}, {Tacchella}, {Wilkins}, {de la Vega}, {Yang}, \& {Zavala}}]{Barro2023}
{Barro}, G., {Perez-Gonzalez}, P.~G., {Kocevski}, D.~D., {et~al.} 2023, arXiv e-prints, arXiv:2305.14418

\bibitem[{{Begelman} \& {Volonteri}(2017)}]{Begelman2017}
{Begelman}, M.~C. \& {Volonteri}, M. 2017, \mnras, 464, 1102

\bibitem[{{Bellovary}(2025)}]{Bellovary2025}
{Bellovary}, J. 2025, arXiv e-prints, arXiv:2501.03309

\bibitem[{Bisigello {et~al.}(2025)Bisigello, Giulietti, Prandoni, Bondi, Bonato, Magliocchetti, Rottgering, Morabito, \& White}]{Bisigello2025a}
Bisigello, L., Giulietti, M., Prandoni, I., {et~al.} 2025, The Open Journal of Astrophysics, 8

\bibitem[{Bradley {et~al.}(2023)Bradley, Sipőcz, Robitaille, Tollerud, Vinícius, Deil, Barbary, Wilson, Busko, Donath, Günther, Cara, Lim, Meßlinger, Conseil, Burnett, Bostroem, Droettboom, Bray, Bratholm, Jamieson, Ginsburg, Barentsen, Craig, Morris, Perrin, Rathi, Pascual, Perren, \& Georgiev}]{Bradley2023}
Bradley, L., Sipőcz, B., Robitaille, T., {et~al.} 2023, astropy/photutils: 1.10.0

\bibitem[{{Carr} \& {Hawking}(1974)}]{Carr1974}
{Carr}, B.~J. \& {Hawking}, S.~W. 1974, \mnras, 168, 399

\bibitem[{{Casey} {et~al.}(2023){Casey}, {Kartaltepe}, {Drakos}, {Franco}, {Harish}, {Paquereau}, {Ilbert}, {Rose}, {Cox}, {Nightingale}, {Robertson}, {Silverman}, {Koekemoer}, {Massey}, {McCracken}, {Rhodes}, {Akins}, {Allen}, {Amvrosiadis}, {Arango-Toro}, {Bagley}, {Bongiorno}, {Capak}, {Champagne}, {Chartab}, {Ch{\'a}vez Ortiz}, {Chworowsky}, {Cooke}, {Cooper}, {Darvish}, {Ding}, {Faisst}, {Finkelstein}, {Fujimoto}, {Gentile}, {Gillman}, {Gould}, {Gozaliasl}, {Hayward}, {He}, {Hemmati}, {Hirschmann}, {Jahnke}, {Jin}, {Khostovan}, {Kokorev}, {Lambrides}, {Laigle}, {Larson}, {Leung}, {Liu}, {Liaudat}, {Long}, {Magdis}, {Mahler}, {Mainieri}, {Manning}, {Maraston}, {Martin}, {McCleary}, {McKinney}, {McPartland}, {Mobasher}, {Pattnaik}, {Renzini}, {Rich}, {Sanders}, {Sattari}, {Scognamiglio}, {Scoville}, {Sheth}, {Shuntov}, {Sparre}, {Suzuki}, {Talia}, {Toft}, {Trakhtenbrot}, {Urry}, {Valentino}, {Vanderhoof}, {Vardoulaki}, {Weaver}, {Whitaker}, {Wilkins}, {Yang}, \& {Zavala}}]{Casey2023}
{Casey}, C.~M., {Kartaltepe}, J.~S., {Drakos}, N.~E., {et~al.} 2023, \apj, 954, 31

\bibitem[{{Dayal}(2024)}]{Dayal2024b}
{Dayal}, P. 2024, \aap, 690, A182

\bibitem[{{Dayal} \& {Ferrara}(2018)}]{Dayal2018}
{Dayal}, P. \& {Ferrara}, A. 2018, \physrep, 780, 1

\bibitem[{{Dayal} {et~al.}(2025){Dayal}, {Volonteri}, {Greene}, {Kokorev}, {Goulding}, {Williams}, {Furtak}, {Zitrin}, {Atek}, {Bezanson}, {Chemerynska}, {Feldmann}, {Glazebrook}, {Labbe}, {Nanayakkara}, {Oesch}, \& {Weaver}}]{Dayal2024}
{Dayal}, P., {Volonteri}, M., {Greene}, J.~E., {et~al.} 2025, \aap, 697, A211

\bibitem[{{Decleir} {et~al.}(2022){Decleir}, {Gordon}, {Andrews}, {Clayton}, {Cushing}, {Misselt}, {Pendleton}, {Rayner}, {Vacca}, \& {Whittet}}]{Decleir2022}
{Decleir}, M., {Gordon}, K.~D., {Andrews}, J.~E., {et~al.} 2022, \apj, 930, 15

\bibitem[{{Delvecchio} {et~al.}(2022){Delvecchio}, {Daddi}, {Sargent}, {Aird}, {Mullaney}, {Magnelli}, {Elbaz}, {Bisigello}, {Ceraj}, {Jin}, {Kalita}, {Liu}, {Novak}, {Prandoni}, {Radcliffe}, {Spingola}, {Zamorani}, {Allevato}, {Rodighiero}, \& {Smol{\v{c}}i{\'c}}}]{Delvecchio2022}
{Delvecchio}, I., {Daddi}, E., {Sargent}, M.~T., {et~al.} 2022, \aap, 668, A81

\bibitem[{{Delvecchio} {et~al.}(2014){Delvecchio}, {Gruppioni}, {Pozzi}, {Berta}, {Zamorani}, {Cimatti}, {Lutz}, {Scott}, {Vignali}, {Cresci}, {Feltre}, {Cooray}, {Vaccari}, {Fritz}, {Le Floc'h}, {Magnelli}, {Popesso}, {Oliver}, {Bock}, {Carollo}, {Contini}, {Le F{\'e}vre}, {Lilly}, {Mainieri}, {Renzini}, \& {Scodeggio}}]{Delvecchio2014}
{Delvecchio}, I., {Gruppioni}, C., {Pozzi}, F., {et~al.} 2014, \mnras, 439, 2736

\bibitem[{{DESI Collaboration} {et~al.}(2024){DESI Collaboration}, {Adame}, {Aguilar}, {Ahlen}, {Alam}, {Aldering}, {Alexander}, {Alfarsy}, {Allende Prieto}, {Alvarez}, {Alves}, {Anand}, {Andrade-Oliveira}, {Armengaud}, {Asorey}, {Avila}, {Aviles}, {Bailey}, {Balaguera-Antol{\'\i}nez}, {Ballester}, {Baltay}, {Bault}, {Bautista}, {Behera}, {Beltran}, {BenZvi}, {Beraldo e Silva}, {Bermejo-Climent}, {Berti}, {Besuner}, {Beutler}, {Bianchi}, {Blake}, {Blum}, {Bolton}, {Brieden}, {Brodzeller}, {Brooks}, {Brown}, {Buckley-Geer}, {Burtin}, {Cabayol-Garcia}, {Cai}, {Canning}, {Cardiel-Sas}, {Carnero Rosell}, {Castander}, {Cervantes-Cota}, {Chabanier}, {Chaussidon}, {Chaves-Montero}, {Chen}, {Chen}, {Chuang}, {Claybaugh}, {Cole}, {Cooper}, {Cuceu}, {Davis}, {Dawson}, {de Belsunce}, {de la Cruz}, {de la Macorra}, {Della Costa}, {de Mattia}, {Demina}, {Demirbozan}, {DeRose}, {Dey}, {Dey}, {Dhungana}, {Ding}, {Ding}, {Doel}, {Doshi}, {Douglass}, {Edge}, {Eftekharzadeh}, {Eisenstein}, {Elliott}, {Ereza}, {Escoffier},
  {Fagrelius}, {Fan}, {Fanning}, {Fawcett}, {Ferraro}, {Flaugher}, {Font-Ribera}, {Forero-Romero}, {Forero-S{\'a}nchez}, {Frenk}, {G{\"a}nsicke}, {Garc{\'\i}a}, {Garc{\'\i}a-Bellido}, {Garcia-Quintero}, {Garrison}, {Gil-Mar{\'\i}n}, {Golden-Marx}, {Gontcho A Gontcho}, {Gonzalez-Morales}, {Gonzalez-Perez}, {Gordon}, {Graur}, {Green}, {Gruen}, {Guy}, {Hadzhiyska}, {Hahn}, {Han}, {Hanif}, {Herrera-Alcantar}, {Honscheid}, {Hou}, {Howlett}, {Huterer}, {Ir{\v{s}}i{\v{c}}}, {Ishak}, {Jacques}, {Jana}, {Jiang}, {Jimenez}, {Jing}, {Joudaki}, {Joyce}, {Jullo}, {Juneau}, {Kara{\c{c}}ayl{\i}}, {Karim}, {Kehoe}, {Kent}, {Khederlarian}, {Kim}, {Kirkby}, {Kisner}, {Kitaura}, {Kizhuprakkat}, {Kneib}, {Koposov}, {Kov{\'a}cs}, {Kremin}, {Krolewski}, {L'Huillier}, {Lahav}, {Lambert}, {Lamman}, {Lan}, {Landriau}, {Lang}, {Lange}, {Lasker}, {Leauthaud}, {Le Guillou}, {Levi}, {Li}, {Linder}, {Lyons}, {Magneville}, {Manera}, {Manser}, {Margala}, {Martini}, {McDonald}, {Medina}, {Medina-Varela}, {Meisner}, {Mena-Fern{\'a}ndez},
  {Meneses-Rizo}, {Mezcua}, {Miquel}, {Montero-Camacho}, {Moon}, {Moore}, {Moustakas}, {Mueller}, {Mundet}, {Mu{\~n}oz-Guti{\'e}rrez}, {Myers}, {Nadathur}, {Napolitano}, {Neveux}, {Newman}, {Nie}, {Nikutta}, {Niz}, {Norberg}, {Noriega}, {Paillas}, {Palanque-Delabrouille}, {Palmese}, {Pan}, {Parkinson}, {Penmetsa}, {Percival}, {P{\'e}rez-Fern{\'a}ndez}, {P{\'e}rez-R{\`a}fols}, {Pieri}, {Poppett}, {Porredon}, \& {Pothier}}]{DESI_EDR}
{DESI Collaboration}, {Adame}, A.~G., {Aguilar}, J., {et~al.} 2024, \aj, 168, 58

\bibitem[{{Donley} {et~al.}(2012){Donley}, {Koekemoer}, {Brusa}, {Capak}, {Cardamone}, {Civano}, {Ilbert}, {Impey}, {Kartaltepe}, {Miyaji}, {Salvato}, {Sanders}, {Trump}, \& {Zamorani}}]{Donley2012}
{Donley}, J.~L., {Koekemoer}, A.~M., {Brusa}, M., {et~al.} 2012, \apj, 748, 142

\bibitem[{{Euclid Collaboration: Aussel} {et~al.}(2025){Euclid Collaboration: Aussel}, {Tereno}, {Schirmer}, {et~al.}}]{Q1-TP001}
{Euclid Collaboration: Aussel}, H., {Tereno}, I., {Schirmer}, M., {et~al.} 2025, A\&A, submitted (Euclid Q1 SI), arXiv:2503.15302

\bibitem[{{Euclid Collaboration: Bisigello} {et~al.}(2024){Euclid Collaboration: Bisigello}, {Massimo}, {Tortora}, {et~al.}}]{EP-Bisigello}
{Euclid Collaboration: Bisigello}, L., {Massimo}, M., {Tortora}, C., {et~al.} 2024, \aap, 691, A1

\bibitem[{{Euclid Collaboration: Cropper} {et~al.}(2025){Euclid Collaboration: Cropper}, {Al-Bahlawan}, {Amiaux}, {et~al.}}]{EuclidSkyVIS}
{Euclid Collaboration: Cropper}, M., {Al-Bahlawan}, A., {Amiaux}, J., {et~al.} 2025, A\&A, 697, A2

\bibitem[{{Euclid Collaboration: Jahnke} {et~al.}(2025){Euclid Collaboration: Jahnke}, {Gillard}, {Schirmer}, {et~al.}}]{EuclidSkyNISP}
{Euclid Collaboration: Jahnke}, K., {Gillard}, W., {Schirmer}, M., {et~al.} 2025, A\&A, 697, A3

\bibitem[{{Euclid Collaboration: Margalef-Bentabol} {et~al.}(2025){Euclid Collaboration: Margalef-Bentabol}, {Wang}, {La Marca}, {et~al.}}]{Q1-SP015}
{Euclid Collaboration: Margalef-Bentabol}, B., {Wang}, L., {La Marca}, A., {et~al.} 2025, A\&A, submitted (Euclid Q1 SI), arXiv:2503.15318

\bibitem[{{Euclid Collaboration: Matamoro Zatarain} {et~al.}(2025){Euclid Collaboration: Matamoro Zatarain}, {Fotopoulou}, {Ricci}, {et~al.}}]{Q1-SP027}
{Euclid Collaboration: Matamoro Zatarain}, T., {Fotopoulou}, S., {Ricci}, F., {et~al.} 2025, A\&A, submitted (Euclid Q1 SI), arXiv:2503.15320

\bibitem[{{Euclid Collaboration: McCracken} {et~al.}(2025){Euclid Collaboration: McCracken}, {Benson}, {Dolding}, {et~al.}}]{Q1-TP002}
{Euclid Collaboration: McCracken}, H.~J., {Benson}, K., {Dolding}, C., {et~al.} 2025, A\&A, accepted (Euclid Q1 SI), arXiv:2503.15303

\bibitem[{{Euclid Collaboration: McPartland} {et~al.}(2025){Euclid Collaboration: McPartland}, {Zalesky}, {Weaver}, {et~al.}}]{EP-McPartland}
{Euclid Collaboration: McPartland}, C.~J.~R., {Zalesky}, L., {Weaver}, J.~R., {et~al.} 2025, \aap, 695, A259

\bibitem[{{Euclid Collaboration: Mellier} {et~al.}(2025){Euclid Collaboration: Mellier}, {Abdurro'uf}, {Acevedo~Barroso}, {et~al.}}]{EuclidSkyOverview}
{Euclid Collaboration: Mellier}, Y., {Abdurro'uf}, {Acevedo~Barroso}, J., {et~al.} 2025, A\&A, 697, A1

\bibitem[{{Euclid Collaboration: Moneti} {et~al.}(2022){Euclid Collaboration: Moneti}, {McCracken}, {Shuntov}, {et~al.}}]{Moneti-EP17}
{Euclid Collaboration: Moneti}, A., {McCracken}, H.~J., {Shuntov}, M., {et~al.} 2022, \aap, 658, A126

\bibitem[{{Euclid Collaboration: Polenta} {et~al.}(2025){Euclid Collaboration: Polenta}, {Frailis}, {Alavi}, {et~al.}}]{Q1-TP003}
{Euclid Collaboration: Polenta}, G., {Frailis}, M., {Alavi}, A., {et~al.} 2025, A\&A, accepted (Euclid Q1 SI), arXiv:2503.15304

\bibitem[{{Euclid Collaboration: Romelli} {et~al.}(2025){Euclid Collaboration: Romelli}, {K\"ummel}, {Dole}, {et~al.}}]{Q1-TP004}
{Euclid Collaboration: Romelli}, E., {K\"ummel}, M., {Dole}, H., {et~al.} 2025, A\&A, in press (Euclid Q1 SI), \url{https://doi.org/10.1051/0004-6361/202554586}, arXiv:2503.15305

\bibitem[{{Euclid Collaboration: Roster} {et~al.}(2025){Euclid Collaboration: Roster}, {Salvato}, {Buchner}, {et~al.}}]{Q1-SP003}
{Euclid Collaboration: Roster}, W., {Salvato}, M., {Buchner}, J., {et~al.} 2025, A\&A, accepted (Euclid Q1 SI), arXiv:2503.15316

\bibitem[{{Euclid Collaboration: Tucci} {et~al.}(2025){Euclid Collaboration: Tucci}, {Paltani}, {Hartley}, {et~al.}}]{Q1-TP005}
{Euclid Collaboration: Tucci}, M., {Paltani}, S., {Hartley}, W.~G., {et~al.} 2025, A\&A, in press (Euclid Q1 SI), \url{https://doi.org/10.1051/0004-6361/202554588}, arXiv:2503.15306

\bibitem[{{Euclid Collaboration: Zalesky} {et~al.}(2025){Euclid Collaboration: Zalesky}, {McPartland}, {Weaver}, {et~al.}}]{EP-Zalesky}
{Euclid Collaboration: Zalesky}, L., {McPartland}, C.~J.~R., {Weaver}, J.~R., {et~al.} 2025, \aap, 695, A229

\bibitem[{{Euclid Quick Release Q1}(2025)}]{Q1cite}
{Euclid Quick Release Q1}. 2025, \url{https://doi.org/10.57780/esa-2853f3b}

\bibitem[{{Ferrarese}(2002)}]{Ferrarese2002}
{Ferrarese}, L. 2002, \apj, 578, 90

\bibitem[{{Fitzpatrick} {et~al.}(2019){Fitzpatrick}, {Massa}, {Gordon}, {Bohlin}, \& {Clayton}}]{Fitzpatrick2019}
{Fitzpatrick}, E.~L., {Massa}, D., {Gordon}, K.~D., {Bohlin}, R., \& {Clayton}, G.~C. 2019, \apj, 886, 108

\bibitem[{{Fumagalli} {et~al.}(2012){Fumagalli}, {Patel}, {Franx}, {Brammer}, {van Dokkum}, {da Cunha}, {Kriek}, {Lundgren}, {Momcheva}, {Rix}, {Schmidt}, {Skelton}, {Whitaker}, {Labbe}, \& {Nelson}}]{Fumagalli2012}
{Fumagalli}, M., {Patel}, S.~G., {Franx}, M., {et~al.} 2012, \apjl, 757, L22

\bibitem[{{Furtak} {et~al.}(2023){Furtak}, {Zitrin}, {Plat}, {Fujimoto}, {Wang}, {Nelson}, {Labb{\'e}}, {Bezanson}, {Brammer}, {van Dokkum}, {Endsley}, {Glazebrook}, {Greene}, {Leja}, {Price}, {Smit}, {Stark}, {Weaver}, {Whitaker}, {Atek}, {Chevallard}, {Curtis-Lake}, {Dayal}, {Feltre}, {Franx}, {Fudamoto}, {Marchesini}, {Mowla}, {Pan}, {Suess}, {Vidal-Garc{\'\i}a}, \& {Williams}}]{Furtak2023}
{Furtak}, L.~J., {Zitrin}, A., {Plat}, A., {et~al.} 2023, \apj, 952, 142

\bibitem[{{Gaia Collaboration: Vallenari} {et~al.}(2023){Gaia Collaboration: Vallenari}, {Brown}, {Prusti}, {de Bruijne}, {Arenou}, {Babusiaux}, {Biermann}, {Creevey}, {Ducourant}, {Evans}, {Eyer}, {Guerra}, {Hutton}, {Jordi}, {Klioner}, {Lammers}, {Lindegren}, \& et~al.}]{GaiaCollaboration_2023_2023A&A...674A...1G}
{Gaia Collaboration: Vallenari}, A., {Brown}, A.~G.~A., {Prusti}, T., {et~al.} 2023, \aap, 674, A1

\bibitem[{{Gebhardt} {et~al.}(2000){Gebhardt}, {Bender}, {Bower}, {Dressler}, {Faber}, {Filippenko}, {Green}, {Grillmair}, {Ho}, {Kormendy}, {Lauer}, {Magorrian}, {Pinkney}, {Richstone}, \& {Tremaine}}]{Gebhardt2000}
{Gebhardt}, K., {Bender}, R., {Bower}, G., {et~al.} 2000, \apjl, 539, L13

\bibitem[{{Gehrels}(1986)}]{Gehrels1986}
{Gehrels}, N. 1986, \apj, 303, 336

\bibitem[{{Giallongo} {et~al.}(2015){Giallongo}, {Grazian}, {Fiore}, {Fontana}, {Pentericci}, {Vanzella}, {Dickinson}, {Kocevski}, {Castellano}, {Cristiani}, {Ferguson}, {Finkelstein}, {Grogin}, {Hathi}, {Koekemoer}, {Newman}, \& {Salvato}}]{Giallongo2015}
{Giallongo}, E., {Grazian}, A., {Fiore}, F., {et~al.} 2015, \aap, 578, A83

\bibitem[{{Giallongo} {et~al.}(2012){Giallongo}, {Menci}, {Fiore}, {Castellano}, {Fontana}, {Grazian}, \& {Pentericci}}]{Giallongo2012}
{Giallongo}, E., {Menci}, N., {Fiore}, F., {et~al.} 2012, \apj, 755, 124

\bibitem[{{Glikman} {et~al.}(2015){Glikman}, {Simmons}, {Mailly}, {Schawinski}, {Urry}, \& {Lacy}}]{Glikman2015}
{Glikman}, E., {Simmons}, B., {Mailly}, M., {et~al.} 2015, \apj, 806, 218

\bibitem[{{Glikman} {et~al.}(2012){Glikman}, {Urrutia}, {Lacy}, {Djorgovski}, {Mahabal}, {Myers}, {Ross}, {Petitjean}, {Ge}, {Schneider}, \& {York}}]{Glikman2012}
{Glikman}, E., {Urrutia}, T., {Lacy}, M., {et~al.} 2012, \apj, 757, 51

\bibitem[{{Gordon} {et~al.}(2009){Gordon}, {Cartledge}, \& {Clayton}}]{Gordon2009}
{Gordon}, K.~D., {Cartledge}, S., \& {Clayton}, G.~C. 2009, \apj, 705, 1320

\bibitem[{{Gordon} {et~al.}(2023){Gordon}, {Clayton}, {Decleir}, {Fitzpatrick}, {Massa}, {Misselt}, \& {Tollerud}}]{Gordon2023}
{Gordon}, K.~D., {Clayton}, G.~C., {Decleir}, M., {et~al.} 2023, \apj, 950, 86

\bibitem[{{Gordon} {et~al.}(2021){Gordon}, {Misselt}, {Bouwman}, {Clayton}, {Decleir}, {Hines}, {Pendleton}, {Rieke}, {Smith}, \& {Whittet}}]{Gordon2021}
{Gordon}, K.~D., {Misselt}, K.~A., {Bouwman}, J., {et~al.} 2021, \apj, 916, 33

\bibitem[{{Grazian} {et~al.}(2024){Grazian}, {Giallongo}, {Boutsia}, {Cristiani}, {Fontanot}, {Bischetti}, {Bisigello}, {Bongiorno}, {Calderone}, {Chiti Tegli}, {Cupani}, {De Lucia}, {D'Odorico}, {Feruglio}, {Fiore}, {Gandolfi}, {Girardi}, {Guarneri}, {Hirschmann}, {Porru}, {Rodighiero}, {Saccheo}, {Simioni}, {Trost}, \& {Viitanen}}]{Grazian2024}
{Grazian}, A., {Giallongo}, E., {Boutsia}, K., {et~al.} 2024, \apj, 974, 84

\bibitem[{{Greene} {et~al.}(2024){Greene}, {Labbe}, {Goulding}, {Furtak}, {Chemerynska}, {Kokorev}, {Dayal}, {Volonteri}, {Williams}, {Wang}, {Setton}, {Burgasser}, {Bezanson}, {Atek}, {Brammer}, {Cutler}, {Feldmann}, {Fujimoto}, {Glazebrook}, {de Graaff}, {Khullar}, {Leja}, {Marchesini}, {Maseda}, {Matthee}, {Miller}, {Naidu}, {Nanayakkara}, {Oesch}, {Pan}, {Papovich}, {Price}, {van Dokkum}, {Weaver}, {Whitaker}, \& {Zitrin}}]{Greene2024}
{Greene}, J.~E., {Labbe}, I., {Goulding}, A.~D., {et~al.} 2024, \apj, 964, 39

\bibitem[{Grogin {et~al.}(2011)Grogin, Kocevski, Faber, Ferguson, Koekemoer, Riess, Acquaviva, Alexander, Almaini, Ashby, Barden, Bell, Bournaud, Brown, Caputi, Casertano, Cassata, Castellano, Challis, Chary, Cheung, Cirasuolo, Conselice, Cooray, Croton, Daddi, Dahlen, Dav{\'{e}}, de~Mello, Dekel, Dickinson, Dolch, Donley, Dunlop, Dutton, Elbaz, Fazio, Filippenko, Finkelstein, Fontana, Gardner, Garnavich, Gawiser, Giavalisco, Grazian, Guo, Hathi, Häussler, Hopkins, Huang, Huang, Jha, Kartaltepe, Kirshner, Koo, Lai, Lee, Li, Lotz, Lucas, Madau, McCarthy, McGrath, McIntosh, McLure, Mobasher, Moustakas, Mozena, Nandra, Newman, Niemi, Noeske, Papovich, Pentericci, Pope, Primack, Rajan, Ravindranath, Reddy, Renzini, Rix, Robaina, Rodney, Rosario, Rosati, Salimbeni, Scarlata, Siana, Simard, Smidt, Somerville, Spinrad, Straughn, Strolger, Telford, Teplitz, Trump, van~der Wel, Villforth, Wechsler, Weiner, Wiklind, Wild, Wilson, Wuyts, Yan, \& Yun}]{Grogin2011}
Grogin, N.~A., Kocevski, D.~D., Faber, S.~M., {et~al.} 2011, The Astrophysical Journal Supplement Series, 197, 35

\bibitem[{{Gruppioni} {et~al.}(2020){Gruppioni}, {B{\'e}thermin}, {Loiacono}, {Le F{\`e}vre}, {Capak}, {Cassata}, {Faisst}, {Schaerer}, {Silverman}, {Yan}, {Bardelli}, {Boquien}, {Carraro}, {Cimatti}, {Dessauges-Zavadsky}, {Ginolfi}, {Fujimoto}, {Hathi}, {Jones}, {Khusanova}, {Koekemoer}, {Lagache}, {Lemaux}, {Oesch}, {Pozzi}, {Riechers}, {Rodighiero}, {Romano}, {Talia}, {Vallini}, {Vergani}, {Zamorani}, \& {Zucca}}]{Gruppioni2020}
{Gruppioni}, C., {B{\'e}thermin}, M., {Loiacono}, F., {et~al.} 2020, \aap, 643, A8

\bibitem[{{G{\"u}ltekin} {et~al.}(2009){G{\"u}ltekin}, {Richstone}, {Gebhardt}, {Lauer}, {Tremaine}, {Aller}, {Bender}, {Dressler}, {Faber}, {Filippenko}, {Green}, {Ho}, {Kormendy}, {Magorrian}, {Pinkney}, \& {Siopis}}]{Gultekin2009}
{G{\"u}ltekin}, K., {Richstone}, D.~O., {Gebhardt}, K., {et~al.} 2009, \apj, 698, 198

\bibitem[{{Habouzit} {et~al.}(2021){Habouzit}, {Li}, {Somerville}, {Genel}, {Pillepich}, {Volonteri}, {Dav{\'e}}, {Rosas-Guevara}, {McAlpine}, {Peirani}, {Hernquist}, {Angl{\'e}s-Alc{\'a}zar}, {Reines}, {Bower}, {Dubois}, {Nelson}, {Pichon}, \& {Vogelsberger}}]{Habouzit2021}
{Habouzit}, M., {Li}, Y., {Somerville}, R.~S., {et~al.} 2021, \mnras, 503, 1940

\bibitem[{{Habouzit} {et~al.}(2020){Habouzit}, {Pisani}, {Goulding}, {Dubois}, {Somerville}, \& {Greene}}]{Habouzit2020}
{Habouzit}, M., {Pisani}, A., {Goulding}, A., {et~al.} 2020, \mnras, 493, 899

\bibitem[{{Hainline} {et~al.}(2024){Hainline}, {Maiolino}, {Juodzbalis}, {Scholtz}, {Ubler}, {D'Eugenio}, {Helton}, {Sun}, {Sun}, {Robertson}, {Tacchella}, {Bunker}, {Carniani}, {Charlot}, {Curtis-Lake}, {Egami}, {Johnson}, {Lin}, {Lyu}, {Perez-Gonzalez}, {Rinaldi}, {Silcock}, {Venturi}, {Williams}, {Willmer}, {Willott}, {Zhang}, \& {Zhu}}]{Hainline2024}
{Hainline}, K.~N., {Maiolino}, R., {Juodzbalis}, I., {et~al.} 2024, arXiv e-prints, arXiv:2410.00100

\bibitem[{{Harikane} {et~al.}(2023){Harikane}, {Zhang}, {Nakajima}, {Ouchi}, {Isobe}, {Ono}, {Hatano}, {Xu}, \& {Umeda}}]{Harikane2023}
{Harikane}, Y., {Zhang}, Y., {Nakajima}, K., {et~al.} 2023, \apj, 959, 39

\bibitem[{{Hawking}(1971)}]{Hawking1971}
{Hawking}, S. 1971, \mnras, 152, 75

\bibitem[{{Hickox} \& {Alexander}(2018)}]{Hickox2018}
{Hickox}, R.~C. \& {Alexander}, D.~M. 2018, \araa, 56, 625

\bibitem[{{Inayoshi} {et~al.}(2016){Inayoshi}, {Haiman}, \& {Ostriker}}]{Inayoshi2016}
{Inayoshi}, K., {Haiman}, Z., \& {Ostriker}, J.~P. 2016, \mnras, 459, 3738

\bibitem[{{Inayoshi} \& {Maiolino}(2024)}]{Inayoshi2024}
{Inayoshi}, K. \& {Maiolino}, R. 2024, arXiv e-prints, arXiv:2409.07805

\bibitem[{{Juod{\v{z}}balis} {et~al.}(2024){Juod{\v{z}}balis}, {Ji}, {Maiolino}, {D'Eugenio}, {Scholtz}, {Risaliti}, {Fabian}, {Mazzolari}, {Gilli}, {Prandoni}, {Arribas}, {Bunker}, {Carniani}, {Charlot}, {Curtis-Lake}, {de Graaff}, {Hainline}, {Parlanti}, {Perna}, {P{\'e}rez-Gonz{\'a}lez}, {Robertson}, {Tacchella}, {{\"U}bler}, {Williams}, {Willott}, \& {Witstok}}]{Juodzbalis2024}
{Juod{\v{z}}balis}, I., {Ji}, X., {Maiolino}, R., {et~al.} 2024, \mnras, 535, 853

\bibitem[{{Killi} {et~al.}(2024){Killi}, {Watson}, {Brammer}, {McPartland}, {Antwi-Danso}, {Newshore}, {Coe}, {Allen}, {Fynbo}, {Gould}, {Heintz}, {Rusakov}, \& {Vejlgaard}}]{Killi2024}
{Killi}, M., {Watson}, D., {Brammer}, G., {et~al.} 2024, \aap, 691, A52

\bibitem[{{Kocevski} {et~al.}(2024){Kocevski}, {Finkelstein}, {Barro}, {Taylor}, {Calabr{\`o}}, {Laloux}, {Buchner}, {Trump}, {Leung}, {Yang}, {Dickinson}, {P{\'e}rez-Gonz{\'a}lez}, {Pacucci}, {Inayoshi}, {Somerville}, {McGrath}, {Akins}, {Bagley}, {Bisigello}, {Bowler}, {Carnall}, {Casey}, {Cheng}, {Cleri}, {Costantin}, {Cullen}, {Davis}, {Donnan}, {Dunlop}, {Ellis}, {Ferguson}, {Fujimoto}, {Fontana}, {Giavalisco}, {Grazian}, {Grogin}, {Hathi}, {Hirschmann}, {Huertas-Company}, {Holwerda}, {Illingworth}, {Juneau}, {Kartaltepe}, {Koekemoer}, {Li}, {Lucas}, {Magee}, {Mason}, {McLeod}, {McLure}, {Napolitano}, {Papovich}, {Pirzkal}, {Rodighiero}, {Santini}, {Wilkins}, \& {Yung}}]{Kocevski2024}
{Kocevski}, D.~D., {Finkelstein}, S.~L., {Barro}, G., {et~al.} 2024, arXiv e-prints, arXiv:2404.03576

\bibitem[{{Kocevski} {et~al.}(2023){Kocevski}, {Onoue}, {Inayoshi}, {Trump}, {Arrabal Haro}, {Grazian}, {Dickinson}, {Finkelstein}, {Kartaltepe}, {Hirschmann}, {Aird}, {Holwerda}, {Fujimoto}, {Juneau}, {Amor{\'\i}n}, {Backhaus}, {Bagley}, {Barro}, {Bell}, {Bisigello}, {Calabr{\`o}}, {Cleri}, {Cooper}, {Ding}, {Grogin}, {Ho}, {Hutchison}, {Inoue}, {Jiang}, {Jones}, {Koekemoer}, {Li}, {Li}, {McGrath}, {Molina}, {Papovich}, {P{\'e}rez-Gonz{\'a}lez}, {Pirzkal}, {Wilkins}, {Yang}, \& {Yung}}]{Kocevski2023}
{Kocevski}, D.~D., {Onoue}, M., {Inayoshi}, K., {et~al.} 2023, \apjl, 954, L4

\bibitem[{{Koekemoer} {et~al.}(2011){Koekemoer}, {Faber}, {Ferguson}, {Grogin}, {Kocevski}, {Koo}, {Lai}, {Lotz}, {Lucas}, {McGrath}, {Ogaz}, {Rajan}, {Riess}, {Rodney}, {Strolger}, {Casertano}, {Castellano}, {Dahlen}, {Dickinson}, {Dolch}, {Fontana}, {Giavalisco}, {Grazian}, {Guo}, {Hathi}, {Huang}, {van der Wel}, {Yan}, {Acquaviva}, {Alexander}, {Almaini}, {Ashby}, {Barden}, {Bell}, {Bournaud}, {Brown}, {Caputi}, {Cassata}, {Challis}, {Chary}, {Cheung}, {Cirasuolo}, {Conselice}, {Roshan Cooray}, {Croton}, {Daddi}, {Dav{\'e}}, {de Mello}, {de Ravel}, {Dekel}, {Donley}, {Dunlop}, {Dutton}, {Elbaz}, {Fazio}, {Filippenko}, {Finkelstein}, {Frazer}, {Gardner}, {Garnavich}, {Gawiser}, {Gruetzbauch}, {Hartley}, {H{\"a}ussler}, {Herrington}, {Hopkins}, {Huang}, {Jha}, {Johnson}, {Kartaltepe}, {Khostovan}, {Kirshner}, {Lani}, {Lee}, {Li}, {Madau}, {McCarthy}, {McIntosh}, {McLure}, {McPartland}, {Mobasher}, {Moreira}, {Mortlock}, {Moustakas}, {Mozena}, {Nandra}, {Newman}, {Nielsen}, {Niemi}, {Noeske}, {Papovich},
  {Pentericci}, {Pope}, {Primack}, {Ravindranath}, {Reddy}, {Renzini}, {Rix}, {Robaina}, {Rosario}, {Rosati}, {Salimbeni}, {Scarlata}, {Siana}, {Simard}, {Smidt}, {Snyder}, {Somerville}, {Spinrad}, {Straughn}, {Telford}, {Teplitz}, {Trump}, {Vargas}, {Villforth}, {Wagner}, {Wandro}, {Wechsler}, {Weiner}, {Wiklind}, {Wild}, {Wilson}, {Wuyts}, \& {Yun}}]{Koekemoer2011}
{Koekemoer}, A.~M., {Faber}, S.~M., {Ferguson}, H.~C., {et~al.} 2011, \apjs, 197, 36

\bibitem[{{Kokorev} {et~al.}(2024{\natexlab{a}}){Kokorev}, {Caputi}, {Greene}, {Dayal}, {Trebitsch}, {Cutler}, {Fujimoto}, {Labb{\'e}}, {Miller}, {Iani}, {Navarro-Carrera}, \& {Rinaldi}}]{Kokorev2024}
{Kokorev}, V., {Caputi}, K.~I., {Greene}, J.~E., {et~al.} 2024{\natexlab{a}}, \apj, 968, 38

\bibitem[{{Kokorev} {et~al.}(2024{\natexlab{b}}){Kokorev}, {Chisholm}, {Endsley}, {Finkelstein}, {Greene}, {Akins}, {Bromm}, {Casey}, {Fujimoto}, {Labb{\'e}}, \& {Larson}}]{Kokorev2024b}
{Kokorev}, V., {Chisholm}, J., {Endsley}, R., {et~al.} 2024{\natexlab{b}}, \apj, 975, 178

\bibitem[{{Kokubo} \& {Harikane}(2024)}]{Kokubo2024}
{Kokubo}, M. \& {Harikane}, Y. 2024, arXiv e-prints, arXiv:2407.04777

\bibitem[{{Kulkarni} {et~al.}(2019){Kulkarni}, {Worseck}, \& {Hennawi}}]{Kulkarni2019}
{Kulkarni}, G., {Worseck}, G., \& {Hennawi}, J.~F. 2019, \mnras, 488, 1035

\bibitem[{{Labbe} {et~al.}(2023){Labbe}, {Greene}, {Bezanson}, {Fujimoto}, {Furtak}, {Goulding}, {Matthee}, {Naidu}, {Oesch}, {Atek}, {Brammer}, {Chemerynska}, {Coe}, {Cutler}, {Dayal}, {Feldmann}, {Franx}, {Glazebrook}, {Leja}, {Marchesini}, {Maseda}, {Nanayakkara}, {Nelson}, {Pan}, {Papovich}, {Price}, {Suess}, {Wang}, {Whitaker}, {Williams}, \& {Zitrin}}]{Labbe2023b}
{Labbe}, I., {Greene}, J.~E., {Bezanson}, R., {et~al.} 2023, arXiv e-prints, arXiv:2306.07320

\bibitem[{{Labbe} {et~al.}(2025){Labbe}, {Greene}, {Bezanson}, {Fujimoto}, {Furtak}, {Goulding}, {Matthee}, {Naidu}, {Oesch}, {Atek}, {Brammer}, {Chemerynska}, {Coe}, {Cutler}, {Dayal}, {Feldmann}, {Franx}, {Glazebrook}, {Leja}, {Maseda}, {Marchesini}, {Nanayakkara}, {Nelson}, {Pan}, {Papovich}, {Price}, {Suess}, {Wang}, {Weaver}, {Whitaker}, {Williams}, \& {Zitrin}}]{Labbe2025}
{Labbe}, I., {Greene}, J.~E., {Bezanson}, R., {et~al.} 2025, \apj, 978, 92

\bibitem[{{Labb{\'e}} {et~al.}(2023){Labb{\'e}}, {van Dokkum}, {Nelson}, {Bezanson}, {Suess}, {Leja}, {Brammer}, {Whitaker}, {Mathews}, {Stefanon}, \& {Wang}}]{Labbe2023}
{Labb{\'e}}, I., {van Dokkum}, P., {Nelson}, E., {et~al.} 2023, \nat, 616, 266

\bibitem[{{Langeroodi} \& {Hjorth}(2023)}]{Langeroodi2023}
{Langeroodi}, D. \& {Hjorth}, J. 2023, \apjl, 957, L27

\bibitem[{{Leung} {et~al.}(2024){Leung}, {Finkelstein}, {P{\'e}rez-Gonz{\'a}lez}, {Morales}, {Taylor}, {Barro}, {Kocevski}, {Akins}, {Carnall}, {Ch{\'a}vez Ortiz}, {Cleri}, {Cullen}, {Donnan}, {Dunlop}, {Ellis}, {Grogin}, {Hirschmann}, {Koekemoer}, {Kokorev}, {Lucas}, {McLeod}, {Papovich}, \& {Yung}}]{Leung2024}
{Leung}, G. C.~K., {Finkelstein}, S.~L., {P{\'e}rez-Gonz{\'a}lez}, P.~G., {et~al.} 2024, arXiv e-prints, arXiv:2411.12005

\bibitem[{{Lin} {et~al.}(2024){Lin}, {Zheng}, {Jiang}, {Yuan}, {Ho}, {Wang}, {Jiang}, {Rhoads}, {Malhotra}, {Barrientos}, {Wold}, {Infante}, {Zhu}, {Ji}, \& {Fu}}]{Lin2024}
{Lin}, R., {Zheng}, Z.-Y., {Jiang}, C., {et~al.} 2024, arXiv e-prints, arXiv:2412.08396

\bibitem[{{Madau} \& {Dickinson}(2014)}]{Madau2014}
{Madau}, P. \& {Dickinson}, M. 2014, \araa, 52, 415

\bibitem[{{Madau} {et~al.}(2024){Madau}, {Giallongo}, {Grazian}, \& {Haardt}}]{Madau2024}
{Madau}, P., {Giallongo}, E., {Grazian}, A., \& {Haardt}, F. 2024, \apj, 971, 75

\bibitem[{Magorrian {et~al.}(1998)Magorrian, Tremaine, Richstone, Bender, Bower, Dressler, Faber, Gebhardt, Green, Grillmair, Kormendy, \& Lauer}]{Magorrian1998}
Magorrian, J., Tremaine, S., Richstone, D., {et~al.} 1998, \apj, 115, 2285

\bibitem[{{Maiolino} {et~al.}(2024{\natexlab{a}}){Maiolino}, {Risaliti}, {Signorini}, {Trefoloni}, {Juodzbalis}, {Scholtz}, {Uebler}, {D'Eugenio}, {Carniani}, {Fabian}, {Ji}, {Mazzolari}, {Bertola}, {Brusa}, {Bunker}, {Charlot}, {Comastri}, {Cresci}, {DeCoursey}, {Egami}, {Fiore}, {Gilli}, {Perna}, {Tacchella}, \& {Venturi}}]{Maiolino2024}
{Maiolino}, R., {Risaliti}, G., {Signorini}, M., {et~al.} 2024{\natexlab{a}}, arXiv e-prints, arXiv:2405.00504

\bibitem[{{Maiolino} {et~al.}(2024{\natexlab{b}}){Maiolino}, {Scholtz}, {Curtis-Lake}, {Carniani}, {Baker}, {de Graaff}, {Tacchella}, {{\"U}bler}, {D'Eugenio}, {Witstok}, {Curti}, {Arribas}, {Bunker}, {Charlot}, {Chevallard}, {Eisenstein}, {Egami}, {Ji}, {Jones}, {Lyu}, {Rawle}, {Robertson}, {Rujopakarn}, {Perna}, {Sun}, {Venturi}, {Williams}, \& {Willott}}]{Maiolino2024b}
{Maiolino}, R., {Scholtz}, J., {Curtis-Lake}, E., {et~al.} 2024{\natexlab{b}}, \aap, 691, A145

\bibitem[{{Maiolino} {et~al.}(2024{\natexlab{c}}){Maiolino}, {Scholtz}, {Witstok}, {Carniani}, {D'Eugenio}, {de Graaff}, {{\"U}bler}, {Tacchella}, {Curtis-Lake}, {Arribas}, {Bunker}, {Charlot}, {Chevallard}, {Curti}, {Looser}, {Maseda}, {Rawle}, {Rodr{\'\i}guez del Pino}, {Willott}, {Egami}, {Eisenstein}, {Hainline}, {Robertson}, {Williams}, {Willmer}, {Baker}, {Boyett}, {DeCoursey}, {Fabian}, {Helton}, {Ji}, {Jones}, {Kumari}, {Laporte}, {Nelson}, {Perna}, {Sandles}, {Shivaei}, \& {Sun}}]{Maiolino2024c}
{Maiolino}, R., {Scholtz}, J., {Witstok}, J., {et~al.} 2024{\natexlab{c}}, \nat, 627, 59

\bibitem[{{Matthee} {et~al.}(2024){Matthee}, {Naidu}, {Brammer}, {Chisholm}, {Eilers}, {Goulding}, {Greene}, {Kashino}, {Labbe}, {Lilly}, {Mackenzie}, {Oesch}, {Weibel}, {Wuyts}, {Xiao}, {Bordoloi}, {Bouwens}, {van Dokkum}, {Illingworth}, {Kramarenko}, {Maseda}, {Mason}, {Meyer}, {Nelson}, {Reddy}, {Shivaei}, {Simcoe}, \& {Yue}}]{Matthee2024}
{Matthee}, J., {Naidu}, R.~P., {Brammer}, G., {et~al.} 2024, \apj, 963, 129

\bibitem[{{Mazzolari} {et~al.}(2024){Mazzolari}, {Gilli}, {Maiolino}, {Prandoni}, {Delvecchio}, {Norman}, {Jimenez-Andrade}, {Belladitta}, {Vito}, {Momjian}, {Chiaberge}, {Trefoloni}, {Signorini}, {Ji}, {D'Amato}, {Risaliti}, {Baldi}, {Fabian}, {{\"U}bler}, {D'Eugenio}, {Scholtz}, {Juod{\v{z}}balis}, {Mignoli}, {Brusa}, {Murphy}, \& {Muxlow}}]{Mazzolari2024}
{Mazzolari}, G., {Gilli}, R., {Maiolino}, R., {et~al.} 2024, arXiv e-prints, arXiv:2412.04224

\bibitem[{{McGreer} {et~al.}(2013){McGreer}, {Jiang}, {Fan}, {Richards}, {Strauss}, {Ross}, {White}, {Shen}, {Schneider}, {Myers}, {Brandt}, {DeGraf}, {Glikman}, {Ge}, \& {Streblyanska}}]{McGreer2013}
{McGreer}, I.~D., {Jiang}, L., {Fan}, X., {et~al.} 2013, \apj, 768, 105

\bibitem[{{Mezcua} {et~al.}(2024){Mezcua}, {Pacucci}, {Suh}, {Siudek}, \& {Natarajan}}]{Mezcua2024}
{Mezcua}, M., {Pacucci}, F., {Suh}, H., {Siudek}, M., \& {Natarajan}, P. 2024, \apjl, 966, L30

\bibitem[{{Mullaney} {et~al.}(2012){Mullaney}, {Daddi}, {B{\'e}thermin}, {Elbaz}, {Juneau}, {Pannella}, {Sargent}, {Alexander}, \& {Hickox}}]{Mullaney2012}
{Mullaney}, J.~R., {Daddi}, E., {B{\'e}thermin}, M., {et~al.} 2012, \apjl, 753, L30

\bibitem[{{Netzer}(2015)}]{Netzer2015}
{Netzer}, H. 2015, \araa, 53, 365

\bibitem[{{Niida} {et~al.}(2020){Niida}, {Nagao}, {Ikeda}, {Akiyama}, {Matsuoka}, {He}, {Matsuoka}, {Toba}, {Onoue}, {Kobayashi}, {Taniguchi}, {Furusawa}, {Harikane}, {Imanishi}, {Kashikawa}, {Kawaguchi}, {Komiyama}, {Shirakata}, {Terashima}, \& {Ueda}}]{Niida2020}
{Niida}, M., {Nagao}, T., {Ikeda}, H., {et~al.} 2020, \apj, 904, 89

\bibitem[{{Noboriguchi} {et~al.}(2023){Noboriguchi}, {Inoue}, {Nagao}, {Toba}, \& {Misawa}}]{Noboriguchi2023}
{Noboriguchi}, A., {Inoue}, A.~K., {Nagao}, T., {Toba}, Y., \& {Misawa}, T. 2023, \apjl, 959, L14

\bibitem[{{Oke} \& {Gunn}(1983)}]{Oke1983}
{Oke}, J.~B. \& {Gunn}, J.~E. 1983, \apj, 266, 713

\bibitem[{{Pacucci} \& {Loeb}(2022)}]{Pacucci2022}
{Pacucci}, F. \& {Loeb}, A. 2022, \mnras, 509, 1885

\bibitem[{{Pacucci} {et~al.}(2017){Pacucci}, {Natarajan}, {Volonteri}, {Cappelluti}, \& {Urry}}]{Pacucci2017}
{Pacucci}, F., {Natarajan}, P., {Volonteri}, M., {Cappelluti}, N., \& {Urry}, C.~M. 2017, \apjl, 850, L42

\bibitem[{{Pan} {et~al.}(2022){Pan}, {Jiang}, {Fan}, {Wu}, \& {Yang}}]{Pan2022}
{Pan}, Z., {Jiang}, L., {Fan}, X., {Wu}, J., \& {Yang}, J. 2022, \apj, 928, 172

\bibitem[{{P{\'e}rez-Gonz{\'a}lez} {et~al.}(2024){P{\'e}rez-Gonz{\'a}lez}, {Barro}, {Rieke}, {Lyu}, {Rieke}, {Alberts}, {Williams}, {Hainline}, {Sun}, {Pusk{\'a}s}, {Annunziatella}, {Baker}, {Bunker}, {Egami}, {Ji}, {Johnson}, {Robertson}, {Rodr{\'\i}guez Del Pino}, {Rujopakarn}, {Shivaei}, {Tacchella}, {Willmer}, \& {Willott}}]{PerezGonzalez2024}
{P{\'e}rez-Gonz{\'a}lez}, P.~G., {Barro}, G., {Rieke}, G.~H., {et~al.} 2024, \apj, 968, 4

\bibitem[{{Reis} {et~al.}(2019){Reis}, {Baron}, \& {Shahaf}}]{Reis_2019_2019AJ....157...16R}
{Reis}, I., {Baron}, D., \& {Shahaf}, S. 2019, \aj, 157, 16

\bibitem[{{Richards} {et~al.}(2003){Richards}, {Hall}, {Vanden Berk}, {Strauss}, {Schneider}, {Weinstein}, {Reichard}, {York}, {Knapp}, {Fan}, {Ivezi{\'c}}, {Brinkmann}, {Budav{\'a}ri}, {Csabai}, \& {Nichol}}]{Richards2003}
{Richards}, G.~T., {Hall}, P.~B., {Vanden Berk}, D.~E., {et~al.} 2003, \aj, 126, 1131

\bibitem[{Rigby {et~al.}(2023)Rigby, Perrin, McElwain, Kimble, Friedman, Lallo, Doyon, Feinberg, Ferruit, Glasse, Rieke, Rieke, Wright, Willott, Colon, Milam, Neff, Stark, Valenti, Abell, Abney, Abul-Huda, Scott~Acton, Adams, Adler, Aguilar, Ahmed, Albert, Alberts, Aldridge, Allen, Altenburg, Álvarez Márquez, Alves~de Oliveira, Andersen, Anderson, Anderson, Argyriou, Armstrong, Arribas, Artigau, Arvai, Atkinson, Bacon, Bair, Banks, Barrientes, Barringer, Bartosik, Bast, Baudoz, Beatty, Bechtold, Beck, Bergeron, Bergkoetter, Bhatawdekar, Birkmann, Blazek, Blome, Boccaletti, Böker, Boia, Bonaventura, Bond, Bosley, Boucarut, Bourque, Bouwman, Bower, Bowers, Boyer, Bradley, Brady, Braun, Breda, Bresnahan, Bright, Britt, Bromenschenkel, Brooks, Brooks, Brown, Brown, Brown, Bunker, Burger, Bushouse, Cale, Cameron, Cameron, Canipe, Caplinger, Caputo, Cara, Carey, Carniani, Carrasquilla, Carruthers, Case, Catherine, Chance, Chapman, Charlot, Charlow, Chayer, Chen, Cherinka, Chichester, Chilton, Chonis, Clampin,
  Clark, Clark, Coe, Coleman, Comber, Comeau, Connolly, Cooper, Cooper, Coppock, Correnti, Cossou, Coulais, Coyle, Cracraft, Curti, Cuturic, Davis, Davis, Dean, DeLisa, deMeester, Dencheva, Dencheva, DePasquale, Deschenes, Hunor~Detre, Diaz, Dicken, DiFelice, Dillman, Dixon, Doggett, Donaldson, Douglas, DuPrie, Dupuis, Durning, Easmin, Eck, Edeani, Egami, Ehrenwinkler, Eisenhamer, Eisenhower, Elie, Elliott, Elliott, Ellis, Engesser, Espinoza, Etienne, Etxaluze, Falini, Feeney, Ferry, Filippazzo, Fincham, Fix, Flagey, Florian, Flynn, Fontanella, Ford, Forshay, Fox, Franz, Fu, Fullerton, Galkin, Galyer, García~Marín, Gardner, Gardner, Garland, Garrett, Gasman, Gaspar, Gaudreau, Gauthier, Geers, Geithner, Gennaro, Giardino, Girard, Giuliano, Glassmire, Glauser, Glazer, Godfrey, Golimowski, Gollnitz, Gong, Gonzaga, Gordon, Gordon, Goudfrooij, Greene, Greenhouse, Grimaldi, Groebner, Grundy, Guillard, Gutman, Ha, Haderlein, Hagedorn, Hainline, Haley, Hami, Hamilton, Hammel, Hansen, Harkins, Harr, Hart, Hart,
  Hartig, Hashimoto, Haskins, Hathaway, Havey, Hayden, Hecht, Heller-Boyer, Henriques, Henry, Hermann, Hernandez, Hesman, Hicks, Hilbert, Hines, Hoffman, Holfeltz, Holler, Hoppa, Hott, Howard, Howard, Hunter, Hunter, Hurst, Husemann, Hustak, Ilinca~Ignat, Illingworth, Irish, Jackson, Jahromi, Jakobsen, James, James, Januszewski, Jenkins, Jirdeh, Johnson, Johnson, Jones, Jones, Jones, Jones, Jordan, Jordan, Jurczyk, Jurling, Kaleida, Kalmanson, Kammerer, Kang, Kao, Karakla, Kavanagh, Kelly, Kendrew, Kennedy, Kenny, Keski-kuha, Keyes, Kidwell, Kinzel, Kirk, Kirkpatrick, Kirshenblat, Klaassen, Knapp, Scott~Knight, Knollenberg, Koehler, Koekemoer, Kovacs, Kulp, Kumari, Kyprianou, La~Massa, Labador, Labiano, Lagage, Lajoie, Lallo, Lam, Lamb, Lambros, Lampenfield, Langston, Larson, Law, Lawrence, Lee, Leisenring, Lepo, Leveille, Levenson, Levine, Levy, Lewis, Lewis, Libralato, Lightsey, Link, Liu, Lo, Lockwood, Logue, Long, Long, Loomis, Lopez-Caniego, Lorenzo~Alvarez, Love-Pruitt, Lucy, Luetzgendorf, Maghami,
  Maiolino, Major, Malla, Malumuth, Manjavacas, Mannfolk, Marrione, Marston, Martel, Maschmann, Masci, Masciarelli, Maszkiewicz, Mather, McKenzie, McLean, McMaster, Melbourne, Meléndez, Menzel, Merz, Meyett, Meza, Miskey, Misselt, Moller, Morrison, Morse, Moseley, Mosier, Mountain, Mueckay, Mueller, Mullally, Murphy, Murray, Murray, Mustelier, Muzerolle, Mycroft, Myers, Myrick, Nanavati, Nance, Nayak, Naylor, Nelan, Nickson, Nielson, Nieto-Santisteban, Nikolov, Noriega-Crespo, O’Shaughnessy, O’Sullivan, Ochs, Ogle, Oleszczuk, Olmsted, Osborne, Ottens, Owens, Pacifici, Pagan, Page, Park, Parrish, Patapis, Paul, Pauly, Pavlovsky, Pedder, Peek, Pena-Guerrero, Penanen, Perez, Perna, Perriello, Phillips, Pietraszkiewicz, Pinaud, Pirzkal, Pitman, Piwowar, Platais, Player, Plesha, Pollizi, Polster, Pontoppidan, Porterfield, Proffitt, Pueyo, Pulliam, Quirt, Quispe~Neira, Ramos~Alarcon, Ramsay, Rapp, Rapp, Rauscher, Ravindranath, Rawle, Regan, Reichard, Reis, Ressler, Rest, Reynolds, Rhue, Richon, Rickman,
  Ridgaway, Ritchie, Rix, Robberto, Robinson, Robinson, Robinson, Rock, Rodriguez, Rodriguez Del~Pino, Roellig, Rohrbach, Roman, Romelfanger, Rose, Roteliuk, Roth, Rothwell, Rowlands, Roy, Royer, Royle, Rui, Rumler, Runnels, Russ, Rustamkulov, Ryden, Ryer, Sabata, Sabatke, Sabbi, Samuelson, Sapp, Sappington, Sargent, Sauer, Scheithauer, Schlawin, Schlitz, Schmitz, Schneider, Schreiber, Schulze, Schwab, Scott, Sembach, Shanahan, Shaughnessy, Shaw, Shawger, Shay, Sheehan, Shen, Sherman, Shiao, Shih, Shivaei, Sienkiewicz, Sing, Sirianni, Sivaramakrishnan, Skipper, Sloan, Slocum, Slowinski, Smith, Smith, Smith, Smith, Snyder, Soh, Tony~Sohn, Soto, Spencer, Stallcup, Stansberry, Starr, Starr, Stewart, Stiavelli, Straughn, Strickland, Stys, Summers, Sun, Sunnquist, Swade, Swam, Swaters, Swoish, Taylor, Taylor, Te~Plate, Tea, Teague, Telfer, Temim, Thatte, Thompson, Thompson, Thomson, Tikkanen, Tippet, Todd, Toolan, Tran, Trejo, Truong, Tsukamoto, Tustain, Tyra, Ubeda, Underwood, Uzzo, Van~Campen, Vandal,
  Vandenbussche, Vila, Volk, Wahlgren, Waldman, Walker, Wander, Warfield, Warner, Wasiak, Watkins, Weaver, Weilert, Weiser, Weiss, Weissman, Welty, West, Wheate, Wheatley, Wheeler, White, Whiteaker, Whitehouse, Whiteleather, Whitman, Williams, Willmer, Willoughby, Wilson, Wirth, Wislowski, Wolf, Wolfe, Wolff, Workman, Wright, Wu, Wu, Wymer, Yates, Yeager, Yeates, Yerger, Yoon, Young, Yu, Zak, Zeidler, Zhou, Zielinski, Zincke, \& Zonak}]{Rigby_2023}
Rigby, J., Perrin, M., McElwain, M., {et~al.} 2023, Publications of the Astronomical Society of the Pacific, 135, 048001

\bibitem[{{Rowan-Robinson}(1968)}]{Rowan-Robinson1968}
{Rowan-Robinson}, M. 1968, \mnras, 138, 445

\bibitem[{{Salvato} {et~al.}(2022){Salvato}, {Wolf}, {Dwelly}, {Georgakakis}, {Brusa}, {Merloni}, {Liu}, {Toba}, {Nandra}, {Lamer}, {Buchner}, {Schneider}, {Freund}, {Rau}, {Schwope}, {Nishizawa}, {Klein}, {Arcodia}, {Comparat}, {Musiimenta}, {Nagao}, {Brunner}, {Malyali}, {Finoguenov}, {Anderson}, {Shen}, {Ibarra-Medel}, {Trump}, {Brandt}, {Urry}, {Rivera}, {Krumpe}, {Urrutia}, {Miyaji}, {Ichikawa}, {Schneider}, {Fresco}, {Boller}, {Haase}, {Brownstein}, {Lane}, {Bizyaev}, \& {Nitschelm}}]{Salvato2022}
{Salvato}, M., {Wolf}, J., {Dwelly}, T., {et~al.} 2022, \aap, 661, A3

\bibitem[{{Schmidt}(1968)}]{Schmidt1968}
{Schmidt}, M. 1968, \apj, 151, 393

\bibitem[{{Silk} \& {Rees}(1998)}]{Silk1998}
{Silk}, J. \& {Rees}, M.~J. 1998, \aap, 331, L1

\bibitem[{{Simmonds} {et~al.}(2024){Simmonds}, {Tacchella}, {Hainline}, {Johnson}, {Pusk{\'a}s}, {Robertson}, {Baker}, {Bhatawdekar}, {Boyett}, {Bunker}, {Cargile}, {Carniani}, {Chevallard}, {Curti}, {Curtis-Lake}, {Ji}, {Jones}, {Kumari}, {Laseter}, {Maiolino}, {Maseda}, {Rinaldi}, {Stoffers}, {{\"U}bler}, {Villanueva}, {Williams}, {Willott}, {Witstok}, \& {Zhu}}]{Simmonds2024}
{Simmonds}, C., {Tacchella}, S., {Hainline}, K., {et~al.} 2024, \mnras, 535, 2998

\bibitem[{{Stepney} {et~al.}(2024){Stepney}, {Banerji}, {Tang}, {Hewett}, {Temple}, {Wethers}, {Puglisi}, \& {Molyneux}}]{Stepney2024}
{Stepney}, M., {Banerji}, M., {Tang}, S., {et~al.} 2024, \mnras, 533, 2948

\bibitem[{{Tacconi} {et~al.}(2020){Tacconi}, {Genzel}, \& {Sternberg}}]{Tacconi2020}
{Tacconi}, L.~J., {Genzel}, R., \& {Sternberg}, A. 2020, \araa, 58, 157

\bibitem[{{Tanaka} {et~al.}(2024){Tanaka}, {Silverman}, {Shimasaku}, {Arita}, {Akins}, {Inayoshi}, {Ding}, {Onoue}, {Liu}, {Casey}, {Lambrides}, {Kokorev}, {Jin}, {Faisst}, {Drakos}, {Shen}, {Li}, {Zhuang}, {Fei}, {Ito}, {Ren}, {Matsui}, {Ando}, {Hatano}, {Fujii}, {Kartaltepe}, {Koekemoer}, {Liu}, {McCracken}, {Rhodes}, {Robertson}, {Franco}, {Andika}, {Cloonan}, {Fan}, {Gozaliasl}, {Harish}, {Hayward}, {Huertas-Company}, {Kakkad}, {Kinugawa}, {Roy}, {Shuntov}, {Talia}, {Toft}, {Vijayan}, \& {Zhang}}]{Tanaka2024}
{Tanaka}, T.~S., {Silverman}, J.~D., {Shimasaku}, K., {et~al.} 2024, arXiv e-prints, arXiv:2412.14246

\bibitem[{{Tee} {et~al.}(2024){Tee}, {Fan}, {Wang}, \& {Yang}}]{Tee2024}
{Tee}, W.~L., {Fan}, X., {Wang}, F., \& {Yang}, J. 2024, arXiv e-prints, arXiv:2412.05242

\bibitem[{{Traina} {et~al.}(2024){Traina}, {Gruppioni}, {Delvecchio}, {Calura}, {Bisigello}, {Feltre}, {Magnelli}, {Schinnerer}, {Liu}, {Adscheid}, {Behiri}, {Gentile}, {Pozzi}, {Talia}, {Zamorani}, {Algera}, {Gillman}, {Lambrides}, \& {Symeonidis}}]{Traina2024}
{Traina}, A., {Gruppioni}, C., {Delvecchio}, I., {et~al.} 2024, \aap, 681, A118

\bibitem[{{Urrutia} {et~al.}(2008){Urrutia}, {Lacy}, \& {Becker}}]{Urrutia2008}
{Urrutia}, T., {Lacy}, M., \& {Becker}, R.~H. 2008, \apj, 674, 80

\bibitem[{van~der Wel {et~al.}(2014)van~der Wel, Franx, van Dokkum, Skelton, Momcheva, Whitaker, Brammer, Bell, Rix, Wuyts, Ferguson, Holden, Barro, Koekemoer, Chang, McGrath, Häussler, Dekel, Behroozi, Fumagalli, Leja, Lundgren, Maseda, Nelson, Wake, Patel, Labbé, Faber, Grogin, \& Kocevski}]{Wel2014}
van~der Wel, A., Franx, M., van Dokkum, P.~G., {et~al.} 2014, \apj, 788, 28

\bibitem[{Virtanen {et~al.}(2020)Virtanen, Gommers, Oliphant, Haberland, Reddy, Cournapeau, Burovski, Peterson, Weckesser, Bright, {van der Walt}, Brett, Wilson, Millman, Mayorov, Nelson, Jones, Kern, Larson, Carey, Polat, Feng, Moore, {VanderPlas}, Laxalde, Perktold, Cimrman, Henriksen, Quintero, Harris, Archibald, Ribeiro, Pedregosa, {van Mulbregt}, \& {SciPy 1.0 Contributors}}]{SciPy}
Virtanen, P., Gommers, R., Oliphant, T.~E., {et~al.} 2020, Nature Methods, 17, 261

\bibitem[{Wang {et~al.}(2024)Wang, Leja, de~Graaff, Brammer, Weibel, van Dokkum, Baggen, Suess, Greene, Bezanson, Cleri, Hirschmann, Labbé, Matthee, McConachie, Naidu, Nelson, Oesch, Setton, \& Williams}]{Wang_2024}
Wang, B., Leja, J., de~Graaff, A., {et~al.} 2024, \apjl, 969, L13

\bibitem[{{Webster} {et~al.}(1995){Webster}, {Francis}, {Petersont}, {Drinkwater}, \& {Masci}}]{Webster1995}
{Webster}, R.~L., {Francis}, P.~J., {Petersont}, B.~A., {Drinkwater}, M.~J., \& {Masci}, F.~J. 1995, \nat, 375, 469

\bibitem[{{Wethers} {et~al.}(2018){Wethers}, {Banerji}, {Hewett}, {Lemon}, {McMahon}, {Reed}, {Shen}, {Abdalla}, {Benoit-L{\'e}vy}, {Brooks}, {Buckley-Geer}, {Capozzi}, {Carnero Rosell}, {CarrascoKind}, {Carretero}, {Cunha}, {D'Andrea}, {da Costa}, {DePoy}, {Desai}, {Doel}, {Flaugher}, {Fosalba}, {Frieman}, {Garc{\'\i}a-Bellido}, {Gerdes}, {Gruen}, {Gruendl}, {Gschwend}, {Gutierrez}, {Honscheid}, {James}, {Jeltema}, {Kuehn}, {Kuhlmann}, {Kuropatkin}, {Lima}, {Maia}, {Marshall}, {Martini}, {Menanteau}, {Miquel}, {Nichol}, {Nord}, {Plazas}, {Romer}, {Sanchez}, {Scarpine}, {Schindler}, {Schubnell}, {Sevilla-Noarbe}, {Smith}, {Smith}, {Soares-Santos}, {Sobreira}, {Suchyta}, {Tarle}, \& {Walker}}]{Wethers2018}
{Wethers}, C.~F., {Banerji}, M., {Hewett}, P.~C., {et~al.} 2018, \mnras, 475, 3682

\bibitem[{{Williams} {et~al.}(2024){Williams}, {Alberts}, {Ji}, {Hainline}, {Lyu}, {Rieke}, {Endsley}, {Suess}, {Sun}, {Johnson}, {Florian}, {Shivaei}, {Rujopakarn}, {Baker}, {Bhatawdekar}, {Boyett}, {Bunker}, {Cameron}, {Carniani}, {Charlot}, {Curtis-Lake}, {DeCoursey}, {de Graaff}, {Egami}, {Eisenstein}, {Gibson}, {Hausen}, {Helton}, {Maiolino}, {Maseda}, {Nelson}, {P{\'e}rez-Gonz{\'a}lez}, {Rieke}, {Robertson}, {Saxena}, {Tacchella}, {Willmer}, \& {Willott}}]{Williams2024}
{Williams}, C.~C., {Alberts}, S., {Ji}, Z., {et~al.} 2024, \apj, 968, 34

\bibitem[{{Wyithe} \& {Loeb}(2012)}]{Wyithe2012}
{Wyithe}, J. S.~B. \& {Loeb}, A. 2012, \mnras, 425, 2892

\bibitem[{{Yue} {et~al.}(2024){Yue}, {Eilers}, {Ananna}, {Panagiotou}, {Kara}, \& {Miyaji}}]{Yue2024}
{Yue}, M., {Eilers}, A.-C., {Ananna}, T.~T., {et~al.} 2024, \apjl, 974, L26

\bibitem[{{Zhang} {et~al.}(2024){Zhang}, {Jiang}, {Liu}, \& {Ho}}]{Zhang2024}
{Zhang}, Z., {Jiang}, L., {Liu}, W., \& {Ho}, L.~C. 2024, arXiv e-prints, arXiv:2411.02729

\end{thebibliography}

\begin{appendix}
\section{IRAC photometry comparison}\label{sec:IRAC_comparison}
In this Section we show the comparison between the IRAC photometry derived in this work (see \cref{sec:IRAC}) and the one from the Cosmic Dawn Survey Catalogue \citep{EP-Zalesky}. This comparison is possible only in the EDF-N and EDF-F, as the Cosmic Dawn Catalogues are not available for the EDF-S. Sources are matched considering a $1\arcsec$ radius, but we verify that the difference does not change significantly considering a smaller matching radius. 

\begin{figure}[h!]
    \centering
    \includegraphics[width=\linewidth]{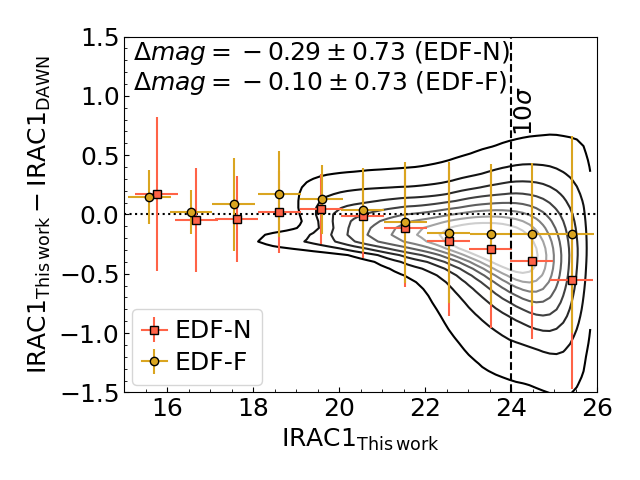}
    \includegraphics[width=\linewidth]{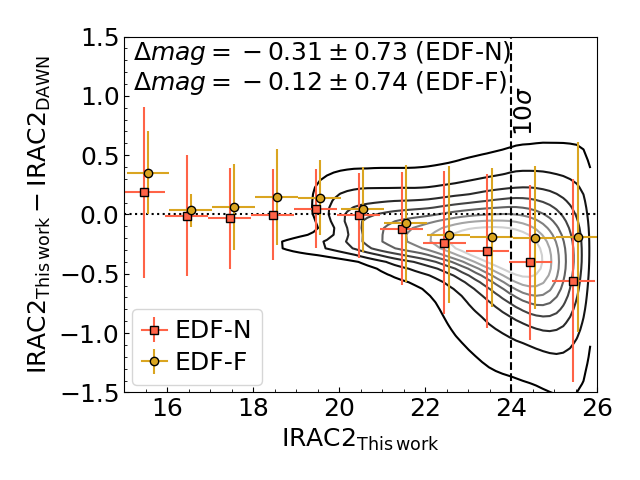}
    \caption{Difference between the magnitudes derived in this work and the ones in the Cosmic Dawn Catalogue. The black contours show the distribution of all sources in common between the two catalogues (from 10\% to 90\% of the sample), while red squares and yellow circles show the average values in magnitude bins for the EDF-N and EDF-F, respectively. We report in the top left the mean and standard deviation of the difference in magnitude in the two fields. We report results for the IRAC1 filter on the top and the IRAC2 filter on the bottom panel.  }
    \label{fig:IRAC_comp}
\end{figure}

As visible in \cref{fig:IRAC_comp}, the magnitude estimated in our work are, on average, slightly underestimated by $-0.10$ magnitudes in the EDF-F and $-0.30$ magnitudes in the EDF-N. These differences are mainly dominated by faint sources as they decrease to $\Delta \rm IRAC1=0.07\pm0.34$ and $\Delta \rm IRAC2=0.07\pm0.35$ in the EDF-F and $\Delta \rm IRAC1=0.01\pm0.34$ and $\Delta \rm IRAC2=0.01\pm0.35$ in the EDF-N when we limit the analysis to sources brighter than 21 in the respective bands.

\section{Redshift recovery}\label{sec:z_pipeline}
The \Euclid pipeline, at the moment, lacks LRD templates. This can have a direct impact on the redshift estimation for these objects. We therefore decided to test the redshift recovery for LRDs by simulating \Euclid observations of LRDs and input this mock photometry to the \Euclid pipeline. 

In particular, we start by simulating simple LRD spectra, using two power laws, with red and blue continuum slopes at rest-frame optical and UV wavelengths. For the $\beta_{\rm UV}$ and $\beta_{\rm opt}$ continuum slopes, we considered the values measured by \citet{Kocevski2024} for a sample of 341 LRDs at $z=2$--$11$, but with a median redshift of $z=6.4$. We used these mock LRDs as templates, moving them from $z=1$ to $z=7.6$, where we have enough bands to measure the two slopes in all EDFs. We scale their absolute UV magnitudes ($M_{\rm UV}$) and compare these with the \Euclid observational depths. We then derive photometric redshift, considering the same setup used in the \Euclid pipeline.

\begin{figure}[h!]
    \centering
    \includegraphics[width=\linewidth]{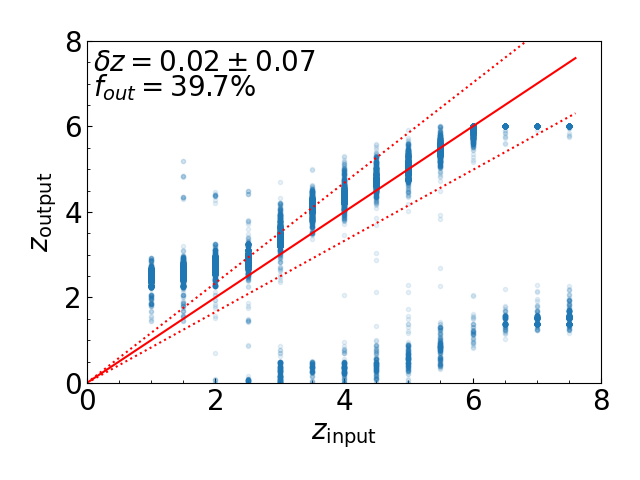}
    \includegraphics[width=\linewidth]{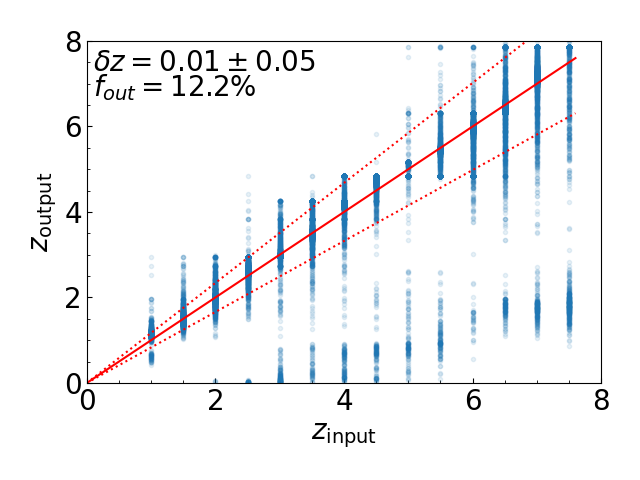}
    \caption{Redshift recovery using the \Euclid pipeline (top) and with the double power-law fit improvement (bottom). In the upper right of each panel we report the fraction of outliers, defined as objects with $|\delta z|=|z_{\rm output}-z_{\rm input}|/(1+z_{\rm input})>0.15$, as well as the mean and standard deviation of $\delta z$, measured after removing the outliers.}
    \label{fig:zout}
\end{figure}

In the top panel of \cref{fig:zout} we report the comparison between the recovered and input redshifts. As can be seen, the first effect is a bimodal distribution, with some galaxies that are wrongly placed at $z<2$. In addition, the output redshift tends to be overestimated for galaxies at $z_{\rm input}<2$, while the upper limit at $z=6$ naturally underestimate the redshift of $z>6$ sources. Overall, we have around $40\%$ of outliers ($f_{\rm out}$), defined as galaxies with $|\delta z|=|z_{\rm output}-z_{\rm input}|/(1+z_{\rm input})>0.15$. If we remove these outliers, we obtain a distribution consistent with no redshift bias, since $\delta z=0.02\pm0.07$.

In the bottom panel of \cref{fig:zout} we instead show the comparison between the true redshift and the one recovered by fitting the mock photometry with a double power law and leaving the redshift free to vary (see \cref{sec:selection}). In this fit, we considered the pipeline redshift as a starting point and the uncertainties as limits. We verified that leaving the redshift totally free improves the results. This refinement in the redshift produces a substantial reduction in the outlier fraction, more than halving it ($f_{\rm out}=12\%$). At the same time, the biases at the different redshifts are similar and the $z=6$ upper limit is removed. The expected mean redshift bias, after removing the outliers, is also slightly improved, becoming $\delta z=0.01\pm0.05$. 

By comparing the intrinsic redshift distribution and the one derived after the two power-law fit, we derived an average redshift correction to apply to the estimated luminosity function (\cref{sec:LF}). These redshift corrections are derived in redshift bins and are: 0.60 at $z=1.0$--$1.5$; 0.91 at $z=1.5$--$2.0$; 1.32 at $z=2.0$--$2.5$; 1.09 at $z=2.5$--$4.0$; and 1.09 at $z=4.0$--$6.0$ 

\FloatBarrier
\section{Impact of slope uncertainties on sample selection}\label{sec:rand_slopes}
In this Section we focus on the impact of the uncertainties of the rest-frame UV and optical slopes on the sample selection. In particular, we start from the subsamples of sources with ${\rm S/N>3}$ in more than four filters. We then randomise the rest-frame UV and optical slopes 100 times, using a normal distribution centred on the best value and with standard deviation equal to the respective uncertainties. After this randomization, we select for each iteration the number of v-shape sources, checking then the subsample of these objects that is also compact, is not contaminated by nebular emission lines and has a $\chi^{2}<100$. We do not perform the visual check of all sources selected in the different iterations. 

In \cref{tab:Nscatter} we report the 16\% and 84\% of the distribution of the number of selected objects in the different fields. As visible, the values are larger than the one reported in \cref{tab:numbers}, indicating that the number of LRD candidates may be underestimated due to the uncertainties on the rest-frame UV and optical slopes. However, given that the number of potential contaminants is expected to be larger than the number of LRD, we prefer to keep the conservative estimates derived in the main text of this work. 

\begin{table}[]
\caption{16\% and 84\% percentages of the number of selected objects when randomising the slope measurements.}
    \centering
    \resizebox{\linewidth}{!}{
    \begin{tabular}{lccc}
    \hline
    \hline
    \noalign{\vskip 2pt}
     & EDF-F & EDF-N & EDF-S \\
     & \multicolumn{3}{c}{IRAC} \\
     \hline
     \noalign{\vskip 2pt}
v-shape continuum   &  (3700, 3824) & (1997, 2071) & (12\,625 12\,839) \\
Compact  & (228, 253) & (100, 120) & (670, 716) \\
No emission lines  & (163, 190) & (49, 62) & (378, 414) \\
$\chi^{2}<100$  & (151, 170) & (38, 49) & (357, 391)  \\
\hline
\noalign{\vskip 2pt}
& \multicolumn{3}{c}{No-IRAC, $z\leq2.1$} \\
 \hline     \noalign{\vskip 2pt}
v-shape continuum   & (112\,290, 112\,823) & (106\,490, 106\,997) & (226\,002, 226\,804)\\
Compact  & (3439, 3533) & (3092, 3189) & (7010, 7144)\\
No emission lines & (2845, 2936) & (2375,   2451) & (5764, 5890) \\
$\chi^{2}<100$ & (2806, 2900) & (2211, 2287) & (5674, 5798) \\
     \hline
    \end{tabular}}
    \label{tab:Nscatter}
\end{table}

\FloatBarrier
\section{Complete sample}
In \cref{tab:prop} we include the final list of LRD candidates and their properties, derived using the double power law fit described in \cref{sec:selection} and whose performance is shown with mock data in \cref{sec:z_pipeline}.

\begin{table*}[]
    \caption{Properties of LRD candidates.}
    \centering
    \resizebox{\textwidth}{!}{
    \begin{tabular}{ccccccccccc}
\hline
\hline
\noalign{\vskip 2pt}
  \multicolumn{1}{c}{ID} &
  \multicolumn{1}{c}{RA [deg]} &
  \multicolumn{1}{c}{Dec [deg]} &
  \multicolumn{1}{c}{$z$} &
  \multicolumn{1}{c}{$\sigma_{z}$} &
  \multicolumn{1}{c}{$M_{\rm UV}$} &
  \multicolumn{1}{c}{$\sigma_{M_{\rm UV}}$} &
  \multicolumn{1}{c}{$\beta_{\rm UV}$} &
  \multicolumn{1}{c}{$\sigma_{\beta_{\rm UV}}$} &
  \multicolumn{1}{c}{$\beta_{\rm opt}$} &
  \multicolumn{1}{c}{$\sigma_{\beta_{\rm opt}}$} \\
\hline
\noalign{\vskip 2pt}
  $-508115529279019500$ & 50.811552 & $-27.901950$ & 1.73 & 0.04 & $-19.3$ & 0.4 & $-2.5$ & 0.8 & 1.3 & 0.7\\
  $-509746984272137369$ & 50.974698 & $-27.213736$ & 1.43 & 0.01 & $-19.1$ & 0.4 & $-2.3$ & 0.6 & 0.4 & 0.4\\
  $-506780226277952382$ & 50.678022 & $-27.795238$ & 1.84 & 0.09 & $-19.2$ & 0.2 & $-2.0$ & 0.5 & 1.0 & 0.5\\
  $-508524594269841903$ & 50.852459 & $-26.984190$ & 2.08 & 0.05 & $-19.4$ & 0.2 & $-1.2$ & 0.5 & 0.2 & 0.7\\
  $-508984186271393227$ & 50.898418 & $-27.139322$ & 2.04 & 0.01 & $-19.1$ & 0.4 & $-1.3$ & 0.7 & 0.8 & 0.9\\
  $-509604641270953419$ & 50.960464 & $-27.095341$ & 1.56 & 0.07 & $-18.5$ & 0.5 & $-1.7$ & 0.8 & 0.8 & 0.5\\
  $-508222098269760365$ & 50.822209 & $-26.976036$ & 1.93 & 0.02 & $-19.6$ & 0.2 & $-1.5$ & 0.4 & 0.2 & 0.5\\
  $-508762123270516640$ & 50.876212 & $-27.051664$ & 1.56 & 0.01 & $-19.4$ & 0.3 & $-2.6$ & 0.6 & 0.1 & 0.5\\
  $-508397504271496709$ & 50.839750 & $-27.149670$ & 2.04 & 0.07 & $-19.1$ & 0.4 & $-1.8$ & 0.9 & 1.4 & 1.2\\
  $-509723625268249072$ & 50.972362 & $-26.824907$ & 1.67 & 0.19 & $-18.4$ & 0.5 & $-1.3$ & 0.6 & 0.4 & 0.6\\
  \hline
\end{tabular}
    }
    \tablefoot{Here we report the first 10 objects, while the complete list is available online.}
    \label{tab:prop}
\end{table*}

\FloatBarrier
\section{Observed properties of the LRD candidates}\label{sec:prop}
\begin{figure}
    \centering
    \includegraphics[width=\linewidth,trim={0 0 0 0.5cm},clip,keepaspectratio]{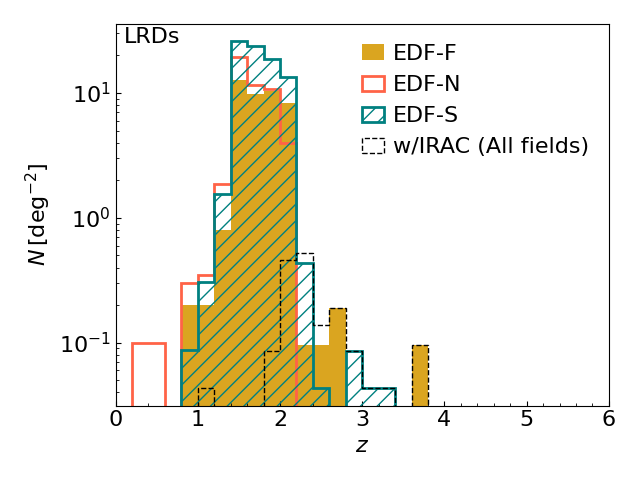}
    \caption{Photometric redshift distribution of the samples of LRD candidates in the three EDFs. The black dotted histogram shows the overall distribution of sources detected in IRAC.}
    \label{fig:zdist}    
\end{figure}

In  \cref{fig:zdist} we report the photometric redshift distribution for the three fields. The average redshift is similar among the three fields and is $z=1.7$. The average redshift of IRAC-detected LRD candidates is larger, that is $z=2.4$, with no LRD candidates at $z\geq4$.

\begin{figure*}
    \centering
    \includegraphics[width=0.49\linewidth]{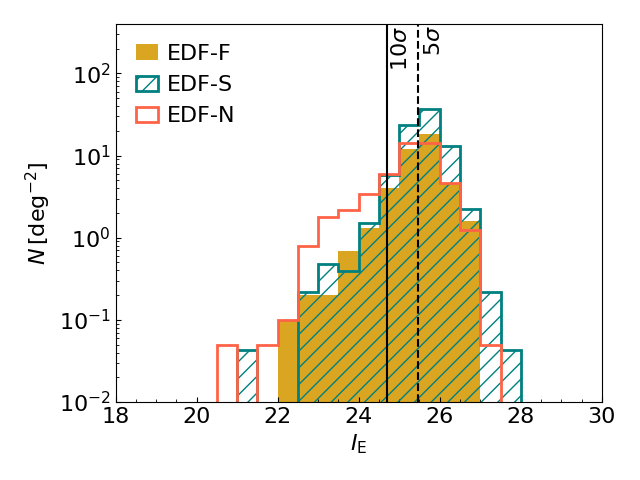}
    \includegraphics[width=0.49\linewidth]{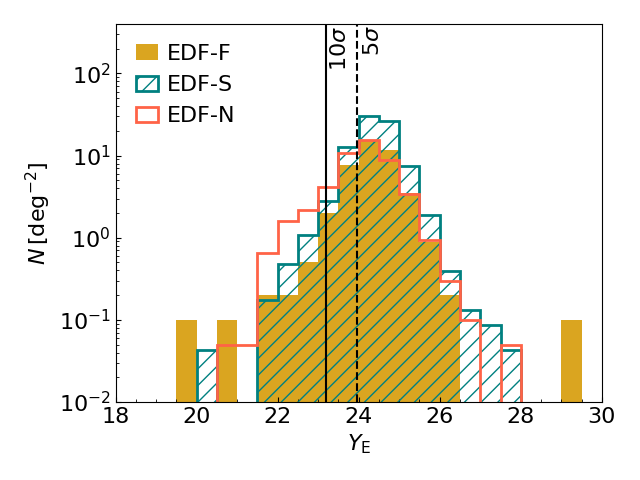}
    \includegraphics[width=0.49\linewidth]{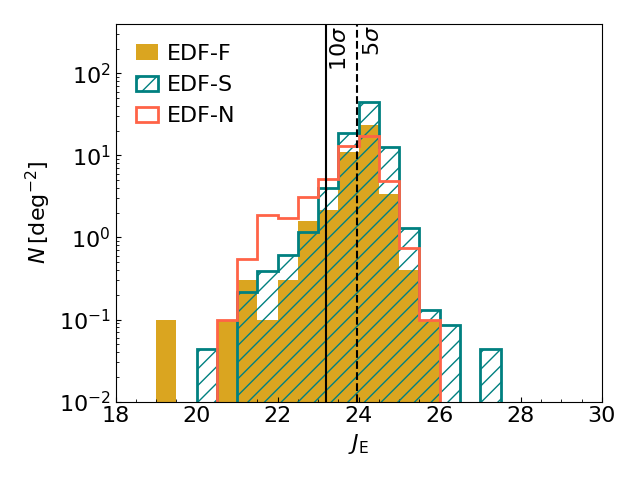}
    \includegraphics[width=0.49\linewidth]{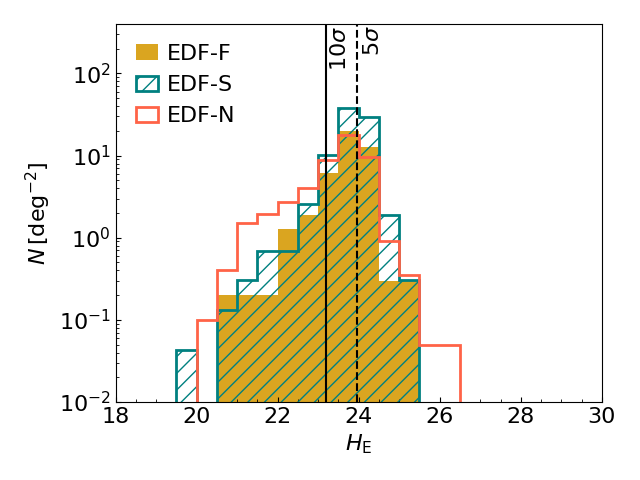}
    \caption{Magnitude density distribution of the LRD candidates in the three fields in the four \Euclid filters. The black vertical solid and dashed lines indicate the $10\sigma$ and $5\sigma$ depths, respectively. }
    \label{fig:mag_dist}
\end{figure*}

In \cref{fig:mag_dist} we report the magnitude distributions in the four \Euclid filters, showing that they are similar for all three fields, with the EDF-N having a light excess at bright magnitudes with respect of the other two fields. Indeed, the mean magnitudes in the EDF-F and EDF-S differ by less than 0.1 mag, while the EDF-N shows slightly brighter magnitudes than the other two fields, with a difference below 0.3 magnitudes. The similarities in the magnitude distributions of the three fields reassures us that the selection is reasonably uniform across these fields.

LRD candidates have mean magnitudes of $\IE=25.5$, $\YE=24.3$, $\JE=24.0$, and $\HE=23.7$ and hence are relatively faint. Indeed, only 11\% and 8\% of them have magnitudes brighter than the $10\sigma$ depth in the \HE and \IE filters, respectively, while these fractions decrease to 7\% and 3\% in the \JE and \YE filters. The sources detected in IRAC have a mean magnitudes of $\rm IRAC1=22.2$ and $\rm IRAC2=21.5$, showing that they are relatively bright in these two filters. In general, the fit is based only on four filters with ${\rm S/N}>3$ for 20\% of the sample, while the remaining sources have from five to 10 filters with ${\rm S/N}>3$.

\FloatBarrier
\section{Comparison with other \Euclid AGN catalogues}\label{sec:QSO-appendix}

As mentioned in \cref{sec:QSO}, we verified the overlap between our sample of LRD candidates and other AGN catalogue derived using Q1 data. We give here more information on the criteria used in these other paper, their completeness, purity and level of overlap with our sample.

\subsection{Optical photometric selections}

The first selection presented in \citet{Q1-SP027}, named B24A, is based only on two \Euclid colours and was presented in \citet{EP-Bisigello}. Taking into account the limitations of this colour selection, since it has both a low purity ($P=0.166\pm0.015$) and a low completeness ($C = 0.347 \pm 0.004$), 254 (8\%) of our LRD candidates could be classified as QSO candidates. The selection is indeed based on the $\IE-\JE$ and $\IE-\YE$ colours, so it could trace the blue UV rest-frame slope, which is present in both LRDs and blue QSOs, but missing the red optical rest-frame slope. 

The presence of observations in the $u$ band in the EDF-N allows us to use an additional colour selection from \citet{EP-Bisigello}, applied again by \citet{Q1-SP027}. The selection, which we refer to as B24B, is based on the $u-z$ and $\IE-\HE$ colours, corresponding to a completeness of $C=0.861\pm0.004$ and a purity $P=0.992\pm0.017$. Of the entire sample of LRD candidates in the EDF-N, 258 (26\%) are selected by the B24B selection.

In addition, \citet{Q1-SP027} identified two new colour selections fine-tuned based on the colours occupied by DESI QSO. A first selection is based on \Euclid colours ($\IE-\YE$ and $\JE-\HE$) and a compactness criteria, while the second include also ground-based ancillary colours ($g-z$ and $\IE-\HE$). We refer to this selections as MZ25a and MZ25b, respectively. The first selection incudes 1686 QSO candidates, corresponding to 50\% of our catalogue, while the second criteria selects 1939 QSO candidates, corresponding to 58\% of our catalogue.  

The official \Euclid catalogue \citep{Q1-TP005} includes also a probability of being a QSO, obtained by performing a supervised machine learning method called \texttt{Probabilistic Random Forest} \citep[PRF,][]{Reis_2019_2019AJ....157...16R}, trained using photometric data only. Using this classification, \citet{Q1-SP027} identified QSO as objects having a QSO probability above 85\% and not being classified as stars. Of our sample, 34 (1\%) LRD candidates are also classified as QSO using this method.

None of our LRD candidates are in the purified subsets of the \textit{Gaia} QSO candidates catalogue \citep[see Sect. 3.2.4. in][]{Q1-SP027} based on the Data Release 3 \citep[DR3,][]{GaiaCollaboration_2023_2023A&A...674A...1G} or in the machine-learning classification based on \Euclid images by \citet{Q1-SP015}.

\subsection{Near-IR photometric selections}

Taking advantage of the WISE coverage, \citet{Q1-SP027} also applied the two photometric selections by \citet{Assef2018}. With the selection corresponding to 75\% completeness, called A18C75, 35 (1\%) of the LRD galaxies are identified as potential QSOs. With the selection corresponding to 90\% reliability, called A18R90, only four LRD candidates are also QSO candidates.

In addition, given that IRAC observations are available for at least part of the sample, we verify if any of our LRD candidates satisfy the AGN selection by \citet{Donley2012}. Unfortunately, in the EDF-S this selection can not be applied, as there are observations only on two IRAC bands. In the other two fields, only two LRD candidates has a {\rm S/N>3} in all four IRAC filters and these objects are inside the selection criteria by \citet{Donley2012}.

\subsection{Spectroscopic selections}

The EDF-N is partially covered by DESI EDR \citep{DESI_EDR}, allowing for the classification of QSOs using spectroscopic data. This classification is also included in \citet[][Sect. 3.3.1]{Q1-SP027} and two LRD candidates (0.2\% of our sample in the EDF-N) are indeed classified as QSOs both by the DESI spectral type classification and looking at the DESI spectra, having $\rm FWHM \geq 1200\, \rm km\,s^{-1}$ in one of the hydrogen lines. \citet{Q1-SP027} also includes other diagnostic based on spectroscopic DESI spectra, but no LRD candidates are selected by them. 

\subsection{X-ray AGN}
We now compare our catalogue with the \Euclid X-ray selected AGN catalogue \citep{Q1-SP003}. This catalogue lists the most likely \Euclid counterparts to the 4XMM DR13, \textit{Chandra} CSC 2.0, and eROSITA DR1 catalogue. While the former two cover the three \Euclid field in a few, deep pointed observations, the latter covers EDF-F and EDF-S homogeneously but at a shallow depth \citep[see figures 1 and 2 of][]{Q1-SP003}. 

Out of the entire sample of LRD candidates, three sources (0.09\%) are present in the \Euclid X-ray selected AGN catalogue, two in the EDF-S and one in the EDF-F. The two sources in the EDF-S are matched with sources in the eROSITA DR1, with a probability of being the right counterparts of $p_{\rm any}=0.16$ and 0.97. The source in the EDF-F is matched with a X-ray source in the \textit{Chandra} CSC 2.0 catalogue with a probability of being the right counterparts of $p_{\rm any}=0.98$. For these sources the chance association can be higher than 80\% \citep[see Fig. 7 in][]{Q1-SP003}. Out of the three matches, only one has ${\rm S/N}>3$ at (0.5--2.3)\,keV, that is $f_{0.5-2.3\,\rm kev}=(3.92\pm1.08)\times10^{-14}\,\rm erg\,s^{-1}\,cm^{-2}$, corresponding to a luminosity of $2.61\times10^{44}\,\rm erg\,s^{-1}$ at of $z_{\rm phot}=1.1$. A second source has instead a ${\rm S/N}=2$ at (0.5--2.3)\,keV, that is $f_{0.5-2.3\,\rm kev}=(1.24\pm0.62)\times10^{-14}\,\rm erg\,s^{-1}\,cm^{-2}$, while the third one is below ${\rm S/N}<1$, with a $3\sigma$ upper limit of $f_{0.5-2.3\,\rm kev}=5.06\times10^{-16}\,\rm erg\,s^{-1}\,cm^{-2}$. 

\citet{Q1-SP003} also identify potential X-ray emitters, using a combination of Bayesian statistics and machine learning. This classification allows us to identify potential X-ray emitters outside the area covered by X-ray observations, which may be useful, for example, for follow up studies. There are 73 LRD candidates that are classified as potential X-ray emitters. 

Overall, the small fraction of X-ray LRD candidates of this work is in line with previous works, showing that the majority of LRD are X-ray weak \citep[e.g.,][]{Yue2024}, but it is necessary to consider that the X-ray surveys considered here are shallower than the one used to matched previous LRDs selected with JWST. Further analysis of the X-ray emission of these sources in the future could shed light on the AGN contribution in LRDs.

 A reassuring result is that all X-ray LRDs have a probability of less than 1\% of being Galactic sources. This probability was derived by \citet{Q1-SP003} using a random forest algorithm trained on the methodology described in \citet{Salvato2022}.

\end{appendix}
\end{document}